\documentclass[reprint,showpacs,aps,pre]{revtex4-1}

\usepackage{graphicx}

\usepackage{amsmath}
\usepackage{bm}
\usepackage{gensymb} 

\usepackage[T1]{fontenc} 
\usepackage{lmodern}
\usepackage{color}

\begin{document}

\title{The Physics of Biofilms -- An Introduction}
\author{Marco G. Mazza}
\email{marco.mazza@ds.mpg.de}
\affiliation{Max Planck Institute for Dynamics and Self-Organization, Am Fa{\ss}berg 17,
37077 G\"ottingen, Germany}

\date{\today}

\begin{abstract}
Biofilms are complex, self-organized consortia of microorganisms that produce a functional, protective matrix of biomolecules.  Physically, the structure of a biofilm can be described as an   entangled polymer network which grows and changes
under the effect of gradients of nutrients, cell differentiation, quorum sensing, bacterial motion, and
interaction with the environment. Its development is complex, and constantly adapting to environmental
stimuli.  Here, we review the fundamental physical processes the govern the inception, growth and development of a biofilm. Two important mechanisms  guide
the initial phase in a biofilm life cycle: (\emph{i}) the cell motility near or at a solid interface, and (\emph{ii}) the cellular adhesion. Both processes are crucial for initiating the colony and for ensuring its stability.
A mature biofilm behaves as a viscoelastic fluid with a complex, history-dependent dynamics. We discuss progress and challenges in the determination of its physical properties. Experimental and theoretical methods are now available that aim at integrating the biofilm's hierarchy of interactions, and the heterogeneity of composition and spatial structures.  
We also discuss important directions in which future work should be directed.
\end{abstract}
\maketitle

\section{Introduction}
\label{sec:Introduction}

Biofilms are among the most ancient evidence of life on Earth: they appear as fossils of microbial mats and stromatolites from western Australia dating back to about 3.5 billion years ago~\cite{noffkeAstrobiology2013,VanKranendonkPreRea2008}. Thus, it should come as no surprise that biofilms are the most widespread form of life~\cite{flemmingNatRM2010}. They form extremely diverse and complex structures, are capable of colonizing almost every environment, and have evolved a vast arsenal of biological responses to environmental stimuli. 
Today we know that all three domains of life can produce biofilms: bacteria~\cite{hallNatRM2004}, archaea~\cite{orellARMB2013}, microalgae~\cite{Leadbeater1992} and  fungi~\cite{fanningPLOSPATHOG2012,ramageCRMB2009} all produce biofilms. But the understanding that most microorganisms live in aggregates rather than in a planktonic state has only recently emerged. Antonie van Leeuwenhoek observed ``\emph{animalcules}'' in the plaque of his own teeth under a microscope of his  fabrication. His report in $1676$ marks the discovery of biofilms~\cite{porterBacterRev1976}. However, only in the first half of the twentieth century with the work of Henrici~\cite{henriciJBacter1933} and Zobel~\cite{zobellJBacter1943} was the import of biofilms really appreciated by the scientific community, and finally the word biofilm was first used in 1981~\cite{mccoyCanJMB1981}, which marks the recognition of its function and organization.

Biofilms are structured, self-organized communities of microorganisms  that synthesize a protective matrix and adhere to each other and/or to an interface~\cite{otooleARMB2000,vertPAChem2012}. They can be populated by a single species but more often multiple species are present.
The level of complexity and specialization present in biofilms brings to mind the analogy to a large, bustling city~\cite{watnickJBacter2000} with different levels of organization~\cite{berkScience2012}. Biofilms can grow on solid surfaces in the presence of water virtually anywhere, for example on pebbles in a river bed (the periphyton), in deep-sea hydrothermal systems~\cite{reysenbachTrendsMB2001} and vents~\cite{taylorAEMB1999}, or on boat hulls; biofilms can also grow at the air-water interface of limnic or marine environments~\cite{wotton2005surface}.  Like the inhabitants of a city, the microorganisms in a biofilm can modify their surroundings. Cellular metabolism can either increase the porosity of geologic media by the dissolution of minerals, or reduce the porosity by precipitating secondary minerals or by clogging the pores with biomass. Microbial activity can change the electrochemical properties, surface roughness, elastic moduli, stiffness, as well as seismic and magnetic properties of minerals through the precipitation of bacterial magnetosomes. In general, biofilms play a role in so many geophysical processes that a new discipline is now emerging: biogeophysics~\cite{atekwanaRGeophysics2009}.

 Biofilms play a role also in human activities. Their growth has an enormous impact in biomedical sciences~\cite{SihorkarPharmaReas2001}. Microbial ecosystems can grow on the surface of teeth and on open wounds; they can infest the respiratory mucous and the airways of patients with cystic fibrosis pneumonia. {\it Pseudomonas aeruginosa},  for example, lives normally (and harmlessly) on our skin but if it enters the blood circulation of immunocompromised individuals it can infect organs of the urinary and respiratory system, bones and joints. Medical devices such as catheters, prosthetic heart valves, cardiac pacemakers, skull implants, cerebrospinal fluid shunts and orthopedic devices can harbor 
pathogens~\cite{pavithraBiomedMaterials2008,hallNatRM2004}, and due to the intimate contact of these medical tools with the human body biofilms can grow and trigger virulent infections.

Biofilms cause billions of dollars in damage to metal pipes in the oil and gas industry, and metal structures in water treatment plants. Sulfate-reducing bacteria, for example members of the \emph{Acidithiobacillus} genus, transforms molecular hydrogen into hydrogen sulfide which, in turn, produces sulfuric acid that corrodes metal surfaces causing catastrophic failures~\cite{haoCREST1996,hamiltonAnnRevMB1985,muyzerNatRevMB2008}.

Biofilms have, however, also found useful employment in environmental biotechnology. They are used in a well-established technique for wastewater treatment~\cite{nicolellaJBiotech2000,duBiotechAdv2007,DeBeer2013}, or for \emph{in situ} immobilization of heavy metals in soil~\cite{dielsRevEnvSciTech2002,gaddGeoder2004}. Biofilms allow to process large volumes of liquids, and they naturally grow by converting organic materials. Furthermore, microorganisms (typically bacteria and fungi) can be used for microbial leaching, that is, as a way to extract metals from ores~\cite{sandApplMBBiotech1995,suzukiBiotechAdv2001}. Copper, uranium and gold are examples of metals commercially recovered by microorganisms~\cite{boseckerFEMSMBRev1997}.

\begin{figure*}
\centering
\includegraphics[width=1.\textwidth]{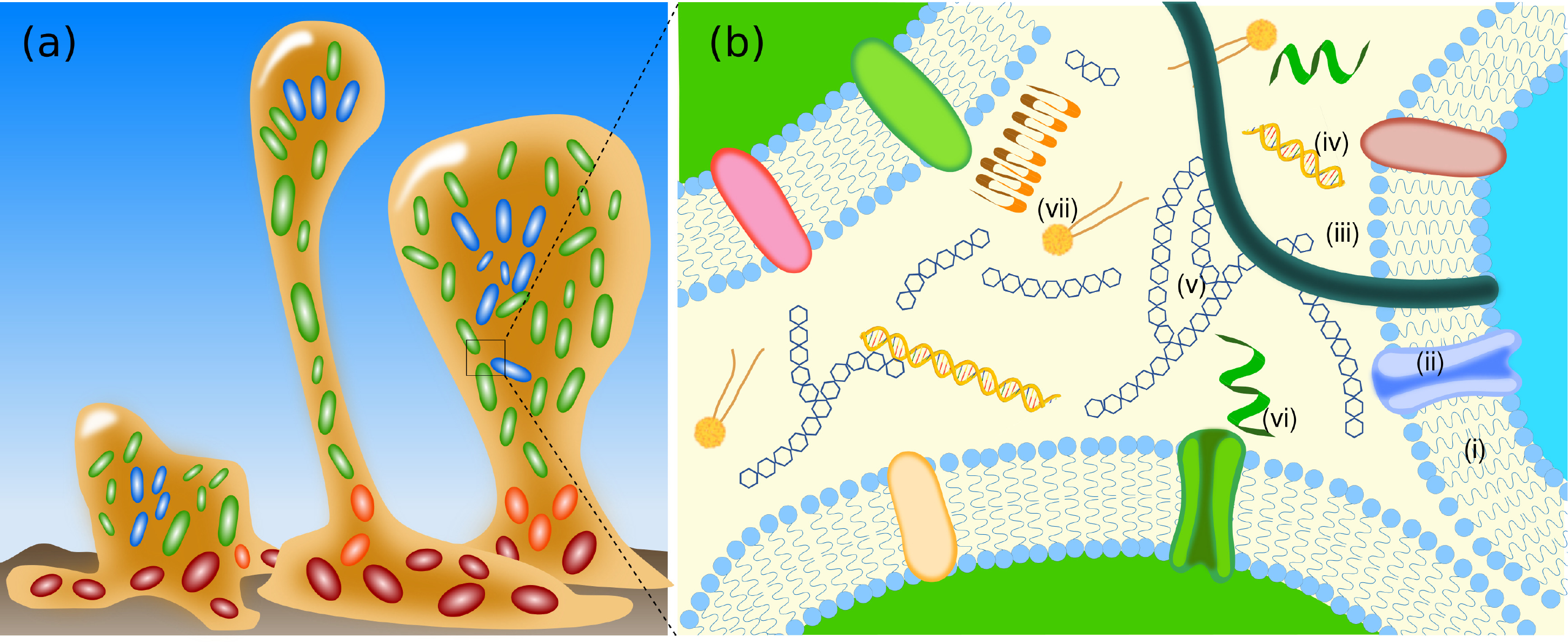}
\caption{(a) Pictorial representation of a biofilm produced by a community of microorganisms.   Different cellular shapes represent different species of microorganisms coexisting in the same biofilm, while different  colors represent phenotypic variations within the same species.
(b) Zoomed-in view of the main components of a biofilm, underlying the complexity of its physico-chemical properties: the extracellular polymeric substance and cells. 
We highlight (\emph{i}) the cellular membrane, (\emph{ii}) proteins embedded in the membrane, (\emph{iii}) pili, (\emph{iv}) extracellular DNA, (\emph{v}) polysaccharides, (\emph{vi}) proteins, (\emph{vii}) lipids. Inspired by a sketch in Ref.~\cite{flemmingNatRM2010}.}
\label{fig:biofilm-sketch}
\end{figure*}

Numerous experimental studies have determined that biofilms have a heterogeneous structure and composition~\cite{lawrenceJBact1991,caldwellJApplBact1993,costertonJBact1994,gjaltemaBiotechBioeng1994,stoodleyBiotechBioeng1999,stoodleyAnnRevMB2002,blauertBiotechBioeng2015} which  vary greatly among different species. But we also know that they share some fundamental elements.  Microorganisms colonizing an interface produce a mass of biopolymers, collectively termed \emph{extracellular polymeric substance} (EPS), that provides protection and structural stability to the cells. The EPS also performs numerous functions specific to the  biofilm state, such as creating its three-dimensional architecture, protection against physical, chemical and biological agents, providing an external digestive system, facilitating cell-cell signaling and horizontal gene transfer~\cite{madsenFEMSImmun2012}.

The EPS matrix  represents up to $90\%$ of the dry mass of the biofilm (the remaining being the cells) and is composed of polysaccharides, proteins, humic acids, DNA, lipids, and remnants of lysed cells~\cite{flemmingNatRM2010}.  (However, there are exceptions: some strains of \emph{Staphylococcus epidermidis} and \emph{Staphylococcus aureus} produce smaller amounts of EPS.) Figure~\ref{fig:biofilm-sketch} shows a sketch of a biofilm with its primary elements.  Below, we briefly describe these components and highlight their functions in the biofilm.

{\boldmath$\cdot$} \textsc{Polysaccharides} are long, linear or branched, polymeric chains that are ubiquitously found in biofilms, and represent their largest component~\cite{wingenderMEthEnzym2001,flemmingJBact2007}. 
The number and diversity of these molecules is stunning. Very little can be said in general about them as the number of chemical and structural possibilities is virtually infinite~\cite{christensenJBiotech1989,sutherland1977}. Gram-negative bacteria typically produce polyanionic polysaccharides, or  in some cases neutral ones, due to the presence of uronic acids or ketal-linked pyruvates; some chemical linkages ($1,3$- or $1,4$-${\beta}$-linked hexose) provide backbone rigidity to the polymers, while other linkages produce more flexible polymers~\cite{sutherlandMB2001}. 

Diverse  microscopy methods~\cite{wingender2012,lawrenceAEMB2003} show that polysaccharides form a complex network of fine strands  linking cells to each other and to the substratum. Polysaccharides are involved in most processes that take place in a biofilm~\cite{flemmingNatRM2010}: (\emph{i}) \emph{adhesion}; for example, the polysaccharides Pel and Psl (named after the operons involved in pellicle formation and polysaccharide synthesis locus, respectively~\cite{ryderCurrOpMB2007}) are essential for the adhesion of \emph{Pseudomonas aeruginosa} to a variety of substrata and they provide redundant mechanisms  for this task~\cite{colvinEnvirMB2012,maJBacter2006,cooleySoftMatt2013}. (\emph{ii}) \emph{Structural stability and cohesion}; the polysaccharides of algae form rigid structures because the binding (chelation) of Ca$^{2+}$ or Sr$^{2+}$ ions forms cross-linked polymer networks~\cite{donlanEID2002,sutherlandMB2001}.  Because bacterial polysaccharides are often acetylated, the interaction between polymers is reduced and so is  the resulting network-forming ability~\cite{sutherlandMB2001}; for example, {\it Pseudomonas fluorescens} requires an acetylated form of cellulose for the effective colonization of the air-liquid interface~\cite{spiersMolMB2003}, and N-acetylglucosamines are crucial for intercellular adhesion in the biofilm of \emph{Staphylococcus epidermidis}~\cite{gotzMolMB2002,mackJBacter1996}, a major cause of nosocomial infections. (\emph{iii}) \emph{Hydration}; polysaccharides can bind water molecules (for example, hyaluronic acid can bind the considerable amount of $1$~kg of water per $1$ g of saccharide~\cite{sutherlandMB2001}; however, most molecules will probably bind less). Because the polysaccharides form an  entangled
 polymer network they are subject to an osmotic pressure $\Pi$ such that will swell with water until the shear modulus of the network $G\approx\Pi$~\cite{seminaraPNAS2012,wilkingMRS2011,rubinstein2003}. A hydraulic decoupling between unsaturated soil bacteria  and the vadose zone has been hypothesized to protect the cells from cycles of wetting and drying events~\cite{orVadose2007}. (\emph{iv}) \emph{Storage of nutrients};  both the polysaccharide network itself and absorbed organic matter serve as a source of carbon during periods of nutrient deprivation~\cite{freemanLimnolOcean1995}. (\emph{v}) \emph{Protection from toxic ions}; the exopolysaccharides can adsorb toxic ions such as Cd, Zn, Pb, Cu, and Sr~\cite{norbergBiotechBioeng1984,wingender2012}; for example, \emph{Pseudomonas aeruginosa} in a biofilm is up to $600$ times more resistant to Zn, CU and Pb than in the planktonic state~\cite{teitzelApplEnvMB2003}. (\emph{vi}) \emph{Antibiotic protection};  \emph{Pseudomonas aeruginosa} infecting cystic fibrosis patients can change to a mucoid phenotype characterized by increased production of  the polysaccharide alginate, but the biofilm contains also Pel and Psl. All three components have been found to have a protective role against antibiotics~\cite{hentzer2001alginate,colvinPLOS2011,billingsPLOS2013}.
(\emph{vii}) \emph{Reservoir of enzymes}; enzymes are stored, accumulated and stabilized by the polysaccharide network~\cite{wingender1999}; for example, it has been suggested~\cite{lockOikos1984} that epilithic biofilms of microorganisms growing on river beds (a highly changeable environment) benefit from the accumulation of enzymes because, firstly,  the products of enzymatic activity are readily available to the cells, secondly, they may additionally trigger new enzymatic activity in adjacent cells, and, thirdly,  new generations do not need to spend energy on enzyme synthesis. 
(\emph{viii}) \emph{Sink for excess carbon}; when supplied with excess C, the production of polysaccharides is greatly enhanced~\cite{oteroJPhyco2004}

\begin{figure*}
\centering
\includegraphics[width=0.45\textwidth, height=0.278\textwidth]{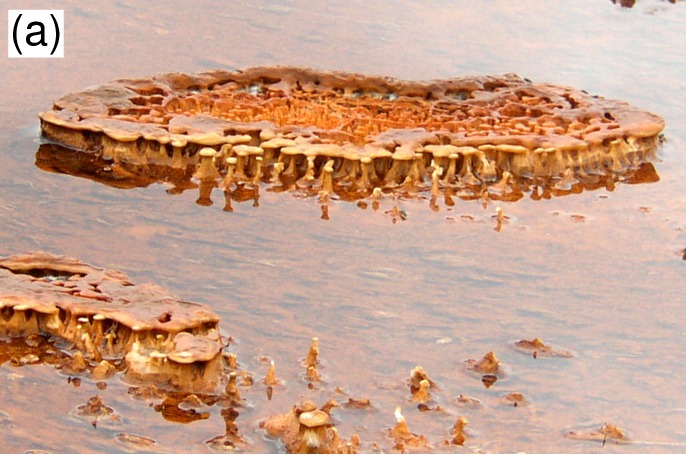}
\includegraphics[width=0.45\textwidth, height=0.278\textwidth]{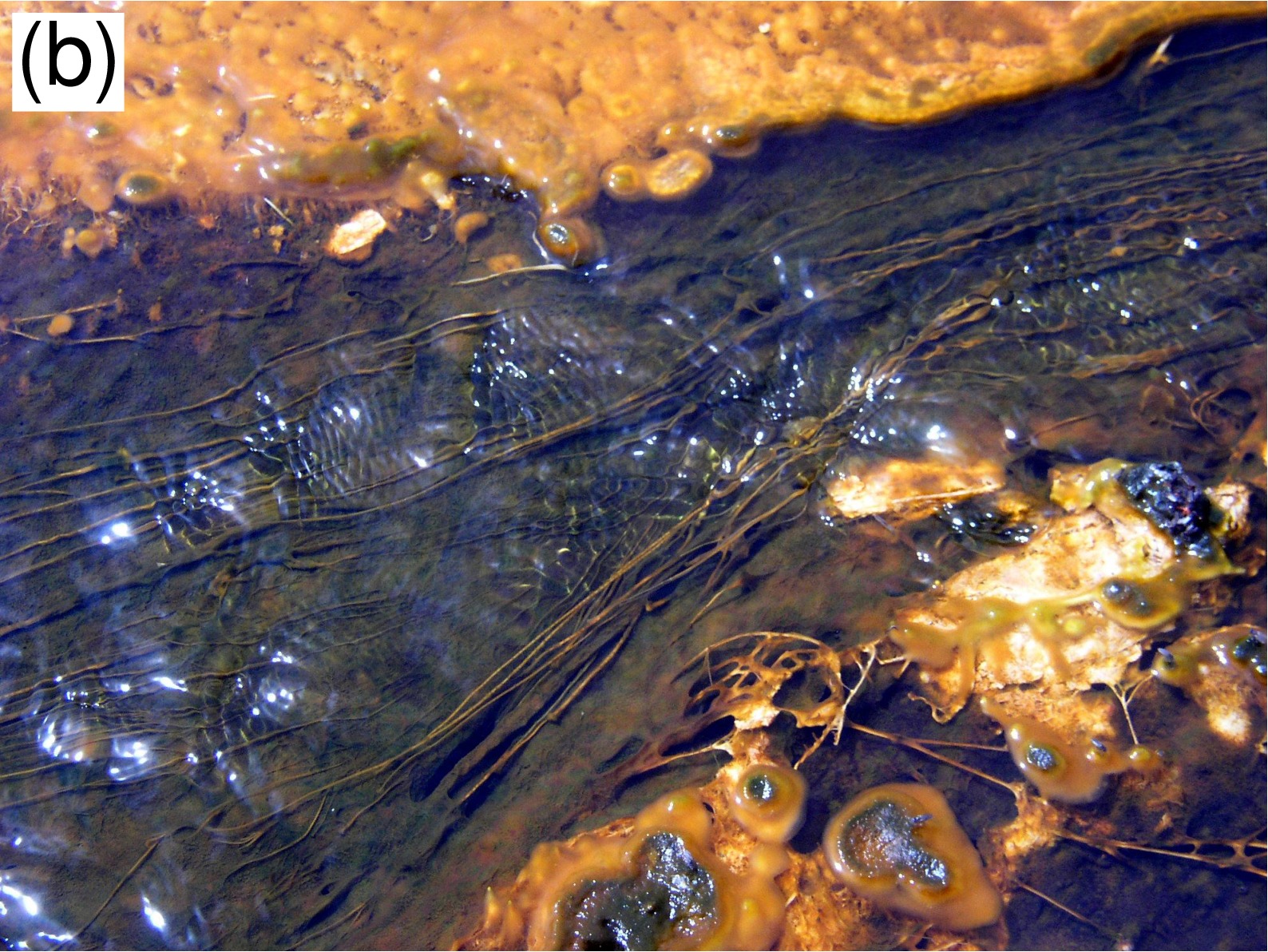}
\includegraphics[width=0.45\textwidth, height=0.278\textwidth]{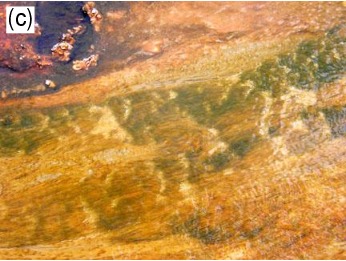}
\includegraphics[width=0.45\textwidth, height=0.278\textwidth]{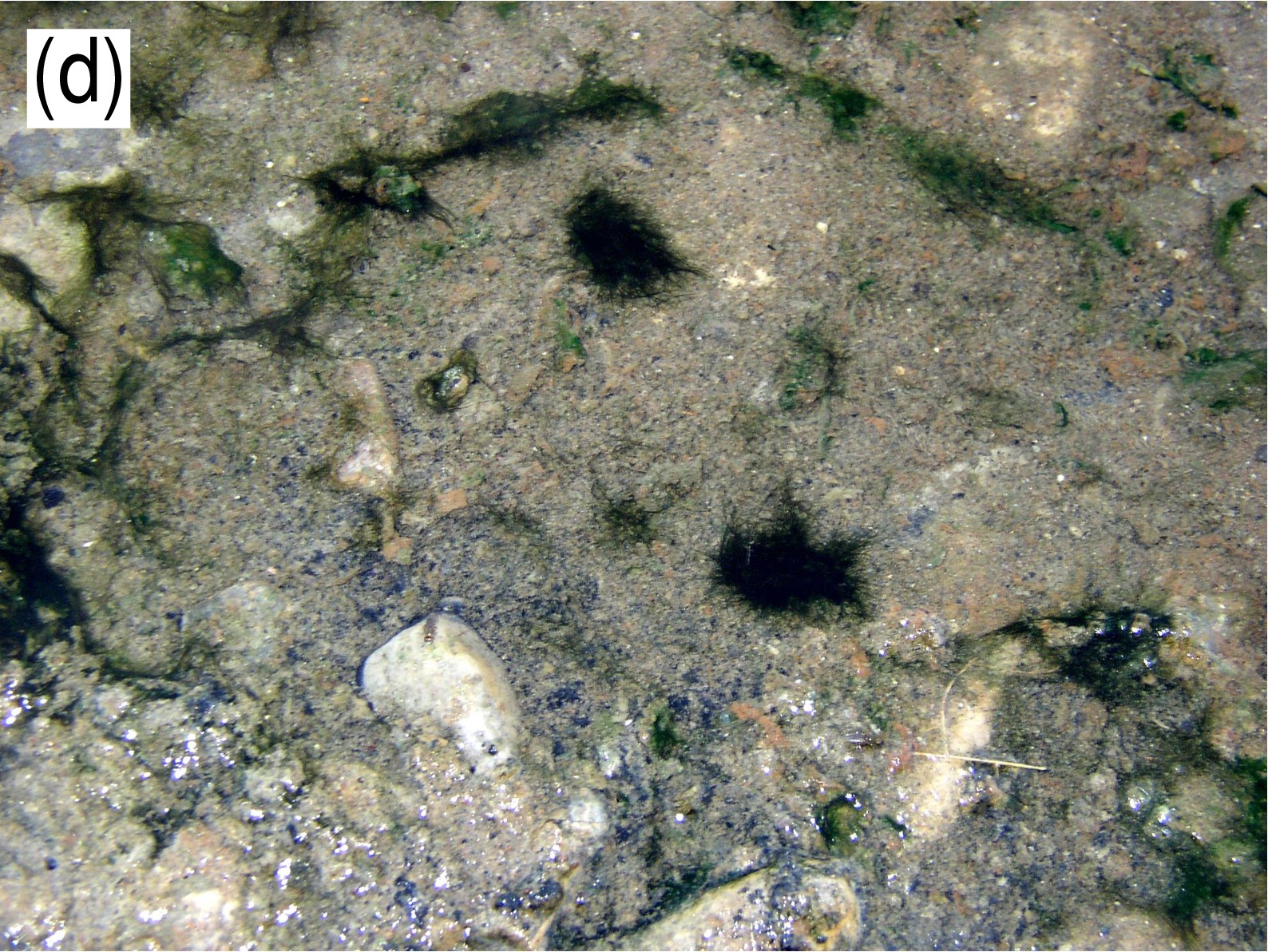}
\includegraphics[width=0.45\textwidth, height=0.278\textwidth]{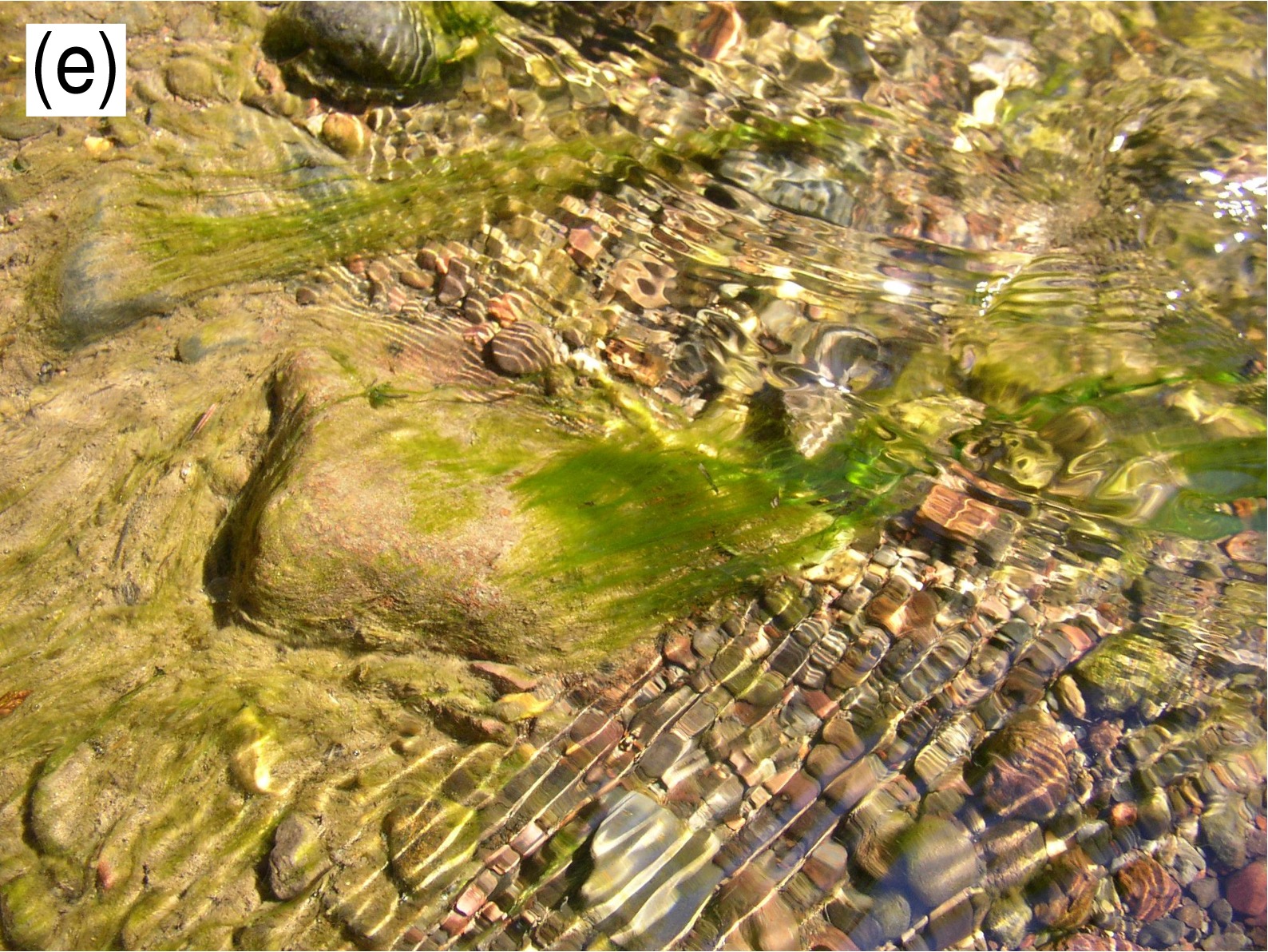}
\includegraphics[width=0.45\textwidth, height=0.278\textwidth]{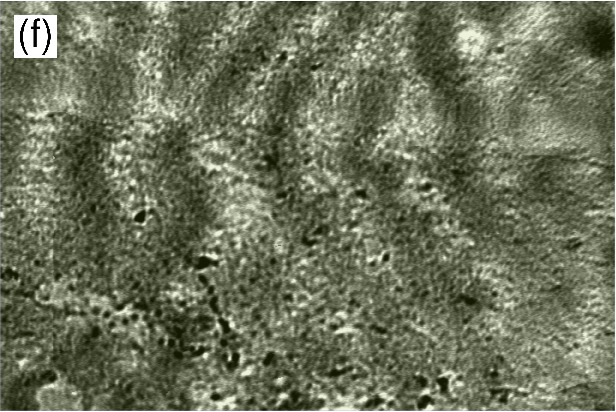}
\includegraphics[width=0.8\textwidth,trim={0 7cm 0 6cm},clip]{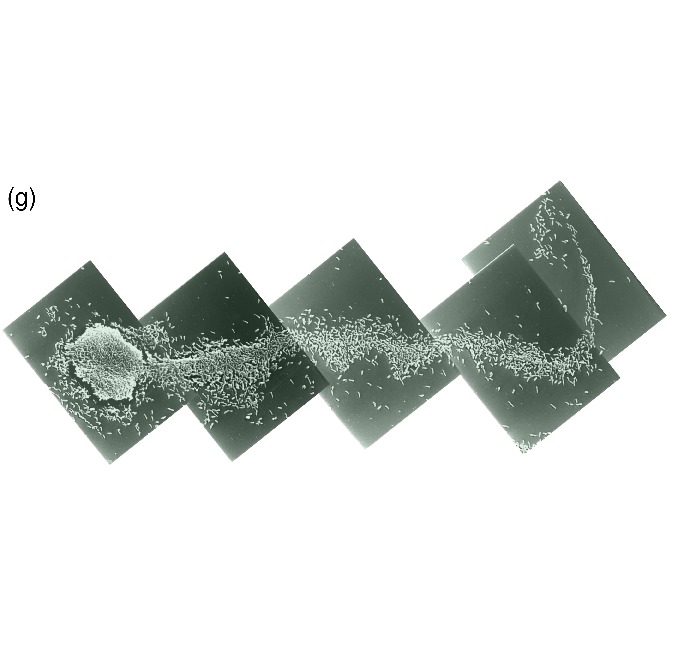}
\caption{Examples of biofilms growing in different environments. (a-c) Biofilms growing in hydrothermal hot springs from the Biscuit Basin thermal area, Yellowstone National Park, USA; (d) Gardener River, Yellowstone National Park, USA; (e) Hyalite Creek, Bozeman, Montana, USA; (f-g) \emph{Pseudomonas aeruginosa} PAN067 biofilm grown in a flow cell
with a flow of $1$~m/s. Biofilms growing in stagnant waters tend to form round, mushroom-like structures, as in (a) and (d), while biofilms growing under hydrodynamic flow form ripples, as in (c) and (f), and streamers, as in  (b), (e) and (g). 
Images courtesy of Dr.~P.~Stoodley. Reprinted with permission from Ref.~\cite{hallNatRM2004}.}
\label{fig:biofilm-examples}
\end{figure*}

{\boldmath$\cdot$} \textsc{Extracellular DNA} has been found not to be  merely debris left from lysed cells but an actively produced component of the EPS that plays a crucial role in the formation and structural integrity of the biofilms of \emph{Variovorax paradoxus} and \emph{Rhodococcus erythropolis}~\cite{steinbergerApplEnvMB2005}. 
 Extracellular DNA functions as an intercellular connector in \emph{Pseudomonas aeruginosa}'s biofilms~\cite{yangMB2007}; also, if these biofilms are younger than $60$ hours they can be dissolved by enzymatic degradation of the extracellular DNA~\cite{whitchurchScience2002}. The Gram-positive \emph{Bacillus cereus} employs extracellular DNA as an adhesin~\cite{vilainApplEnvMB2009}. Recently, a novel function of extracellular DNA has been recognized: antimicrobial activity. By binding and sequestering cations, extracellular DNA induces physical alterations in the bacterial outer membrane~\cite{mulcahyPLOSPath2008}.

{\boldmath$\cdot$} \textsc{Lipids} or \textsc{biosurfactants} modify the hydrophobic character of microbial cells and therefore can modify their ease of adhesion to surfaces. For example, lipopeptides produced by \emph{Bacillus subtilis} change the hydrophobicity of the cells depending on the initial level of hydrophobicity~\cite{ahimouEnzMBTech2000}. These functions are also used to protect the biofilm from invading microorganisms. In the late stages of its biofilm, \emph{Pseudomonas aeruginosa} produces rhamnolipids that can disrupt biofilms formed by \emph{Bordetella bronchiseptica}~\cite{irie2005pseudomonas} or inhibit the attachment of a variety of bacteria and yeasts~\cite{rodriguesJAMB2006}. \emph{Bacillus subtilis} and \emph{licheniformis} produce biorsurfactants that selectively prevent the adhesion of pathogenic cells~\cite{RivardoApplMBBiotech2009}. 

{\boldmath$\cdot$} \textsc{Proteins} are a large component of the EPS and have diverse functions. As enzymes they form an external digestive system for the microorganisms in the biofilm as they break down biopolymers to low-molecular-mass products that can be utilized by the cells~\cite{flemmingNatRM2010}. Enzymes can degrade the EPS matrix~\cite{romaniMBEcol2008,zhangChemosph2003} or protect against oxidizers~\cite{elkinsApplEnvironMB1999}.

  Finally, non-enzymatic proteins contribute to the structural stability of the EPS network by attaching the cellular surface to the polysaccharides. Lectins are one such class~\cite{barondesScience1984}. For example, lectins contribute to thicker biofilms in different environmental conditions in \emph{Pseudomonas aeruginosa}~\cite{diggleEnvMB2006}; the presence of a plant lectin can greatly influence the biofilm of a rhizobium~\cite{perezIntJMB2009}.

 In addition to their composition, also the temporal development of biofilms is structured.  Five stages have been recognized in the life-cycle of a biofilm~\cite{stoodleyAnnRevMB2002}: (\emph{i}) the initial attachment of cells to a surface or substratum; (\emph{ii}) production of EPS and irreversible attachment; (\emph{iii}) development of the biofilm architecture; (\emph{iv}) maturation of the structure and composition of the biofilm; (\emph{v}) dispersal of individual cells from the biofilm that revert to the planktonic phenotype to find a new niche and that close in this way the life-cycle of the biofilm.
During dispersal the EPS matrix immobilizing the cells is broken down. This last stage of the biofilm's life-cycle allows the colonization of new interfaces and it may be necessary when the microenvironment becomes  unfavorable to  sessile microorganisms. Dispersal is a highly complex process~\cite{karatanMBMolBiolRev2009,donlanEID2002} involving the production of enzymes to degrade the EPS network~\cite{sauerJBact2004}, production of surfactants~\cite{bolesMolMB2005}, and induction of motility~\cite{jacksonJBact2002}. 
For more details regarding the important phase of dispersal we refer the reader to a useful review~\cite{mcdougaldNatRevMB2012}.

The possibility that the vast majority of microorganisms can form biofilms points to a strong evolutionary incentive for this cooperative behavior. Biofilms growing in different environments show strikingly similar  responses to similar physical stimuli. As Fig.~\ref{fig:biofilm-examples} shows, fast-moving waters produce similar morphologies called ``streamers'' in very different environments. Quiescent waters induce instead biofilms with more rounded morphologies~\cite{hallNatRM2004}. These facts point to convergent survival strategies within a biofilm.
Indeed, social interactions in microorganisms have been recognized~\cite{crespiTrendsEcolEvol2001}. The advantages of a biofilm mode of life consists not only of the protection from toxins, dehydration, antibiotics, starvation, but also of a synergetic consortium that enhances immensely the chances of survival by sharing resources and allowing differentiation, for example. 

For all the reasons mentioned above, it is not surprising that the study of biofilms has attracted enormous interest in microbiology, pharmacology, medicine, industry, and in more recent years also in physics. In fact, the formation, growth, organization and structure of a biofilm are all processes that need to be understood from a physical point of view. In this Review we explore a selection (far from being exhaustive) of key physical processes taking place in a biofilm. The first important step is the motility.  Different kinds of motilities occur in two phases of the biofilm life-cycle: stage (\emph{i}), the attachment, and stage  (\emph{v}), the dispersal. Next, the process of adhesion is decisive for the formation and growth of a stable biofilm. Finally,  the mature biofilm resulting from myriad interactions and history-dependent dynamical responses can be viewed, on a macroscopic level, as a biomaterial, that is susceptible of physical characterization.

The topic we address in this review has a multitude of aspects and spans such a broad range of disciplines (from biochemistry and genetics to hydrodynamics and rheology) that we cannot hope to give a complete depiction of the current knowledge about biofilms. However, we have made efforts to provide a rather general picture that at least points to the most eminent physical aspects. At times, this work might appear as pedagogical. This is intended. We hope in fact that experts in one discipline will find the discussions of other disciplines connected to biofilms helpful. 
For further study we direct to more detailed reviews, such as Ref.~\cite{shapiroAnnRevMB1998} for the concept of multicellularity, and Ref.~\cite{benAdvPhys2000} for the self-organization of microorganisms. One glaring aspect of biofilm that we neglect is the genetic and regulatory network of signals that controls the biofilm's formation, maturation and dispersal. For these topics we recommend the specialized reviews~\cite{otooleARMB2000,karatanMBMolBiolRev2009,kaplanJDentRes2010,mcdougaldNatRevMB2012}.

This work is organized as follows. In Sec.~\ref{sec:motility} we discuss various mechanisms of cellular motility, from the problem of swimming at low Reynolds numbers to surface motility. Section~\ref{sec:adhesion} describes the fundamental physical processes that allow cells to adhere to solid surfaces and thus initiate a biofilm. In Sec.~\ref{sec:viscoelastic} we discuss the viscoelastic behavior of biofilms as materials and the main physical concepts used in the literature. In Sec.~\ref{sec:experim} we describe modern advances  in the experimental investigation of biofilms propelled by the fields of microfluidics and nanofabrication.   In Sec.~\ref{sec:computer} we review the theoretical and computational models of biofilm growth and dynamics. Finally in Sec.~\ref{sec:concl} we summarize our conclusions.

\section{Motility}
\label{sec:motility}

 Why is motility relevant to biofilms? The first, obvious answer is that
the approach of microorganisms to a surface is the first step to the constitution of a biofilm.  This corresponds to stage (\emph{i}) in the life-cycle of a biofilm. For example,  prior to attachment and initiation of a biofilm \emph{Vibrio cholerae} scans a surface by swimming in its close proximity, using two strategies called roaming and orbiting, which differ in the radius of gyration of the trajectories~\cite{teschlerNRMB2015}. Swimming is also relevant in the last stage of a biofilm's life-cycle: the dispersal. Cells of \emph{Pseudomonas aeruginosa} swim away from the internal regions of the biofilm into the bulk liquid to regain access to fresh nutrients~\cite{sauerJBact2002}. It is often claimed that once formed biofilms are communities of sessile microorganisms. A new picture has emerged where motility, at least in some species, plays a significant role in the biofilm's structure formation~\cite{tolkerJBact2000,barkenEnvMB2008}, with spatial and temporal  heterogeneities in phenotypic differentiation and organization~\cite{tolkerJBact2000,vlamakisGeneDevel2008}. It must be said, however, that living organisms exhibit a stunning variety of behaviors that refuse to be easily categorized. In fact, non-swimming species, such as \emph{Staphylococcus
epidermidis} and \emph{Staphylococcus
aureus}  can also form biofilms. Thus, swimming motility is by all means not a necessary condition for biofilm formation. The inevitable generalizations sometimes present in this work should be understood from a physical point of view and not biological: that the physical processes discussed below have been shown to take place in some species does not imply that all biofilm-forming microorganisms display the same characteristics, but rather that they belong to the ``toolbox'' of physical mechanisms available to some species and/or phenotypes, and, as our knowledge of biofilms expands, one should consider that these physical processes might be at play.

\begin{figure}
\centering
\includegraphics[width=1.05\columnwidth]{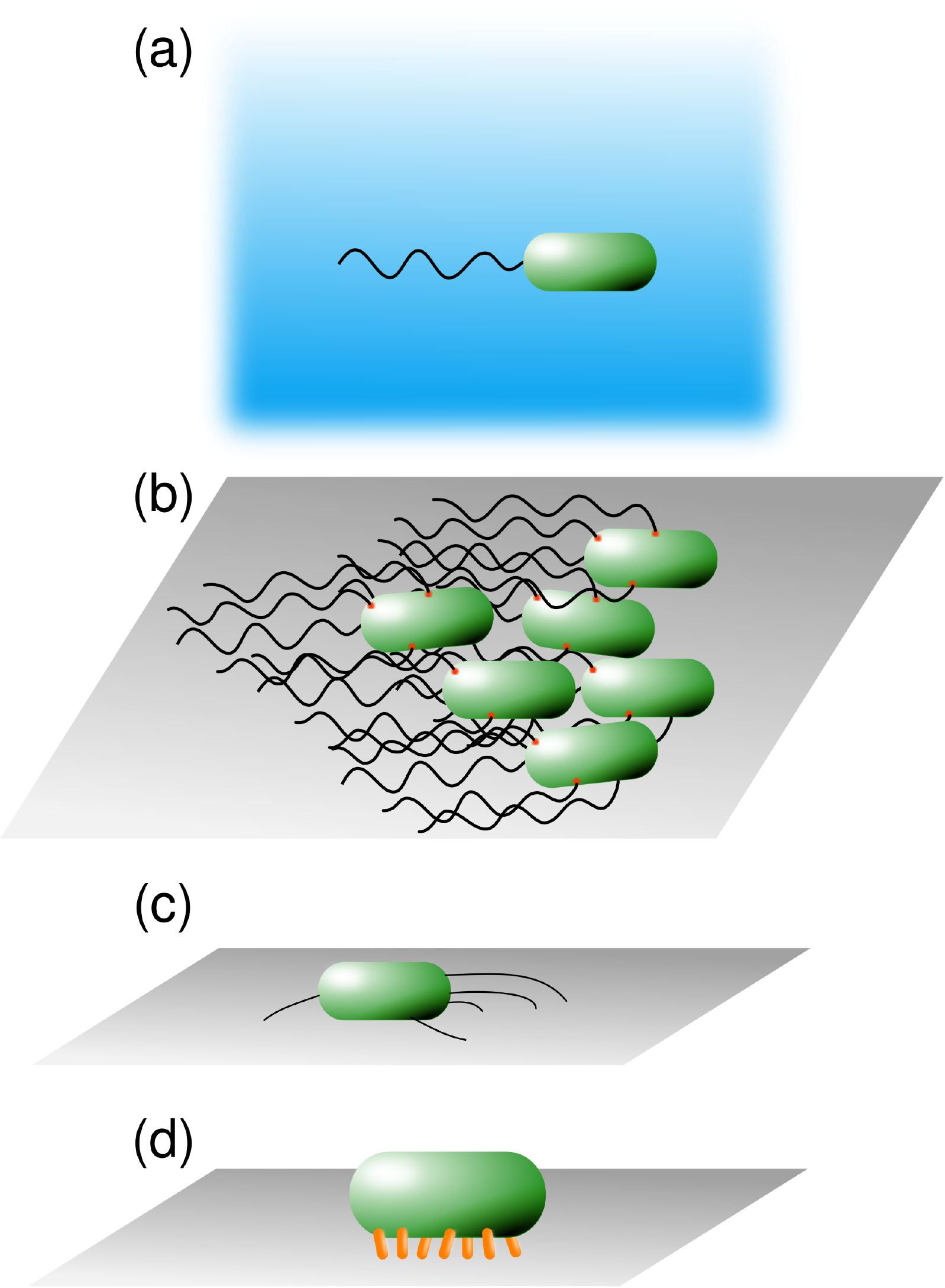}
\caption{Pictorial representation of different motility modes of microorganisms. (a) Swimming in a bulk fluid by means of single polar flagellum; (b) bacteria in the hyperflagellated state swarming on a solid surface; (c) twitching by means of type IV pili; (d) gliding by means of focal adhesion complexes.}
\label{fig:motility-sketch}
\end{figure}

Six different kinds of motility have been identified among $40$ bacterial species belonging to $18$ different genera~\cite{henrichsenBactRev1972}: swimming, swarming, gliding, twitching, sliding and darting (see Fig.~\ref{fig:motility-sketch} for a sketch of the most important motility modes). Except for swimming, all other modes are associated to the presence of a surface. Swimming at or near a surface is actuated through one or more flagella. Let us consider one of the most studied motile bacteria: \emph{Escherichia coli}. It exhibits on average six flagella, $5$ to $10\,\mu$m long and $20$ nm thick, randomly distributed around its body. The flagellum is composed of a basal body (embedded in the cellular membrane), a hook and a filament, that is, a left-handed helix with a pitch of about $2.5\,\mu$m and a diameter of about $0.5\,\mu$m~\cite{turnerJBact2000}. When the flagellar motor, driven by the proton motive force, rotates counterclockwise (CCW) a helical wave travels along the filament. When all flagella rotate CCW mechanical and hydrodynamical forces group the flagella in a single bundle that propels the bacterium in roughly straight trajectories with speeds up to $40\,\mu$m/s. While the bundle rotates CCW at about $100$~Hz the cell body rotates clockwise (CW) at about $25$~Hz~\cite{bergCurrBiol2008} because of conservation of angular momentum. When one or more flagellar motors rotate CW those filaments undergo a polymorphic transition to a right-handed helix. The bundle of flagella divides and a random reorientational motion called ``tumbling phase'' results. The tumbling phase allows the cells to change swimming direction, and it is regulated by the chemotactic signal transduction network~\cite{parkinsonCell1993}.

\emph{Escherichia coli}'s flagellum has a torsional spring constant $k_\theta\approx4\times10^{-12}$ dyne cm rad$^{-1}$~\cite{blockNature1989}, but only up to twist angles of $100\degree$; for larger twist angles a second regime appears with an order of magnitude stiffer response~\cite{blockNature1989}. One possible explanation for these two regimes of the flagellar compliance is that the hook exhibit a soft component, which dominates the initial phase of the twist, and the filament has a stiffer compliance, which dominates larger twist angles. This mechanism has the advantage of removing discontinuities in the motion of the motor, whose dynamics is intrinsically stochastic because it is powered by the passage of protons, which is a Poisson process. Hence the waiting times between events is an exponentially distributed random quantity~\cite{bergAnnRevBiochem2003}.

A microorganism cannot swim like a fish for a simple physical reason: the viscous drag dominates the motion~\cite{laugaRepProgPhys2009}. Motile cells must instead adopt a different strategy that uses drag forces to their advantage. In the next section we discuss the hydrodynamic fundamentals of bodies swimming in a viscous fluid. This is a fascinatingly vast topic. For more details we recommend specialized works on the general problem of swimming  ~\cite{laugaRepProgPhys2009,kochARFM2011,happel2012}, or swarming~\cite{copelandSoftMatt2009}.

\subsection{Swimming at low Reynolds numbers}\label{sec:lowReynolds}

We now discuss the general physical mechanism that produces propulsion in microorganisms that posses flagella or cilia. On general grounds, the motion of a body immersed in an aqueous environment, that is, a Newtonian fluid, is described by the Navier--Stokes equations
\begin{gather}\label{eq:NSmom}
\rho\left(\frac{\partial\vec{v}}{\partial t}+\vec{v}\cdot\nabla\vec{v}\right)=-\nabla p+\eta\nabla^2\vec{v}\,,\\
\label{eq:NSincompress}
\nabla\cdot\vec{v}=0\,,
\end{gather}
where $\rho$ is the fluid density, $\vec{v}$ the flow velocity, $p$ the pressure, $\eta$ the viscosity, $t$ is time, and ``$\cdot$'' is the symbol of scalar product.
Physically, Eq.~\eqref{eq:NSmom} represents conservation of linear momentum in the fluid, and Eq.~\eqref{eq:NSincompress} the condition of incompressibiliy of the fluid. The presence of a body immersed in the fluid is incorporated by the boundary conditions necessary to solve Eq.~\eqref{eq:NSmom}-\eqref{eq:NSincompress}. The no-slip boundary condition (valid for viscous fluids) states that at the boundary $S$ of the body $\vec{v}_{\mathrm{fluid}}(S)=\vec{v}_{\mathrm{solid}}(S)$. Solving the Navier--Stokes equations means obtaining expressions for $\vec{v}$ and $p$ that satisfy Eq.~\eqref{eq:NSmom}-\eqref{eq:NSincompress} and the boundary conditions.

From this knowledge, the stress tensor $\bm{\sigma}$ can be calculated. For a fluid the stress tensor can be decomposed as $\bm{\sigma}=-p\bm{\mathrm{I}}+\bm{\tau}$, where $\bm{\mathrm{I}}$ is the unit tensor, that is, into a term corresponding to the pressure stresses, which are isotropic, and a second term called \emph{deviatoric} stress tensor $\bm{\tau}$ that includes viscous (shear) or other stresses. For a Newtonian fluid the deviatoric stress depends linearly on the instantaneous values of the velocity gradient, so that one can write 
\begin{equation}\label{eq:stressNewton}
\bm{\sigma}=-p\bm{\mathrm{I}}+\eta\left[\nabla\vec{v}+\nabla\vec{v}^{\,\mathrm{T}}\right]\,,
\end{equation}
where the superscript `T' indicates matrix transposition. Once $\bm{\sigma}$ is found, the total hydrodynamic force $\vec{F}$ and torque $\vec{\tau}$ acting on the swimming body can be derived from integrals over its surface
\begin{align}\label{eq:force_torque}
\vec{F}=\int_S\,\bm{\sigma}\cdot \vec{n}\, \mathrm{d}S\,, \quad\quad  \vec{\tau}=\int_S\,\vec{r}\times\bm{\sigma}\cdot\vec{n}\, \mathrm{d}S 
\end{align}
where $\vec{n}$ is a unit vector normal to the surface.

Equation~\eqref{eq:NSmom} represent a balance of inertial forces (the left-hand side) and viscous forces (the term $\eta\nabla^2\vec{v}$), thus it is natural to find a way to quantify the relative importance of inertial to viscous forces. The Reynolds number $\mathcal{R}=\rho\,UL/\eta$, where $U$ is the typical velocity and $L$ the characteristic length scale of the swimmer, gives a dimensionless measure of this balance. For a typical microorganism such as \emph{Escherichia coli} ($L\approx2\,\mu$m) swimming in water ($\rho\approx10^3$ kg/m$^3$, $\eta\approx10^{-3}$ Pa$\cdot$s) with a typical speed $U\approx20\,\mu$m/s the Reynolds number $\mathcal{R}\approx10^{-5}$. It is then clear that to study the swimming of microorganisms we can neglect the inertial term in Eq.~\eqref{eq:NSmom}. In this limit the flow obeys the Stokes equations
\begin{align}\label{eq:Stokes}
\nabla p=\eta\nabla^2\vec{v}\,, \quad\quad  \nabla\cdot\vec{v}=0\,.
\end{align}
Because the Stokes equation is linear and because of Eq.~\eqref{eq:force_torque} the forces acting on the swimming body are directly proportional to the flow velocities~\cite{guyonBookhydrodyn}. Because the displacement of a solid body can be decomposed at any instant of time as the superposition of a translation with velocity $\vec{V}(t)$ and a rotation with angular velocity $\vec{\Omega}(t)$ (this is the Mozzi--Chasles theorem), the instantaneous local velocity of a point can be written as 
\begin{equation}
\vec{v}=\vec{V}+\vec{\Omega}\times\vec{r}\,.
\end{equation}
The hydrodynamic force and torque then obey the equations
\begin{gather}\label{eq:stokes_force}
F_i=-\eta (A_{ij}V_j+B_{ij}\Omega_j)\,,\\
\tau_i=-\eta (C_{ij}V_j+D_{ij}\Omega_j)\,.
\end{gather}
Physically, the matrices $A_{ij}$ and $D_{ij}$ represent the coupling of forces to translations, and of torques to rotations, respectively, while the matrices $B_{ij}$ and $C_{ij}$ represent cross-term couplings between forces and rotations, and between torques and rotations. Dimensionally, the coefficients of $A_{ij}$ have dimensions of length, $B_{ij}$ and $C_{ij}$ of an area, and $D_{ij}$ of a volume. Because of very general arguments~\cite{guyonBookhydrodyn}, which do not depend on the shape of the moving body, these matrices obey the relations
\begin{align}
A_{ij}&=A_{ji}\,\,, & D_{ij}&=D_{ji}\,\,, & B_{ij}=C_{ji}\,\,,
\end{align}
thus, the matrices $B_{ij}$ and $C_{ij}$ are always the transpose of each other. This means that there is a reciprocity principle for the motion in viscous fluids: the force on a rotating body and the torque on a translating body have the same coupling coefficients. 

A moving body with the symmetry of a cube or a sphere will have $B_{ij}=0$, and $A_{ij}=\lambda\delta_{ij}$, where $\delta_{ij}$ is the unit matrix. Physically, the drag force on the body will be collinear with the direction of motion, $\vec{V}$. But, for a general shape, the matrix $A_{ij}$ will not be proportional to the unit matrix; thus, there will be drag anisotropy: $\vec{F}\cdot\vec{V}<0$, that is, force and velocity are not collinear and the sign indicates the energy loss due to viscous dissipation~\cite{guyonBookhydrodyn}. For example, for a prolate ellipsoid of revolution with one axis much larger than the other, $\ell_1\gg\ell_2$, the matrix $\mathbf{A}=A_\|\hat{e}\otimes\hat{e}+A_\perp(\mathbf{I}-\hat{e}\otimes\hat{e})$, with $A_\perp\approx2A_\|$, where $\hat{e}$ is the unit vector (versor) in the direction of the long axis of the ellipsoid, $\otimes$ the tensor product. Physically, this means that the drag force normal to the long axis is twice as large than along it.

Drag anisotropy is the key physical effect that makes locomotion possible at low Reynolds numbers.
Consider for example a thin filament, like a bacterial flagellum, deformed by a periodic wave-form. For simplicity we assume that the cell moves along the $x$ axis, and each point on the flagellum moves only in the normal direction to the $x$ axis. A short segment of the filament can be approximated by a straight, thin rod that experiences a viscous drag $\vec{f}=\vec{f}_\|+\vec{f}_\perp=-\xi_\| \vec{v}_\|-\xi_\perp\vec{v}_\perp$, where $\xi_\perp\approx2\xi_\|$, and $\vec{v}_\|$ and $\vec{v}_\perp$ are the components of the velocity parallel and perpendicular to the segment tangent, that is, $v_\|=v\cos\theta$, $v_\perp=v\sin\theta$. The drag forces have components along the $x$ axis: $f_{\|x}=f_\|\sin\theta$, $f_{\perp x}=f_\perp\cos\theta$. The total force along the x axis, or the propulsion force then results $f_{prop}=(\xi_\|-\xi_\perp)u\sin\theta\cos\theta$~\cite{laugaRepProgPhys2009}. Thus, a periodic change in shape and in the direction of beating can produce a net propulsive force in the direction perpendicular to the beating. Similar calculations show that a helical filament rotating about its axis generates a net propulsive force proportional to the rotational velocity~\cite{guyonBookhydrodyn}. 

The Stokes equation (Eq.~\eqref{eq:Stokes}) is linear and time-independent. An important consequence of this fact is that if the velocity of motion is reversed, the propulsive force also changes sign. Thus, a scallop moving at vanishing Reynolds numbers cannot have any net displacement. This is Purcell's famous \emph{scallop theorem}~\cite{purcellAJP1977}. Note that the theorem is valid only if the sequences of deformations of the moving body is time reversible, nor does it apply close to another surface.

Another consequence of a vanishing Reynolds number is that the dynamics is overdamped: the rate of change of momentum, that is, the acceleration is negligible. Newton's laws then become simple balances of forces and torques, $\vec{F}_{tot}=0$, $\vec{\tau}_{tot}=0$. Thus, the swimming problem of a body immersed in a viscous fluid must be described by two opposite forces, a force dipole. The flow field at position $\vec{r}$ created by a swimmer located at the origin of a coordinate system is
\begin{equation}\label{eq:veloc_field}
\vec{v}(\vec{r})=\frac{\ell f}{8\pi\eta r^2}\left[ 3(\hat{r}\cdot\hat{e})^2-1\right] \hat{r}\,,
\end{equation}
where $\hat{e}$ is the versor in the swimming direction, $r=\vert \vec{r}\vert$, $\hat{r}\equiv \vec{r}/{r}$, $\ell$ the dipole length, and $f$ its force. Depending on the sign of $f$ the swimmer is called ``pusher'' ($f>0$), such as \emph{Escherichia coli}, or ``puller'' ($f<0$), such as algae of the genus \emph{Chlamydomonas}, which swim with characteristic ``breast stroke'' movements of the flagella. 

If we now consider two microorganisms swimming in the same environment, each will be affected by the flow field generated by the other. This interaction transmitted by the fluid is called a hydrodynamic interaction.  Equation~\eqref{eq:veloc_field} has a number of consequences for the mutual interaction of two swimmers: (\emph{i}) two side by side pushers attract each other, while two side by side pullers repel each other; (\emph{ii}) two pushers aligned along the swimming direction repel, while two similarly aligned puller attract each other. Calculations up to the octupole order~\cite{baskaranPNAS2009} show that to leading order the hydrodynamic forces and torques between two swimmers moving in the directions $\hat{e}_1$, $\hat{e}_2$ exhibit a nematic symmetry, that is the interactions are invariant for changes $\hat{e}_1 \to -\hat{e}_1$, $\hat{e}_2 \to -\hat{e}_2$. 

Dense suspensions of swimmers are unstable to long wavelength fluctuations~\cite{SimhaPRL2002,SaintillanPRL2007}, leading to the so-called ``bacterial turbulence''~\cite{kesslerBookChap1997,DombrowskiPRL2004}. There is however an ongoing debate regarding the role of hydrodynamic interactions~\cite{sokolovPRL2012,dunkelPRL2013,zottlPRL2014}. On the one hand, a simple model that includes only deterministic, short-ranged interactions can reproduce the velocity structure functions of dense suspensions of \emph{Bacillus subtilis}~\cite{wensinkPNAS2012}, and experiments on \emph{Escherichia coli} have shown that hydrodynamic interactions are washed out by the stochasticity intrinsic in the bacterial motion~\cite{drescherPNAS2011}. On the other hand, experiments on confined suspensions of \emph{B. subtilis} show that long-range hydrodynamic interactions drive the bacterial collective behavior~\cite{lushiPNAS2014}.

As a swimming microorganism approaches a solid surface it will be increasingly affected by the hydrodynamic interactions with the boundary. Four effects have been identified~\cite{laugaRepProgPhys2009}: (\emph{i}) the swimming speed increases near a solid boundary~\cite{katzJFM1974,brennenARFM1977} because the drag anisotropy increases as the distance to the surface decreases. (\emph{ii}) The swimming trajectory will change shape. For example, the flow field generated by \emph{Escherichia coli} away from any boundary has a cylindrical symmetry, but close to a solid boundary it loses this property. The rotation of the flagellum produces a force normal to it, while an equal and opposite force is produced also on the cell body. The resulting torque is the physical origin of the circular, clockwise motion of \emph{Escherichia coli} (when viewed from the same side of the cell with respect to the surface)~\cite{diluzioNature2005}. Circular trajectories were also seen in bacteria with a single flagellum (monotrichous) such as \emph{Vibrio alginolyticus}~\cite{kudoFEMSMicrobioLett2005,magariyamaBiophysJ2005} and \emph{Caulobacter crescentus}~\cite{liPNAS2008}. (\emph{iii}) Solid walls attract and reorient cells. Mathematically, a solid wall with no-slip boundary conditions can be replaced by a virtual, image cell on the opposite side of the wall, and the two cells will interact hydrodynamically. Thus, a pusher aligns parallel to the wall, and once aligned is attracted to the wall. The higher density of cells close to solid surfaces is experimentally observed~\cite{rothschildNature1963,cossonCellMotCytosk2003,berkePRL2008}. Pullers instead align normal to the walls, and therefore tend to swim against them. (\emph{iv}) The range of the hydrodynamic interactions decreases. Because the image cells represent an opposite flow field, this decays faster in space; for example, the flow field described by Eq.~\eqref{eq:veloc_field} decays as $r^{-2}$ in an infinite fluid but in the presence of a wall it can decay as $r^{-3}$ or 
$r^{-4}$~\cite{blakeJEngMath1974}. Effectively, walls screen the hydrodynamic interactions among swimmers. 

 We have already discussed (see Sec.~\ref{sec:motility}) the relevance of swimming motility to biofilms. However, we note that more work is certainly necessary to firmly establish the connection of the swimming state with the chemical, biological and physical forces acting within a biofilm.


\subsection{Swarming}

Swarming is the multicellular motion of bacteria on a surface powered by flagella~\cite{kearnsNatRevMicrobiol2010}. Under the appropriate circumstances, some microorganisms undergo a phenotypic change and their motion is characterized by spatial and temporal correlations that include billions of individuals~\cite{darntonBiophysJ2010}. 
 At present only three families of bacteria are known to exhibit swarming: the phylum Firmicutes, and the classes Alphaproteobacteria and Gammaproteobacteria~\cite{kearnsNatRevMicrobiol2010}. However, it is likely that the laboratory conditions select against swarming behavior, so that it might be more widespread than currently thought. The concentration of agar on a plate, for example, is important to induce or inhibit swarming. A concentration above $0.3\%$ inhibits swimming and forces the cells to move on the surface, but concentrations above $1\%$ stop the swarming of most species. A commonly used concentration is $1.5\%$. Swarming is a complex phenomenon that requires many concurrent mechanisms. The most important is the increase of the number of flagella. Although some species with a single polar flagellum can both swim and swarm, in most species contact with a surface induces the synthesis of lateral flagella randomly distributed on their bodies~\cite{shinodaJBacter1977}, which are exclusively used for swarming~\cite{mccarterJMMB1999,merinoFEMSMicroLett2006}. The bacteria are said to become \emph{hyperflagellated}. The polar and lateral flagellar systems are driven by separate motors, encoded by different genes, and subject to different regulatory networks~\cite{gavinMolecMB2002,merinoFEMSMicroLett2006,kirovJBact2002}.
   
Swarming behavior also requires a rich medium because the synthesis of flagella and the motion in a highly viscous environment have a high metabolic cost~\cite{merinoFEMSMicroLett2006}, and therefore a medium which supports high growth rates is necessary~\cite{jonesJGMB1967}.

Another (almost) necessary condition for swarming is the production of surfactants by the bacteria that facilitate the spreading of the colony on the solid surface and protects the cells against desiccation~\cite{harsheyAnnRevMB2003}. Mutations inhibiting the synthesis of surfactant stop swarming, but it can be resumed if exogenous surfactants are added~\cite{kearnsMolMB2003,julkowskaJBacter2005}. Synthesis of surfactants is regulated via quorum sensing~\cite{lindumJBacter1998}. Because surfactants are effective in large quantities, it is reasonable to conclude that this quorum sensing mechanism evolved to avoid wasteful synthesis by isolated cells. Interestingly, \emph{Escherichia coli} swarms without any surfactant and it is not known what substance promotes its surface motility~\cite{kearnsNatRevMicrobiol2010}.
   
The collective character of swarming is evident in the formation of \emph{rafts}, dynamical groups of closely associated bacteria with  strong orientational correlations and that move together~\cite{kearnsMolMB2003,eberlJBacter1999}. Interactions among flagella may be responsible for the formation and maintenance of the rafts~\cite{copelandAEnvMB2010}. 
For example, in swarming colonies of \emph{Escherichia coli}  the flagellar bundles may even align at $90\degree$ with respect to the cellular body while remaining intact, and they remain cohesive even during collisions with neighboring cells~\cite{copelandAEnvMB2010};  the cells of \emph{Proteus mirabilis} in the swarming state grow twenty times larger than normally, and move only when they are in contact with other individuals~\cite{morrisonNat1966}; in a raft their flagella are interwoven into helical structures that stabilize the collective motion of adjacent cells~\cite{jonesInfImm2004}.
   
The dynamics of cells in the rafts is out of thermal equilibrium. Clusters exhibit persistent reorganization, they split and merge with other clusters. Individuals constantly leave one raft to join another. Clusters of wild-type \emph{Bacillus subtilis} have many different sizes, and the larger the colony, the more likely it is to find a large cluster. The probability distribution function of the cluster size $P(n)$ obtained from the observations is well described by a power law with an exponential truncation, $P(n)\sim n^{-\alpha}e^{-n/n_c}$, where $\alpha=1.85$ independently of the bacterial density, while $n_c$ does depend on density~\cite{zhangPNAS2010}. That the same dependence of $P(n)$ is predicted by theoretical models and found in fish and African buffaloes~\cite{bonabeauPNAS1999} points to a general mechanism in the collective motion of living organisms. The fluctuations of the bacterial density $\langle\Delta N\rangle$ in a given region grows with the number of cells $N$. However, while for a system in thermal equilibrium it is expected that $\langle\Delta N\rangle\sim\sqrt{N}$, in the swarming colony $\langle\Delta N\rangle\sim N^\beta$, where $\beta\approx0.75$~\cite{zhangPNAS2010}. Density fluctuations with a scaling stronger than the $N^{1/2}$ law are termed ``giant number fluctuations''~\cite{narayanScience2007,dasJTheoBiol2012}.
   
A spectacular feature of many swarming bacterial species is the formation of patterns. As the colony spreads on the solid substratum a pattern of concentric rings, dendrites or vortices emerges on a scale orders of magnitude larger than the cell size. 
Different patterns are largely a result of environmental conditions~\cite{ShimadaJPSJ2004,hiramatsuMBEnv2005}. Physical models of the pattern formation must inevitably consider a simplified  representation of the system, however, they capture the main dynamic of the relevant degrees of freedom. For example, a model using two advection-diffusion-reaction equations that represent the phenotypic cycle of differentiation-dedifferentiation from swimming to swarming state and vice versa can reproduce both the formation of concentric rings and branched structures~\cite{arouhPRE2001}; other reaction-diffusion equations predict the phase diagram of continuous and periodic expansion of the colony~\cite{czirokPRE2001}; an approach that includes lubrication and chemotactic signaling captures many aspects of the branching pattern~\cite{goldingPhysica1998}.
 More detailed equations that include nutrient dynamics, nucleation theory concepts for the cells, and individual-based motion and growth reproduce the knotted-branching pattern of \emph{Bacillus circulans}~\cite{wakanoPRL2003}. A particle-based model reproduces vortices in \emph{Bacillus circulans} by including attractive or repulsive rotational chemotactic response through the term $\tfrac{1}{\vert \vec{v}_i\vert}\vec{v}_i\times(\vec{v}_i\times\nabla C)$, where $C$ is the concentration of chemotactic substance~\cite{benPhysA1997}; however, the role of chemotaxis in swarming remains unclear~\cite{kearnsNatRevMicrobiol2010}. The ``communicating walkers model'' couples a simple diffusion-reaction equation with a random walk and reproduces a variety of achiral~\cite{benNature1994} and chiral~\cite{benPRL1995} branched patterns. A different approach is used in Ref.~\cite{wensinkJPhysCM2012} where a simple model of self-propelled rod-like particles interacting only through steric repulsions is used. This model reproduces swarming for large aspect ratios of the particles, and exhibits giant number fluctuations.

 How is the swarming motility relevant to biofilms? The role of swarming in the early phases of biofilm formation has been recognized only recently. For example, studies of \emph{Pseudomonas aeruginosa} show that surface motility does influence the morphology of the ensuing biofilm. Large surface motility produces flat biofilms,  while limited surface motility produces more corrugated biofilms \cite{shroutMolMB2006}. The stators `MotAB' and `MotCD' involved in the flagellar mechanism of \emph{Pseudomonas aeruginosa} are both involved in the initial, reversible attachment to a substratum~\cite{verstraetenTrendsMB2008}. The connection between swarming and biofilm formation runs deeply, at the level of genetic regulation and quorum sensing; however, its understanding is still in its infancy and more work is required~\cite{parsekTrMB2005}.  
   
\subsection{Twitching}

Twitching motility is a bacterial, surface-related motion actuated by the extension and retraction of type IV pili. These pili are the major virulence factor of \emph{Pseudomonas aeruginosa}~\cite{hahnGene1997} and their twitching motility allows the opportunistic infection of wounds; they are required for the so-called ``social gliding motility'' of \emph{Myxococcus xanthus} and the twitching of \emph{Neisseria gonorrhoeae}. All Gram-negative bacteria have type IV pili~\cite{craigNatRevMB2004}.
Type IV pili are semiflexible homopolymers of the protein pilin, $6-9$~nm in diameter and several microns in length, with a persistent length of about $5~\mu$m and helical conformation~\cite{skerkerPNAS2001,craigCurrOpStrBio2008}. They extend from one or both poles of the bacterium~\cite{mattickARMB2002}. The tip of a type IV pilus contains proteins that can adhere to organic (like mammalian or plant cells) or inorganic surfaces. The twitching motion is produced by the extension, adhesion and retraction of the type IV pili, each tugging the cell body in different directions, which results in the characteristic ``jerky'' motion of the cell. By hydrolyzing ATP within molecular motors the cells dynamically polymerize and depolymerize the pili, actuating in this way the extension and retraction, respectively~\cite{merzNature2000,sunCurrBio2000}. Type IV pili can retract with a speed of about $0.5-1~\mu$m/s~\cite{merzNature2000,skerkerPNAS2001} and exert a force up to $110$~pN for \emph{Neisseria gonorrhoeae}~\cite{maierPNAS2002}, or $150$~pN for \emph{Myxococcus xanthus}~\cite{clausenBiophysJ2009,clausenJBacter2009}.

A number of experimental facts suggest a single scenario that explains the twitching motion. Cells crawl forward by retracting a number of pili at the forward end and unbinding those at the rear end. In \emph{Neisseria gonorrhoeae} the velocity of the crawling motion is about $1.6~\mu$m/s and is lower than the retraction speed of a pilus ($2~\mu$m/s)~\cite{holzPRL2010}. The random motion of cells is not a Brownian motion but has a persistence length of several pili, and this persistence length increases with the number of pili per cell~\cite{holzPRL2010}. A scenario where a ``tug of war'' takes place between the front and rear pili can explain this situation: the pili are under load during the motion, and thus the crawling speed is lower than the retraction speed; the motion is biased towards the direction where several pili are attached close together, and produces a correlation length of several pili~\cite{holzPRL2010,conradResMB2012}. A one-dimensional theoretical model, which assumes only mechanical interactions and no biochemical regulation, is consistent with this scenario~\cite{mullerPNAS2008}. A more precise two-dimensional, stochastic model predicts the existence of directional memory in the retraction and bundling of pili, and is corroborated by experiments~\cite{maratheNatComm2014}.

The crawling of \emph{Pseudomonas aeruginosa} alternates between two different modes: a slow, linear translation of long duration ($0.3-20$~s), and a fast roto-translation of short duration (less than $0.1$~s) and $20$ times as fast as the first mode~\cite{jinPNAS2011}. The second, fast mode results from the sudden release of a single, taut type IV pilus that creates a ``slingshot'' effect with the cell body, and whose large velocity allows the cell to move through shear-thinning fluids such as the EPS of the ensuing biofilm~\cite{jinPNAS2011,conradResMB2012}. 

Type IV pili of \emph{Pseudomonas aeruginosa}  also allow a ``walking'' motion where the cell body is oriented perpendicular to the solid surface~\cite{gibianskyScience2010}. The persistence length of the walking mode $\ell_p^w\approx2~\mu$m is smaller than the persistence length in the crawling mode ($\ell_p^c\approx6~\mu$m), suggesting an uncorrelated motion of the pili~\cite{gibianskyScience2010}. The concerted action of the pili in the walking mode is also evident in the dependence on observation time $\tau$ of the bacterial mean square displacement $\langle(\Delta r)^2(\tau)\rangle\sim \tau^\alpha$. While for the walking mode $\alpha\approx1.1$, so a nearly diffusive motion, $\alpha\approx1.4$ in the crawling mode, indicating a super-diffusive motion that explores space more efficiently~\cite{conradBiophysJ2011,conradResMB2012}.

The twitching motion can be physically described within the broad context of continuous time random walks and L\'evy flights~\cite{zaburdaevRMP2015}. Consider particles that move in a homogeneous, one-dimensional medium and undergo a stochastic motion. The quantity of interest is the probability density of finding the particle at position $x$ at time $t$, $P(x,t)$. The particle can jump a distance $x$ with probability density function $g(x)$, where $\int_{-\infty}^{+\infty}g(x)\mathrm{d}x=1$, and then waits a time $\tau$ before making another jump; the waiting times follow a distribution $f(\tau)$, such that $\int_{0}^{\infty}f(\tau)\mathrm{d}\tau=1$. This is the definition of a continuous time random walk. If we consider the survival probability $F(t)=1-\int_{0}^{t}f(\tau)\mathrm{d}\tau$, that is, the probability of not jumping in the interval $[0, t]$, and given the initial distribution of particles, $P_0(x)$, then we can write the formal solution to the problem: the Montroll--Weiss equation~\cite{montrollJMathPhys1965}
\begin{equation}
P(k,s)=\frac{\hat{F}(s)\tilde{P}_0(k)}{1-\hat{f}(s)\tilde{g}(k)}\,,
\end{equation}
where $\hat{f}(s)\equiv\mathcal{L}\{f(t)\}= \int_0^\infty e^{-st}f(t)\,\mathrm{d}t$ is the Laplace transform of the waiting time distribution, $\hat{F}(s)=\mathcal{L}\{F(t)\}=[1-\hat{f}(s)]/s$, $\tilde{g}(k)\equiv\mathcal{F}\{g(x)\}=\frac{1}{2\pi}\int_{-\infty}^{+\infty}g(x)e^{-ikx}\,\mathrm{d}x$ is the Fourier transform of the jump distribution, and $\tilde{P}_0(k)=\mathcal{F}\{P_0(x)\}$. To proceed further one must specify the waiting time distribution $f(\tau)$ and the jump length distribution $g(x)$. If, in general, the first moment of $f(\tau)$, $\langle \tau\rangle$, and the second moment of $g(x)$, $\langle x^2\rangle$, exist, then one recovers the standard diffusion equation
\begin{equation}
\frac{\partial P(x, t)}{\partial t}=D\nabla^2P(x, t)
\end{equation}
whose solution is the Gaussian distribution
\begin{equation}
P(x, t)=\frac{1}{\sqrt{4\pi Dt}}e^{-x^2/4Dt}\,,
\end{equation}
with $D=\frac{\langle x^2\rangle}{2\langle \tau\rangle}$. Assume now that for large jump lengths and long waiting times the probability density functions scale as 
\begin{align}\label{eq:heavy-tail}
g(x)\sim x^{-(2\beta+1)}\,\,,\quad  f(\tau)\sim \tau^{-(\gamma+1)}\,.
\end{align}
If $\beta>1$ and $\gamma>1$ then the moments $\langle x^2\rangle$, $\langle \tau\rangle$ exist and we have again standard diffusion. For $\beta<1$ or $\gamma<1$, instead, one or both moments diverge.
If $\langle x^2\rangle$ is finite but $\langle \tau\rangle$ diverges then the process is anomalously slow, and the mean square displacement is subdiffusive $\bar{x}=\sqrt{\langle (x-\langle x\rangle)^2\rangle}\sim t^{\gamma/2}$, with $\gamma<1$. If the waiting times are finite but the jumps have no upper bound, then the process is superdiffusive, $\bar{x}\sim t^{1/2\beta}$, with $\beta<1$~\cite{bouchaudPhysRep1990,zaburdaevRMP2015}.

Under conditions of scarcity of resources L\'evy flights are considered the most efficient way of exploring space. The question whether heavy tail distributions (as in Eq.~\eqref{eq:heavy-tail} with $\beta<1$ or $\gamma<1$) and true L\'evy walks are realized by living organisms is passionately debated and source of many controversies. We refer the interested reader to more specialized reviews as Ref.~\cite{viswanathan2011,mendez2013,zaburdaevRMP2015}.
However, the run-and-tumble motion of \emph{Escherichia coli} can be described as a L\'evy walk~\cite{taktikosPLOS2013,zaburdaevRMP2015}: the cell moves in a specific direction with constant speed, and then changes direction after a random waiting time. Recent experiments on individual cells show that the intrinsic molecular noise emerges in a power-law distribution of run times~\cite{korobkovaNat2004}.

How is twitching motility relevant to biofilms? 
In general type IV pili are important adhesins, that is, they can actuate the initial attachment to a surface. Furthermore, the twitching motility that they generate helps in the repositioning of cells with respect to one another and leads to the biofilm differentiation~\cite{burrowsARMB2012}. Complex morphogenetic structures may be directed by regulated, cellular twitching motility (among other factors); for example, the formation of mushroom-like structures in the biofilm of \emph{Pseudomonas aeruginosa} where non-motile mutants form the mushroom stalks, while motile individuals, driven by type IV pili, climb the stalks and aggregate on top to form the mushroom caps~\cite{klausenMolMB2003}.

There is evidence that in \emph{Pseudomonas aeruginosa} type IV pili
and type IV pili-mediated twitching motility play a role in the microcolonies that seed the biofilm. 
Type IV pili play a direct role in stabilizing interactions with the surface and/or in the cell-to-cell interactions
required to form a microcolony. Type IV pili-mediated twitching
motility may also be necessary for cells to migrate along
the surface to form the multicellular aggregates characteristic of
the normal biofilm~\cite{otooleMolecMB1998}.
\emph{Xylella fastidiosa} (a notorious plant pathogen) has two classes of pili: type I and type IV pili that are implicated in surface attachment and in the formation of biofilm~\cite{mengJBacter2005},  whose density seems to be greatly influenced by the presence of type I pili~\cite{liMB2007}. Studies of \emph{Vibrio cholerae} El Tor shows that MSHA type IV pili and flagella accelerate attachment to abiotic surfaces~\cite{watnickMolMB1999}.

\subsection{Gliding and Sliding}

Gliding is a surface-associated movement that does not involve flagella or pili, but instead uses macromolecular structures, known as focal adhesion complexes, that connect the cellular surface to external molecules (for example in the EPS) or surfaces~\cite{kearnsNatRevMicrobiol2010}. Tens of proteins participate in the dynamic assembly and disassembly of these complexes that do not just mediate adhesion but also form a mechanosensing system, since they are coupled to the signal-transduction network of the cell~\cite{geigerNatRevMolCellBio2001}, and they are linked to the cytoskeleton~\cite{nanPNAS2011}, and the actin-myosin network~\cite{zamirJCellSci2001}.

The term gliding is used with different meanings in the literature; for example, the ``social gliding motility'' of \emph{Myxococcus xanthus} is actuated by type IV pili as they collectively spread over a surface, and small rafts of cells move beyond the boundaries of the expanding colony. A second motility mechanism termed ``adventurous gliding motility'' does not require pili~\cite{mcbrideARMB2001}, but rather focal adhesion complexes that assemble at the leading cell pole and disperse at the rear of the cells~\cite{mignotScience2007}.

As the gliding motion does not require appendages and hydrodynamic interactions are negligible simple models that include only steric interactions and active motion are amenable to describe gliding motility.
Experiments on the adventurous gliding motion of \emph{Myxococcus xanthus} show 
that collective motion with cluster formation appears for packing fractions $\phi\gtrsim0.17$ and at the transition the cluster size distribution is a power law~\cite{peruaniPRL2012}. Additionally, the clusters exhibit giant number fluctuations $\langle\Delta N\rangle\sim N^{\beta}$ with $\beta\simeq0.8$ for $\phi\gtrsim0.17$~\cite{peruaniPRL2012}. A similar exponent was found in a simple model of point particles moving with constant speed and aligning nematically with each other~\cite{GinelliPRL2010}. A similar transition to the clustered state was found in models of self-propelled rods~\cite{peruaniPRE2006,yangPRE2010}.


Microorganisms can also move without any active appendages or specialized molecular motors but simply under the influence of the expansive force of a growing colony, and facilitated by the secretion of biosurfactants that reduce the surface tension between cells and surface, such as lipopeptides, lipopolysaccharides and glycolipids~\cite{henrichsenBactRev1972,harsheyAnnRevMB2003}. The fact that groups of cells move as a single unit indicates that this is not an active form of movement.
For example, the Gram-positive \emph{Mycobacterium smegmatis} and \emph{Mycobacterium avium} (a human opportunistic pathogen) spread by forming a monolayer of cells arranged in pseudo-filaments, where the cells passively move along their longitudinal axes~\cite{martinezJBacter1999}. Recently, it was discovered that beside the simple concept of cells pushing each other another physical mechanism affect the sliding motility. The secretion of EPS in the biofilm of \emph{Bacillus subtilis} generates a gradient of osmotic pressure, which causes an uptake of water from the surrounding environment that in turn produces a swelling of the biofilm~\cite{seminaraPNAS2012}. Theoretical calculations show that the osmotically driven biofilm undergoes a transition from an initial vertical swelling of the biofilm to a subsequent horizontal expansion~\cite{seminaraPNAS2012}.
How are gliding and sliding relevant to biofilms? Work on \emph{Mycobacterium smegmatis} and other members of this genus shows that these motility modes are important for the colonization of surfaces, such as during biofilm dispersal, and the restructuring of the biofilm~\cite{rechtJBacter2001}. The glycopeptidolipids synthesized by these microorganisms lower the surface tension and facilitate their surface motility, and there is a strong correlation between the presence of lipids, surface motility and biofilm morphology~\cite{rechtJBacter2001,chenJBacter2006,wuMBPath2009,mayaBioMedResInt2015}.

\section{Adhesion}
\label{sec:adhesion}

The adhesion of cells to a solid substratum is the first, necessary step for the formation of a biofilm.   Contact with a substratum corresponds to
the moment when the regulatory genetic network initiates the production of EPS, induces the genetic expression of polysaccharides~\cite{daviesAppleEnvMB1993} and progressively deactivates flagellar motion~\cite{harsheyAnnRevMB2003}.

Microorganisms can adhere to organic or inorganic surfaces, inert or reactive, living or devitalized~\cite{christensen1985,gristinaScience1987}. The process of adhesion can be divided into two steps~\cite{marshallJGenMB1971}: (\emph{i}) \emph{docking}, that is, the cell comes in close proximity of the substratum thanks to motility or Brownian motion;  this phase of attachment can be easily reverted and is governed by basic physical interactions based on the Coulomb and van der Waals forces. (\emph{ii}) \emph{locking}, that is, the irreversible anchoring of the cell by means of chemical bonding mediated by adhesins~\cite{dunneClinMBRev2002}.

\subsection{Basic Theoretical Framework}

A microbial cell, for example a bacterium, is a micron-sized particle, thus, in the colloidal regime. Most bacteria and natural or artificial surfaces are negatively charged~\cite{carpentierJApplMB1993,juckerJBacter1996,dunneClinMBRev2002}. Consequently, the docking phase of adhesion is possible because of the existence of an equilibrium point between the  electrostatic repulsion and the attractive van der Waals interaction. The mechanism of adhesion of charged colloids is described by the Derjaguin, Landau, Verwey,
and Overbeek (DLVO) theory of colloidal stability~\cite{derjaguin1941,derjaguinProgSurSci1993,verwey1948}, which quantitatively describes the adhesion of cells to solid surfaces. 

We now discuss the fundamental concepts of the DLVO theory. In its classical formulation the DLVO theory includes only the effects of van der Waals and electrostatic interactions, while neglecting other effects, such as steric, solvation, and depletion forces~\cite{israelachviliAccChemRes1987,chandlerNature2005}. Thus, the DLVO interaction energy between two charged materials separated by a distance $d$ is
\begin{equation}
U_{\mathrm{DLVO}}(d)=U_{\mathrm{vdW}}(d)+U_{\mathrm{C}}(d)\,,
\end{equation}
where $U_{\mathrm{vdW}}(d)$ includes the interaction between a permanent dipole and an induced dipole (Debye energy), the interaction between two permanent dipoles (Keesom energy), and the interaction between instantaneous dipoles (London dispersion energy), and $U_{\mathrm{C}}(d)$ is the Coulombic potential energy. While for atoms or molecules $U_{\mathrm{vdW}}(d)=-{C_{\mathrm{vdW}}}/{d^6}$, for larger objects one finds a power law with a smaller exponent; for example, the van der Waals energy between a sphere of radius $R$ and a flat surface is $U_{\mathrm{vdW}}(d)=-R{A}/{6d}$, while between two flat surfaces is $U_{\mathrm{vdW}}(d)=-A/(12\pi d^2)$~\cite{israelachvili2011}, where $A=\pi^2\rho_1\rho_2 C$ is the Hamaker constant, $\rho_1$ and $\rho_2$ are the number densities of the two interacting materials, and  $C$ is the coefficient in the interatomic pair potential which is proportional to the square of the polarizability. In general van der Waals energies (and hence forces) scale as power laws, which means that they  can be long ranged, and therefore the structure and composition of the solid substratum can have surprising effects~\cite{loskillLangmuir2012}.

Surfaces in an aqueous milieu become charged because of the dissociation of surface chemical groups (releasing H$^+$ or Na$^+$) or because of the adsorption of ions from the solvent. The negatively charged surface is balanced by counter-ions which are attracted to it by the electrostatic force. The closest counter-ions are bound to the surface but further away there is a diffuse layer of unbound ions in thermal equilibrium that form the so-called electric double layer. We can derive the equation governing the electrostatic potential $\psi(x)$ for a simple system of only counter-ions in the presence of the surface. The chemical potential $\mu$ of the counter-ions at position $x$ is
\begin{equation}
\mu=ze\psi(x)+k_{\mathrm{B}}T\ln\rho(x)\,,
\end{equation}
where $z$ is the valence, $e$ the absolute value of the electron charge, $k_{\mathrm{B}}$ the Boltzmann constant, and $T$ the temperature of the system. By imposing that in equilibrium the chemical potential is constant one finds the (number) density distribution
\begin{equation}\label{eq:boltzmann}
\rho=\rho_0\exp(-\frac{ze\psi(x)}{k_{\mathrm{B}}T})\,,
\end{equation}
which is just the Boltzmann distribution of the ions. In electrostatics the charge distribution obeys the Poisson equation
\begin{equation}\label{eq:poisson}
\nabla^2\psi=-\frac{ze\rho}{\varepsilon}\,,
\end{equation}
where $\varepsilon$ is the permittivity of the solvent. Inserting Eq.~\eqref{eq:boltzmann} into  Eq.~\eqref{eq:poisson} one arrives at the Poisson--Boltzmann equation
\begin{equation}\label{eq:PB}
\nabla^2\psi=-\frac{ze\rho_0}{\varepsilon}\exp\left(-\frac{ze\psi}{k_{\mathrm{B}}T}\right)\,.
\end{equation}
For a system with more than one species of ions in solution Eq.~\eqref{eq:PB} generalizes in a straightforward manner to a sum on the right-hand side over the different ionic species. The Poisson--Boltzmann equation is nonlinear, and therefore very challenging to solve. In the limit of dilute solutions we can expand in Taylor series the exponential in Eq.~\eqref{eq:PB} (Debye--H\"uckel approximation). The zeroth order term vanishes because of the electroneutrality condition ($\sum_i z_ie\rho_{0,i}=0$). Then, if we stop at the first order, we obtain for the ionic species $i$ 
\begin{equation}\label{eq:helmholtz}
\nabla^2\psi_i=\kappa^2\psi_i\,, \quad \kappa^2\equiv\frac{e^2}{\varepsilon k_{\mathrm{B}}T}\sum_i\rho_{0,i}z_i^2
\end{equation}
where $\lambda_\mathrm{D}=\kappa^{-1}$ is the Debye length, which physically represents the thickness of the electric double layer. Typical solutions of Eq.~\eqref{eq:helmholtz} are of the form $\psi_i(x)=ce^{-\kappa x}/x$, with $c$ a constant~\cite{israelachvili2011}. Thus, putting all together, $U_{\mathrm{DLVO}}(d)$ is the sum of an attractive term $U_{\mathrm{vdW}}(d)\sim -1/d^n$ and a repulsive term $U_{\mathrm{C}}(d)\sim e^{-\kappa d}$. At small enough distances the attraction will always prevail.  It exhibits a deep (primary) minimum at $d=0$, which corresponds to adhesion of the surfaces. At slightly larger distances there is an energy barrier due to the overlap of the electric double layers; the energy barrier increases as the ionic strength decreases. Further away, at typical distances of $5$~nm, there is  a small (secondary) minimum if the ionic strength is large enough.

We can now describe what happens to a bacterium, in the idealized situation described above, when it approaches a surface. As a bacterium moves closer to a surface it will experience an energy barrier a few nm away from the surface that cannot be overcome by active motility or simple Brownian motion. However, depending on the ionic strength the cell will find the secondary minimum and it will be reversibly bound to the surface. In the following step (locking) the biosynthesized pili or EPS surmount the barrier and allow the irreversible adhesion to the surface~\cite{gristinaScience1987,horiBiochemEngJ2010}. \\

\subsection{Complex Role of Biopolymers and Appendages}

The simple picture given by the DLVO theory above describes an idealized situation of homogeneous, charged   surfaces interacting.  However, it leaves out the true complexity of the adhesion process of microorganisms. In general terms, there are three main classes of structures that cause drastic deviations from the predictions of the DLVO theory in real measurements of adhesion forces: (\emph{i}) biopolymers associated with the cellular membrane, (\emph{ii}) the EPS that modifies the physico-chemical properties of surfaces,  (\emph{iii}) appendages, such as pili. The main features of their roles in adhesion will be reviewed below.

The cellular surface is  a physically and chemically heterogeneous structure~\cite{dufreneNatRevMB2008}. For example, the outer membrane of Gram-negative bacteria contains lipopolysaccharides (LPS) which provide protection from many antibiotics. LPS are composed of a lipid A (anchored to the outer membrane), negatively charged core polysaccharides, and the O-antigen, which can extend $30$~nm or more from the outer membrane. Lipopolysaccharides have an inhomogeneous spatial distribution and have been shown to play a key role in the adhesion process~\cite{kotraJACS1999,walkerLangmuir2004,atabekJBacter2007}. A portion of the O-antigen has been shown to form hydrogen bonds with different mineral substrata~\cite{juckerCollSurfB1997}, and $1000$ or fewer of these bonds can firmly anchor the cell to a surface~\cite{juckerCollSurfB1997,horiBiochemEngJ2010}. 
The characteristics of the LPS in \emph{Pseudomonas aeruginosa} are crucial to determine their adhesion to hydrophilic or hydrophobic surfaces, and the phenotypic variations in the expression of LPS can regulate adhesion in their survival strategy~\cite{makinMB1996}. Mutations in \emph{Escherichia coli}'s genes involved in LPS biosynthesis showed decreased adhesion~\cite{genevauxArchMB1999}; removal of up to $80\%$ of LPS molecules also drastically decreases adhesion affinity~\cite{abuEnvSciTech2003}.
Recent work on \emph{Staphylococcus aureus} shows that adhesion initiates already at distances of $50$ to $100$~nm (where the DLVO theory predicts negligible forces), and furthermore the fluctuations of protein density and structure are more important than the details of the binding potential~\cite{thewesSoftMatt2015}.

Another important factor in the adhesion process is the EPS. The polymeric macromolecules form noncovalent bonds, such as electrostatic or hydrogen bonds, with the substratum that lead to the adhesion of the biofilm~\cite{tsunedaFEMSMBLett2003}; EPS contains many polar portions that anchor to a hydrophilic glass and can turn it from hydrophilic to hydrophobic and therefore renders adhesion thermodynamically favorable~\cite{vanLoosdrechtAquaSci1990,azeredoBiofoul2000}. The EPS also produces a ``polymeric interaction'' because the macromolecules bound to the cell membrane effectively increase its size and boost the (attractive) van der Waals interaction with the surface. Additionally, EPS macromolecules around the cell can bind with macromolecules adsorbed on the solid surface, thus bridging the electrostatic barrier~\cite{vanLoosdrechtExperientia1990,azeredoCollSurfB1999}.

The hydrophobic components of the cell membrane also favor adhesion as  has been repeatedly  shown: hydrophobic cells adhere better than hydrophilic cells~\cite{rosenberg1986,vanLoosdrechtApplEnvMB1987,thewesBeilstein2014}. The hydrophobic effect is a consequence of a fundamental aspect of water's molecular structure: the extensive presence of a hydrogen-bond (H-bond) network. When an apolar solute is introduced in bulk water, it will locally disrupt the H-bond network because water molecules cannot bind to it. To minimize this disruption the water molecules surrounding the solute will form a cage-like structure: a clathrate. Because of the clathrates the water molecules have smaller translational and orientational entropy; thus this change is unfavorable. When two apolar moieties are close to each other, it is thermodynamically more favorable to reduce their separation until they are in contact with each other because in this way the surface area of the clathrates is reduced and hence the entropy is maximized~\cite{stillingerJSolChem1973,GiovambattistaPRE2006,giovambattistaPNAS2008}.

Different kinds of pili are also involved in the adhesion process. Type IV pili can bind to inert surfaces bacterial and eukaryotic cells because of adhesins at their tips~\cite{mattickARMB2002}.
The Gram-negative bacterium \emph{Caulobacter crescentus} is among the first to colonize submerged surfaces and in the nonmotile stage of its life cycle produces a stalk tipped by a polysaccharide adhesin that can exert forces in the micronewton range~\cite{tsangPNAS2006}.
Type I (one in Roman numerals) and type P pili are short adhesive structures present in Gram-negative bacteria such as \emph{Escherichia coli} and \emph{Salmonella} species. They are peritrichate and are not associated with twitching motility~\cite{horiBiochemEngJ2010}. Type I pili mediate adhesion to mucosal epithelia and other cells~\cite{ofekNature1977,hansonNature1988}. P pili are composite structures that mediate the adhesion to urinary tract epithelium~\cite{kuehnNature1992}. 
Gram-positive bacteria, such as \emph{Corynebacteria} and \emph{Streptococci}, have also pili, but they are very thin ($2$ to $3$~nm) and have been only recently observed~\cite{kangScience2007}. We now know that \emph{Streptococci} use pili to penetrate mucus (in a human host, for example) or a conditioned surface, and then employ multiple adhesins to build stronger bonds~\cite{nobbsMBMolBiolRev2009}.
\emph{Staphylococcus epidermidis} exhibits two kinds of proteins on fimbria-like appendages protruding from its cell surface that are required for the adhesion function~\cite{veenstraJBActer1996}. Furthermore, polymeric molecules present on the cell wall of \emph{Staphylococcus} have been shown to play a key role in its adhesion capacity and pathogenicity~\cite{heilmannMolMB1997,heilmannMB2003}.

 Loeb and Neihof were the first to report that surfaces immersed in seawater very quickly were coated with organic molecules dissolved in the water before the colonization of microorganisms~\cite{loebAdvChem1975}. We now know that this is a very general process.
 These substances form a so-called ``conditioning layer'' that alters substantially the physicochemical properties of the substratum and its interaction with microorganisms~\cite{gristinaScience1987,bosFEMSMBRev1999,schneiderCollSurfB1994,schneiderJCollInterf1996,deKerchove2007ApplEnvMB,loriteJCollInterfSci2011}.
  A surface may become selectively attractive to some cells. The adhesion of microorganisms might be influenced by hydrophilic and hydrophobic features of the conditioned surface, charge density and exposed chemical groups. The presence of a conditioning layer causes significant deviations from the DLVO theory. In fact, exopolymers may play a dual role: first, by coating the biofilm substratum which strengthen adhesion, and second, by forming polymeric bridges between the EPS-encased cells and the coated substratum~\cite{azeredoBiofoul2000}.\\

\section{Viscoelastic Properties}
\label{sec:viscoelastic}

We now take a larger, almost macroscopic, view of the biofilm and ask the question: how does a biofilm respond to a mechanical perturbation? Clearly a biofilm does not flow as a simple liquid; it requires some structural stability to allow the colony to grow and prosper. It is equally clear that 
a biofilm cannot be as rigid as a crystal because it must allow expansion as cells multiply, and it must deform as a response to changing environmental stresses or stimuli. A biofilm must then be a substance with properties intermediate between a liquid and a solid; thus, it must exhibit both the short-time elastic response of a solid and the long-time viscous response of a liquid. This class of materials is termed viscoelastic~\cite{larson1999}. In fact, measurements confirm that biofilms are viscoelastic polymeric materials~\cite{klapperBiotechBioeng2002,towlerBiofoul2003,vinogradovBiofilms2004,shawPRL2004,ruppApplEnvMB2005,lauBiophysJ2009,hohneLangmuir2009,lielegSoftMatt2011,billingsRepPrgPhys2015}.
This is not surprising as a biofilm consists of a hydrated environment of complex,  entangled polymeric molecules and bacteria that are subject to multiple adhesion and cohesion forces.  In general, a biofilm can be seen as an entangled polymer network where the bonds are non-permanent and physical, rather than chemical~\cite{korstgensJMBMeth2001,wlokaCollPolySci2004}.

\begin{figure}
\centering
\includegraphics[width=0.9\columnwidth]{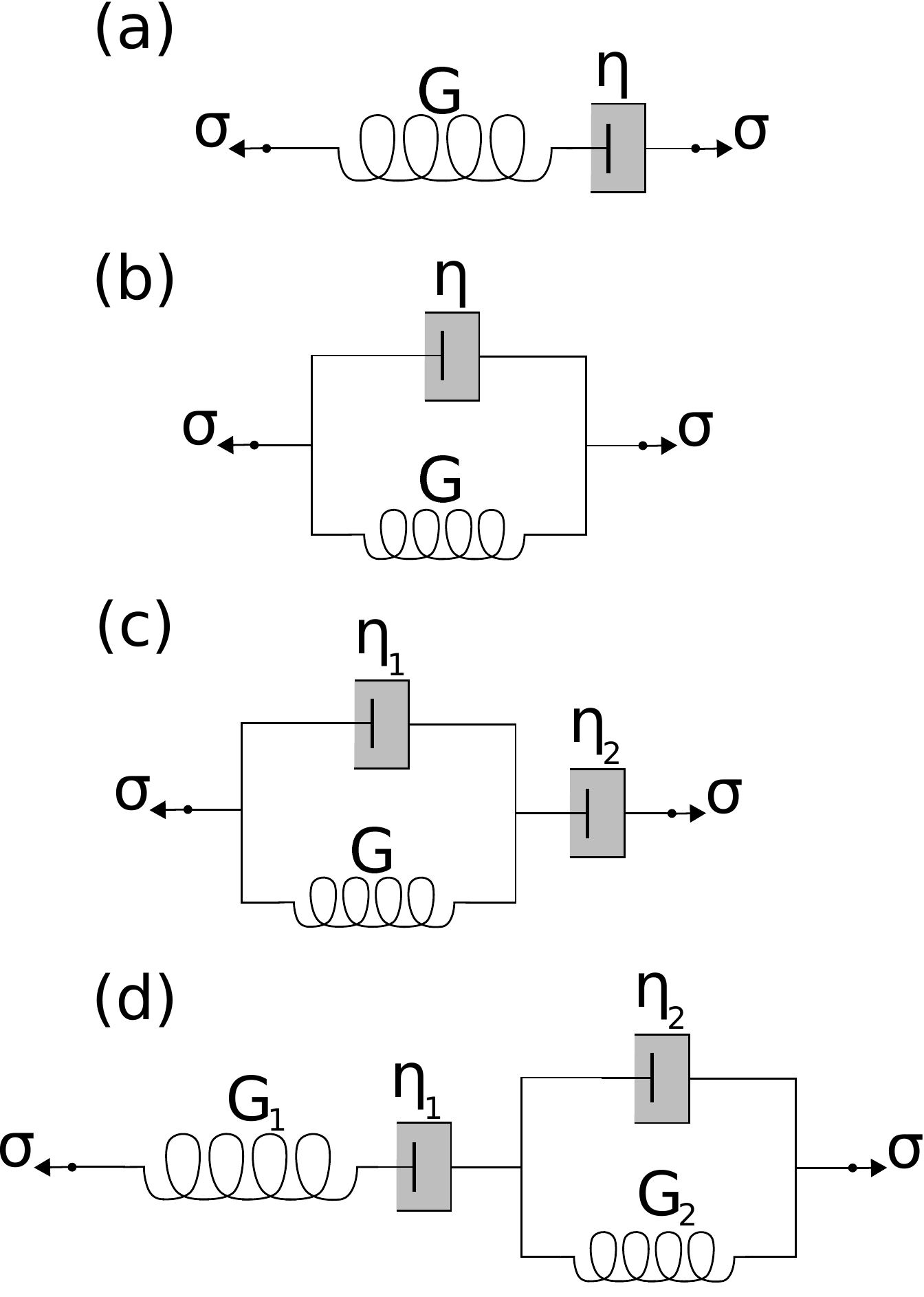}
\caption{Sketch of some mechanical models of viscoelastic fluids that are built from elastic elements (denoted with their Young's modulus $G$) and viscous elements (denoted with their viscosity $\eta$) combined in series or in parallel, and to which a stress $\sigma$ is applied. (a) Maxwell model. (b) Kelvin--Voigt model. (c) Jeffreys model. (d) Burgers model.}
\label{fig:viscoel}
\end{figure}

Viscoelastic materials are not-Newtonian fluids, that is, they do not obey Eq.~\eqref{eq:stressNewton}. The Newtonian expression works remarkably well for simple liquids composed of small, roughly isotropic molecules. This happens because of a separation of times scales: the molecular relaxation time is comparable to the time to diffuse one molecular length, which is $\tau_0\approx\ell^2/D\approx 10^{-13}$~s. The flow velocity can interfere with these relaxation times only in exceptional cases. Consider now a complex fluid comprised of elongated, flexible molecules that possibly form links with each others. The molecular relaxation time is now much larger, and can be matched by realistic shear rates. One is naturally led to define a dimensionless number comparing these time scales, the so-called Weissenberg number $\mathcal{W}\equiv\tau_0\dot{\gamma}$, where $\dot{\gamma}$ is the shear rate. When $\mathcal{W}\ll 1$ one recovers a Newtonian fluid; but for $\mathcal{W} \gtrsim 1$ the shear rate interferes with the molecular relaxation, and therefore one should expect nonlinear or history-dependent terms in the constitutive equation for the stress tensor.

We now discuss the basic physical concepts necessary to describe the viscoelastic behavior of biofilms (see ~\cite{joseph2013,morozov2015} for more details). Any material responds with a deformation $\gamma$ to an applied stress $\sigma$ (force per unit area).  For an elastic solid $\sigma=G\gamma$, where $G$ is the elastic modulus; for a liquid $\sigma=\eta\tfrac{\partial\gamma}{\partial t}$, where $\eta$ is the viscosity. A viscoelastic material must combine both elastic and viscous response. The simplest physical model thereof is the Maxwell element where an elastic spring and a viscous dashpot are combined in series (see Fig.~\ref{fig:viscoel}(a)). For simplicity we consider at first a one-dimensional problem.
For this configuration the stresses on the two elements are equal $\sigma=\sigma_1=\sigma_2$ and the deformations are additive, hence $\tfrac{\partial\gamma}{\partial t}=\tfrac{\partial\gamma_1}{\partial t}+\tfrac{\partial\gamma_2}{\partial t}=\tfrac{1}{G}\tfrac{\partial\sigma}{\partial t}+\tfrac{\sigma}{\eta}$. The 
constitutive equation for the Maxwell model is then 
\begin{equation}\label{eq:maxwell}
\lambda\frac{\partial\sigma}{\partial t}+\sigma=\frac{\partial\gamma}{\partial t}\,,
\end{equation}
where $\lambda\equiv\eta/G$ is a relaxation time. The formal solution of Eq.~\eqref{eq:maxwell} is
\begin{equation}
\sigma(t)=\frac{1}{\lambda}\int_{-\infty}^t e^{-\frac{t-t'}{\lambda}}\eta\frac{\partial\gamma}{\partial t'}\mathrm{d}t'\,.
\end{equation}
While for time scales much shorter than the relaxation ($\lambda\to\infty$) one finds solid-like behavior ($\sigma=G\gamma$), for observation times much larger than the relaxation ($\lambda\to 0$) one finds viscous behavior ($\sigma=\eta\tfrac{\partial\gamma}{\partial t}$). The Maxwell model then behaves as a fluid, and therefore is not suitable for a real viscoelastic material. 

Another, historically important, model is the Kelvin--Voigt element, where the spring and dashpot are arranged in parallel (see Fig.~\ref{fig:viscoel}(b)). The Kelvin--Voigt constitutive equation is 
\begin{equation}\label{eq:kelvinvoigt}
\sigma=G\gamma+\eta\frac{\partial \gamma}{\partial t}\,.
\end{equation}
At long observation times the Kelvin--Voigt material behaves solid-like because it always returns to its equilibrium configuration. 

We can improve our model of viscoelastic materials by adding a dashpot in series with a Kelvin--Voigt element, producing the Jeffreys model (see Fig.~\ref{fig:viscoel}(c)). The strains in the dashpot and in the Kelvin--Voigt element are additive, $\tfrac{\partial\gamma}{\partial t}=\tfrac{\partial\gamma_1}{\partial t}+\tfrac{\partial\gamma_2}{\partial t}$, and the stresses are $\sigma=\eta_1\tfrac{\partial\gamma_1}{\partial t}=G\gamma_2+\eta_2\tfrac{\partial\gamma_2}{\partial t}$. Solving for the total strain $\gamma$ and total stress $\sigma$, one finds the constitutive equation 
\begin{equation}\label{eq:jeffreys}
\lambda_1\frac{\partial \sigma}{\partial t}+\sigma=\eta_1\left(\frac{\partial\gamma}{\partial t}+\lambda_2\frac{\partial^2\gamma}{\partial t^2}\right),
\end{equation}
where $\lambda_1\equiv\left(\tfrac{\eta_1+\eta_2}{G}\right)$ is a relaxation time and $\lambda_2\equiv\tfrac{\eta_2}{G}$ a retardation time. We can obtain a more general equation by considering the deformation field $\xi(x,t)$ and by realizing that $\tfrac{\partial\gamma}{\partial t}=\tfrac{\partial}{\partial t}\tfrac{\partial\xi(x,t)}{\partial x}=\frac{\partial v}{\partial x}$, that is the velocity gradient or rate-of-strain tensor.
In three dimensions, it is $\nabla \vec{v}$. 
Remembering that $\bm{\sigma}=-p\bm{\mathrm{I}}+\bm{\tau}$ (see Sec.~\ref{sec:lowReynolds}), the Jeffreys constitutive equation then becomes 
\begin{equation}
\lambda_1\frac{\partial \bm{\tau}}{\partial t}+\bm{\tau}=2\eta_1\left(\mathbf{D}[\vec{v}]+\lambda_2\frac{\partial \mathbf{D}[\vec{v}]}{\partial t}\right)\,,
\end{equation}
where $\mathbf{D}[\vec{v}]\equiv \tfrac{1}{2}\left(\nabla \vec{v}+\nabla \vec{v}^\mathrm{T}\right)$ is the symmetric part of the rate-of-strain tensor. The formal solution of Jeffreys constitutive equation is
\begin{equation}
\bm{\tau}=2\eta_1\frac{\lambda_2}{\lambda_1}\mathbf{D}[\vec{v}]+2\frac{\eta_1}{\lambda_1}\!\left(1-\frac{\lambda_2}{\lambda_1}\right)\!\int_{-\infty}^t e^{\frac{t-t'}{\lambda_1}}\mathbf{D}[\vec{v}]\mathrm{d}t'.
\end{equation}
The Maxwell model is recovered when $\lambda_2=0$, and the Newtonian fluid when $\lambda_1=\lambda_2$.

Another way of dealing with the rheology of non-Newtonian fluids is to assume that only the rate of dissipation in the fluid changes but not the structure of the stress tensor. One then writes
\begin{equation}
\bm{\sigma}=-p\bm{\mathrm{I}}+\eta(\mathbf{D}[\vec{v}])\mathbf{D}[\vec{v}]\,.
\end{equation}
Because the viscosity cannot change for a change of coordinate system (must be invariant), $\eta(\mathbf{D}[\vec{v}])$ must be a function of the tensorial invariants build with $\mathbf{D}[\vec{v}]$. The lowest order term is $\dot{\gamma}^2\equiv\tfrac{1}{2}\mathbf{D}:\mathbf{D}$. The rheology of the material is then governed by the properties of the function $\eta(\dot{\gamma})$. If $\frac{\partial\eta}{\partial\dot{\gamma}}>0$ the fluid is called \emph{shear-thickening}, that is, it becomes increasingly stiffer as the shear-rate increases; if $\frac{\partial\eta}{\partial\dot{\gamma}}<0$ the fluid is called \emph{shear-thinning}, that is the material becomes less viscous as the shear-rate increases.

Experimentally, a common rheological technique to characterize viscoelastic materials is the measurement of the stress-strain curve. The sample is subject to a shear stress which increases monotonically with time up to a maximum value; this corresponds to the ``loading'' phase. Subsequently, the shear stress is decreased; this is the ``unloading'' phase. The presence of hysteresis is characteristic of viscous dissipation in the material. Another common rheological technique is the creep test, where a stress is applied for a prolonged interval of time and then suddenly released. The temporal evolution of the strain exhibits characteristic regions corresponding to different processes taking place: (\emph{i}) an instantaneous elastic stretching, (\emph{ii}) a viscous flow, (\emph{iii}) after the external stress ceases there is an instantaneous elastic recoil, (\emph{iv}) a time-dependent creep recovery~\cite{stoodleyJIndMBBiotech2002}.

The experimental investigation of biofilms' material properties is still in an early stage, and different models are proposed to explain the experimental data. For example, the biofilms of different \emph{Pseudomonas aeruginosa} strains were subject to strain and creep tests and the results were fitted with a Jeffreys model that includes a slow nonlinearity to account for the shear-thickening behavior~\cite{klapperBiotechBioeng2002}. That theory correctly predicts the relaxation time at high shear rates. However, the biofilm of \emph{Streptococcus mutans} (common in dental plaque) showed creep compliance consistent with a Burgers model~\cite{vinogradovBiofilms2004} (see Fig.~\ref{fig:viscoel}(d)).
Biofilms vary greatly in their composition and structure. The biofilm of \emph{Streptococcus mutans} appears to be shear-thinning~\cite{vinogradovBiofilms2004,cheongRheolActa2009};  the biofilms of the microalga \emph{Chlorella vulgaris} also shows shear-thinning behavior ~\cite{wilemanBioresTech2012}. \emph{Pseudomonas aeruginosa} builds instead a shear-thickening biofilm~\cite{klapperBiotechBioeng2002}. Therefore, it should not be surprising that  measurements based on creep tests of $44$ biofilms determined values of $G$ and $\eta$ spanning eight decades~\cite{shawPRL2004}. However, they also 
found a remarkably small range of variability for the elastic relaxation time $\lambda\approx\eta/G$, with an average of $18$~min~\cite{shawPRL2004}. It is argued that this common relaxation time can hardly be a coincidence, but rather a sign of convergent evolution. The time scale $\lambda$ separates the solid-like from the liquid-like behavior of a biofilm. Short mechanical stresses can be absorbed by an elastic response, but a sustained stress can be deleterious as it could lead to structural failure, The biofilm reacts instead as a viscous fluid for long time-scale stresses. 
The time scale of $18$ min coincides with the time required to elicit a phenotypic response at the cellular level~\cite{vandykApplEnvMB1994,ptitsynApplEnvMB1997}, which is however expensive in terms of cellular resources. Thus, the elastic relaxation time needs to be large enough to avoid unnecessary response to intermittent stresses, but smaller or approximately equal to the biological time to initiate expensive genetically-regulated responses~\cite{shawPRL2004}.

A biofilm growing under flow conditions develops characteristic elongated, filamentous structures called \emph{streamers} (see Fig.~\ref{fig:biofilm-examples}(b), (e) and (g)). They are ubiquitous in natural environments and strongly influence flow through porous materials, medical and industrial devices~\cite{rusconiInterface2010,rusconiBiophysJ2011,drescherPNAS2013}. As the EPS network of streamers percolates through the channels or pores more and more planktonic cells are caught within it. Eventually, the streamers lead to a catastrophic clogging of the pores that stops the flow~\cite{drescherPNAS2013}.
Biofilms grown up to Reynolds numbers of $\mathcal{R}=3600$ produced streamers that behaved as viscoelastic solids for stresses lower than the value at which they were grown, but behaved like viscoelastic fluids at larger stresses~\cite{stoodleyBiotechBioeng1999,stoodleyJIndMBBiotech2002}, that is similar to a Bingham fluid. In the linear regime  streamers exhibited a shear modulus of $27$~N/m$^2$ and a Young modulus  in the range of $17$ to $40$~N/m$^2$~\cite{stoodleyBiotechBioeng1999}.

Strong flow conditions, in the turbulent regime, produce ripple structures in mixed species biofilms that migrate with a speed of $800~\mu$m/h~\cite{stoodleyEnvMB1999}. At $\mathcal{R}=4200$ the ripples have a wavelength of about $75~\mu$m, but it increases to about $200~\mu$m at $\mathcal{R}=1200$. The ripples moves downstream, which points at an important effect for surface colonization. Furthermore the morphology   of the ripples responded within minutes to changes in the flow velocity. This fact indicates that the formation of ripples must have a hydrodynamical origin. Indeed, recent work ~\cite{thomasPhilTransRoySocLond2013} on fossilized microbial mats called Kinneyia confirms this. Kinneyia are sedimentary fossils characterized by ripples with wavelength between $2$ and $20$~mm. Theory and experiments using an artificial biofilm 
shows that the rippled structures can be explained by the ancient flowing of water above the mats that produced a Kelvin--Helmholtz instability~\cite{thomasPhilTransRoySocLond2013}. This hydrodynamic instability occurs at the interface between two fluids with different viscosities and flowing with different velocities and produces ripples with wavelength proportional to the thickness of the biofilm. This theory predicts morphologies, wavelengths and amplitudes consonant with the fossil samples~\cite{thomasPhilTransRoySocLond2013}.

 Two important challenges have been identified for  future investigations of biofilms from the point of view of rheology and complex fluids~\cite{billingsRepPrgPhys2015}: (\emph{i}) the structure and morphology of biofilms depend on their history of growth and on the substratum; for example, the biofilm of \emph{Pseudomonas aeruginosa} growing in human lungs and airways incorporates DNA and the F-actin protein~\cite{walkerInfImm2005}, which modify the material properties of the biofilm with respect to a lab-grown sample. The thickness of the \emph{in vivo} biofilm of the fungus \emph{Candida albicans} is up to four times larger than the \emph{in vitro} biofilm~\cite{andesInfImm2004}.
(\emph{ii}) An integrated model of the biofilm's material properties should guide the understanding of its development and also of its dissolution, which is crucial in clinical settings. Helpful in this regard can be investigations of hydrogels (which are cross-linked networks of hydrophilic polymers) because these well-studied systems can illuminate  different aspects of the biofilm's physics. For example, physical models for the diffusion of macromolecules within gels~\cite{phillipsAICHEJ1989,masaroProgPolySci1999,amsdenMacromol1998} can instruct the understanding of biomolecules regulating the biofilm growth and the mechanism of drug delivery to destroy it. Self-healing polymeric materials~\cite{woolSoftMatt2008,wuProgPolymSci2008} can be a guide to biofilms as adaptive materials.
More progress is expected from the comparison with highly controllable artificial systems~\cite{stewartSciRep2015}.

\section{Integrated Experimental Methods}
\label{sec:experim}

Biofilms are complex systems as they have characteristic length and time scales spanning many orders of magnitude but that crucially interact with each other in a hierarchical way, and produce robust order~\cite{simon1991}. At the nanometer scale molecular processes such as gene expression, biosynthesis and the activity of the signal transduction networks take place. At the micrometer scale flagellar and ciliar motion governs the motility of the cells; diffusion of nutrients, and gradients in the chemoattractants and quorum sensing cues shape the average behavior of a group of cells. At larger scales hundreds of microorganisms form aggregates held together by pili and EPS; they form ripples and streamers under the influence of hydrodynamic forces.  The biofilm undergoes stages of maturation and dispersal, while some cells revert to the motile phenotype and colonize new surfaces~\cite{morgenrothRevEnvSciBiotech2009}. An understanding of the biofilm state must then integrate these different length and time scales involved.

Different ``integrated'', experimental approaches have already emerged. Microfluidic and lab-on-a-chip devices offer the possibility to include several functions in a microscopic environment, offer high throughput, and length and time scales compatible with the ecological conditions of biofilms~\cite{vertesAnalChem2012,karimiLOC2015}. For example, microfluidic devices which mimics xylem vessels were used to measure the adhesion forces of type I and type IV pili for \emph{Xylella fastidiosa}~\cite{delafuente2007}, or to characterize the adhesion, motility and biofilm formation of \emph{Acidovorax citrulli}~\cite{baharFEMSMBLett2010}. A nanoporous microfluidic platform was used to determine the amount of acetic acid necessary to impair the motility of \emph{Listeria monocytogenes}~\cite{wrightLabChip2014}. The biofilm growth and dispersal dynamics of \emph{Staphylococcus epidermidis} was investigated with a microfluidic device that allows to control the local hydrodynamic conditions~\cite{leeBiomedDev2008}. A microfluidic assay was used to develop an \emph{in situ} analytical system to assay the susceptibility of biofilms to antibiotics~\cite{kimLabChip2010}. Microsensors and microfluidics were employed to monitor phenotypic responses of fungal biofilms to changes in shear stress and antifungal drugs~\cite{richterLabChip2007}. A microfluidic biofilm signaling circuit that uses a population-driven quorum sensing switch can form or displace consortia of fungal biofilms~\cite{hongNatCom2012}. This platform allows for non-invasive, robust continuous monitoring of the biofilm conditions.  Microfabricated elastomer chips allow to monitor and study the evolution of bacterial and yeast colonies in a chemically controlled, dynamic environment~\cite{groismanNatMeth2005}. The architecture of the device resembles the complex structures naturally produced by biofilms, that is, clusters of cells separated by channels along which nutrients and waste products flow. A microfluidic porous device demonstrated that the formation of streamers takes place only in a certain flow rate range, and that streamers act as precursors to structures found in biofilm growing in porous media~\cite{valiei2LabChip012}. A microfluidic devices that incorporates a microbial fuel cell was used to generate electricity
and may find application as a small-scale biosensing device that offers great flexibility because of its lab-on-a-chip technology~\cite{liBiotechBioeng2011}. The bacteria used to produce power were the exoelectrogenic \emph{Geobacter sulfurreducens} and \emph{Shewanella oneidensis}.  These Gram-negative bacteria have the capability to exchange electrons with soluble and insoluble electron acceptors as part of their respiratory chain~\cite{nealsonAnnRevMB1994,gorbyPNAS2006}. Recently, a novel method that combines surface enhanced Raman spectroscopy and confocal Raman spectroscopy (SECRaM) was developed~\cite{schkolnikPLOS2015}. This method allows the \emph{in situ} spatial and temporal, analytic investigation of \emph{Shewanella oneidensis} biofilms without disturbing it.

Nanofabrication and microfluidics are excellent tools for manipulating the cells' microenvironment and monitor their responses~\cite{weibelNAtRevMB2007,holScience2014,rusconiAnnRevBiophys2014}. By using surfaces with a pattern of micropillars it was shown that \emph{Escherichia coli} bacteria use their flagella to reach inside the crevices between pillars and adhere to them when the spacing of the pillars is smaller than the cell size~\cite{friedlanderPNAS2013}. Thus, it was shown that flagella in addition to swimming motility create a dense fibrous network that gives structural stability to biofilms~\cite{friedlanderPNAS2013}. Microfluidic methods revealed that there is an underlying hydrodynamic mechanism that produces streamers, that is, the secondary vortical motion that is very pronounced around corners~\cite{rusconiInterface2010,rusconiBiophysJ2011,drescherPNAS2013}.

 Nanofabricated devices have the potential to unravel the complex structural organization of bacterial populations in natural environments, which are chemically and physically heterogeneous. Biofilms commonly 
contain more than one species, and therefore both cooperation and competition shape the community in  dynamic states of organization. Aspects of biofilm sociobiology~\cite{nadellFEMSMBRev2009} are now within the grasp of experimental testing.
\emph{Escherichia coli} colonizing an array of microhabitat patches formed a ``metapopulation'', that is a population of populations, that adapts quickly to habitats with high niche diversity~\cite{keymerPNAS2006}. Well-defined microscale spatial structures are a necessary and sufficient condition for the stability of bacterial communities engaged in ``reciprocal syntrophy'', that is a partition of functions all necessary for the survival of the whole community~\cite{kimPNAS2008}. Thus, the local topography is a crucial factor to account for in the biofilm's growth and viability. There are also efforts to move to three dimensions. A three-dimensional (3D) printing technique was used to study the cell-cell interactions in structured 3D  environments that model bacterial aggregates in the human body~\cite{connellPNAS2013}, for example. It was found that \emph{Staphylococcus aureus}, whose mono-species colonies are susceptible to ampicillin, is granted resistance to it by a community of \emph{Pseudomonas aeruginosa} forming a shell around it~\cite{connellPNAS2013}. However, much still remains to be understood, as it appears that subtle details often  determine the fate of the colony~\cite{holBMCResNot2015}.

These bottom-up approaches aimed at recreating the complex ecological landscape of bacterial communities in a controlled manner hold the promise to illuminate the behavior of single cells and the dynamical organization occurring in a biofilm. Similarly to the revolution in information technology brought forth by the miniaturization of semiconductor electronics, the fields of microfluidics and nanofabrication may usher a new level of understanding of microbiology.

\section{Theoretical and Computational Methods}
\label{sec:computer}

Theoretical models of biofilm growth and dynamics have started since the 1970s with simple one-dimensional differential equations describing a flat layer. In the following we will start by reviewing the basic concepts of that early era that originated from models of bacterial cultures~\cite{herbertJGenMB1956,vanUden1967} because, although our understanding of biofilms has advanced since then, these ideas and equations are still a part of the more complex models and methods considered in the modern scientific debate.

Consider a colony of microorganisms in a well-mixed bath of substrate (for example in a continuous culture apparatus).
The specific growth rate of the colony $\mu$ is written as
$\mu=\tfrac{1}{c}\tfrac{\mathrm{d}c}{\mathrm{d}t}=\frac{\mathrm{d}}{\mathrm{d}t}\ln c$
where $c$ is the concentration of cells. Monod~\cite{monod1942} famously related $\mu$ to the concentration $s$ of an essential growth substrate
\begin{equation}
\mu=\mu_m\left(\frac{s}{K_s+s}\right),
\end{equation}
where $\mu_m$ is the growth rate constant and $K_s$ is the saturation constant, that is, the substrate concentration at which $\mu=\frac{1}{2}\mu_m$. The growth rate is proportional to the substrate concentration when $s\ll K_s$, but it saturates at large values of $s$, $s\gg K_s$. A second fundamental law of Monod relates the growth rate to the substrate consumption rate
\begin{equation}\label{eq:second-monod}
\frac{\mathrm{d}c}{\mathrm{d}t}=-Y\frac{\mathrm{d}s}{\mathrm{d}t},
\end{equation}
where $Y$ is the yield constant, the ratio of mass of cells grown and mass of substrate used. Combining Eq.~\eqref{eq:second-monod} with the definition of $\mu$ yields the basic equations
\begin{equation}\label{eq:basic-monod}
\frac{\mathrm{d}s}{\mathrm{d}t}=-k c\frac{s}{K_s+s}\,,\quad \frac{\mathrm{d}c}{\mathrm{d}t}=\mu_m c\frac{s}{K_s+s}
\end{equation}
where $k\equiv\mu_m/Y$ is the maximum specific rate of substrate consumption. 

Consider now a biofilm immersed in a bulk liquid. For simplicity we assume a biofilm growing on a flat surface that we take as the $xy$-plane and with a $z$-axis normal to it. The substrate, that is, the energy source diffuses from the surface of the biofilm to the deeper layers, and its concentration decays with depth. Thus, there is an additional term in the balance equation for $s$ due to molecular diffusion
\begin{equation}
\frac{\mathrm{d}s}{\mathrm{d}t}=-k\frac{cs}{K_s+s}+D_s\frac{\partial^2s}{\partial z^2}\,,
\end{equation}
where $D_s$ is the diffusivity of substrate within the biofilm.
A biofilm is considered in a steady state if the total biomass is equal to what can be sustained by the flux of substrate~\cite{rittmannBiotechBioeng1980}. It is a statement of conservation of energy.
In a steady state, the net biofilm growth rate of cells  is~\cite{rittmannBiotechBioeng1980}
\begin{equation}
\frac{\mathrm{d}c}{\mathrm{d}t}=\mu_m c\left(\frac{s}{K_s+s}\right)-bc\,,
\end{equation}
where $b$ is the specific decay coefficient.

This and similar approaches~\cite{kisselJEnvEng1984,wannerBiotechBioeng1986,chaudhryChemEng1998} viewed biofilms as simple flat structures, governed by energy balance considerations. However, the development of techniques such as confocal scanning laser microscopy revealed that biofilms have a heterogeneous spatial structure interspersed with channels and chambers~\cite{lawrenceJBact1991,costertonARMB1995}. Thus, the next step in the mathematical modeling of biofilm was the  fully 3D description and the inclusion of a diffusive mechanism for biomass spreading  with a  diffusion coefficient that depends nonlinearly on the biomass concentration. Eberl \emph{et al.}~\cite{eberlJTheoMed2001} proposed the equations
\begin{gather}\label{eq:eberl-NS}
\nabla\cdot\vec{v}=0\,, \quad \frac{\partial\vec{v}}{\partial t}+\vec{v}\cdot\nabla\vec{v}=-\frac{1}{\rho}\nabla p+\nu\nabla^2\vec{v}\,,\\
\label{eq:eberl-nutri}
\frac{\partial s}{\partial t}+\vec{v}\cdot\nabla s=\nabla\cdot\left(D_1(c)\nabla s\right) -f(s,c)\,,\\
\label{eq:eberl-bio}
\frac{\partial c}{\partial t}=\nabla\cdot\left(D_2(c)\nabla c\right) +g(s,c)\,,\\
f(s,c)=\frac{k_1cs}{k_2+s}\,, \quad g(s,c)=k_3(f(s,c)-k_4c),
\end{gather}
where $\vec{v}$ is the flow velocity, $s$ the nutrient concentration, $c$ the biomass density, and $k_1\dots k_4$ are model parameters. Equations~\eqref{eq:eberl-NS} are the Navier--Stokes equations for an incompressible fluid with density $\rho$ and kinematic viscosity $\nu$. Nutrients spread through convection and diffusion and are consumed with a Monod reaction rate $f(s,c)$. The biomass is produced with a rate $g(s,c)$ and spreads through a diffusive flux $D_2\nabla c$ (Eq.~\eqref{eq:eberl-bio}), where $D_2(c)=(\frac{\epsilon}{c_{max}-c})^ac^b$. This functional form produces a biomass with a maximum density $c_{max}$, vanishing diffusion for $c\to0$, and significant biomass spreading only when $c\to c_{max}$.
This set of equations are studied for different values of an important dimensionless number $\mathcal{G}$ that quantifies the ratio of the biomass growth rate and substrate transport rate~\cite{picioreanuBiotechBioeng1998}, and produces heterogeneous, mushroom-like structures similar to what is observed in experiments~\cite{eberlJTheoMed2001,eberlchemEngSc2000}. This approach was generalized to multispecies biofilm with different solutes~\cite{nogueraWatSciTech1999}.

Experimental results in the mechanics of biofilms~\cite{stoodleyJIndMBBiotech2002,fuxJBacter2004} prompted a phenomenological description of the biofilm mechanics in terms of a multi-fluid model~\cite{klapperPRE2006}. Consider a biofilm with volume fraction $\phi_b$ immersed in a solvent with volume fraction $\phi_s$, where $\phi_s+\phi_b=1$. The fundamental quantity of interest is the cohesion energy functional for the biofilm
\begin{equation}\label{eq:enfunctional}
E[\phi_b]=\int\left(f(\phi_b)+\frac{\kappa}{2}|\nabla\phi_b|^2\right)\mathrm{d}V,
\end{equation}
where $f(\phi_b)$ is the energy density of the mixing of biomass and solvent, and the gradient term penalizes short-length scale variations by means of a surface energy. 
The shape of $f(\phi_b)$ can be deduced from general physical arguments.
The function $f(\phi_b)$ should have a minimum at some volume fraction because the cohesive forces of the biomass are balanced by the short-range repulsions. Because attractive forces are weak at long distance $f(\phi_b)\to0$ as $\phi_b\to0$, and because some amount of solvent is necessary, $f(\phi_b)$ must grow at large $\phi_b$. Using mass conservation for the biomass
$\frac{\mathrm{d}\phi_b}{\mathrm{d}t}+\nabla\cdot(\phi_b\vec{u}_b)=0$ one can write the time derivative of the energy
\begin{equation}
\frac{\mathrm{d}E}{\mathrm{d}t}=-\int \left(\phi_b\nabla\cdot\mathbf{\Pi}\right)\cdot\vec{u}_b\mathrm{d}V\,,
\end{equation}
where $\mathbf{\Pi}=-[f'(\phi_b)-\kappa\nabla^2\phi_b]\mathbf{I}$ is the cohesive stress tensor, and  $\vec{u}_b$ the velocity of the biomass. Neglecting the inertial, viscous and viscoelastic terms one finds the evolution equation for $\phi_b$~\cite{klapperPRE2006,klapperSIAMRev2010,wangSolStComm2010}
\begin{multline}\label{eq:cohesive}
\frac{\partial\phi_b}{\partial t}+\nabla\cdot(\phi_b\vec{u}_b)=\nabla\cdot\left[a(\phi_b)f''(\phi_b)\nabla\phi_b\right]-\\
\kappa\nabla\cdot\left[a(\phi_b)\nabla\nabla^2\phi_b\right],
\end{multline}
where $a(\phi_b)=\zeta_0^{-1}\phi_b(1-\phi_b)$, and $\zeta_0$ is a friction constant. This model was studied in one dimension with $\vec{u}_b=0$ (so that Eq.~\eqref{eq:cohesive} becomes a modified Cahn--Hilliard equation) and a free surface film state  is found to form, where spinodal decomposition is a possible mechanism~\cite{klapperPRE2006}. This approach has been extended to include EPS with an additional phase field, which reproduces the cohesive failure of biofilm under shear flow~\cite{tierraJRoySocInterf2015}, and with the inclusion of a viscoelastic model of the EPS~\cite{lindleyPRE2012}.

In contrast to the models based on Eq.~\eqref{eq:enfunctional},
a different paradigm of biofilm growth has been proposed, where the biofilm is modeled with stochastic rules that evolve in a discrete way~\cite{fujikawaJPhysSocJap1989}, as opposed to the continuous changes described by partial differential equations.
In these early computational attempts the growth of a bacterial colony was studied~\cite{matsushitaPhysA1990,fujikawaFEMSMBEco1994,benNature1994}. 
For example, experiments on \emph{Bacillus subtilis} growing on agar plates were compared with three models of stochastic growth: (\emph{i}) diffusion limited aggregation, (\emph{ii}) the Eden model, and (\emph{iii}) the dense branching morphology. Diffusion limited aggregation describes the growth process of small particles in a dilute solution that aggregate when their Brownian trajectories bring them in close contact~\cite{wittenPRL1981}. Coagulated aerosols or zinc ions in solution, for example, exhibit open, randomly branched structures that can be modeled by diffusion limited aggregation. A simple method to reproduce computationally diffusion limited aggregation is as follows. Consider a 2D square lattice at whose center the first seed is planted; next, a new particle is placed far away from the seed; it then undergoes a random walk on the lattice until it reaches a lattice site neighboring the initial seed, at which point they stick together. The procedure is repeated and new particles stick to the cluster whenever they reach one of its neighboring sites. Alternatively, in the Eden growth model new particles are added next to the perimeter of the cluster~\cite{eden1961}. When starting from a seed in a planar geometry, compact, disk-like clusters are obtained. Finally, an addition of a kinetic term to the growth process leads to the increase of branching tips as the radius of the agglomerate increases. This process produces the dense branching morphology, which is characterized by compact branches with a neat spherical envelope~\cite{benPRL1986,meakin1998fractals,stanley2012}. Fujikawa~\cite{fujikawaFEMSMBEco1994} found that at relative low nutrient concentrations the colony formed open fractal branches, similar to the simulations of diffusion limited aggregation, and measured a fractal dimension $D_f=1.73\pm0.02$ which compares well with known theoretical values~\cite{hayakawaPRL1997}. For higher nutrient concentrations, the colony grew compact and round, similarly to an Eden cluster, as the cells do not have to rely on the diffusion of nutrients from the surrounding. At low agar concentrations, instead, patterns similar to the dense branching morphology were found, in which case the reduced viscosity allows a growth driven by the flagellar motion.

These stochastic, discrete approaches eventually produced the widely-used class of cellular automata models~\cite{halleyOikos1994,wimpennyFEMSMBEco1997,hermanowiczWatSciTech1999,picioreanuBiotechBioeng1998,picioreanuBiotechBioeng2000,pizarroJEnvEng2001,tangWatRes2013}.
Generally, cellular automata models consider a 2D space partitioned with a lattice. Every lattice location may be occupied by a parcel of biofilm consisting of cells, EPS, and other molecules. Additionally, fields representing nutrients or waste products can also be distributed on the lattice.  Each lattice location is updated at discrete time steps according to simple rules.
Given a cell in a specific location, the neighboring compartments are searched for empty space, and if the nutrients in that location are sufficient the cell multiplies and occupies that empty location. Nutrients diffuse stochastically through the lattice.
More specific rules informed by biological interactions can be implemented. Although cellular automata models appear as a drastic simplification, they can produce complex structures because of their nonlinear nature~\cite{wolframNature1984}.

A more recent effort at integrating the experimental knowledge about biofilm growth tackles the challenge of modeling its heterogeneous structure and composition~\cite{laspidouWatRes2004a,laspidouWatRes2004b}. They consider three kinds of solid components (the active biomass of microorganisms, the EPS, and the inert biomass) and three kinds of soluble organic compounds (the bacteria's limiting donor substrate, and two types of soluble microbial products, which are excreted by cells or released during cell lysis)~\cite{laspidouWatRes2002b,laspidouWatRes2002a}. The concentrations of all these components evolve in time and space according to partial differential equations that include diffusion, Monod kinetics for the donor substrate, cellular utilization and production, decay of the active biomass and detachment forces. Importantly, biofilm consolidation is included, that is, over time the deeper layers of the biofilm become denser. Coupled to the dynamics of the concentrations of the different biofilm components, the mechanical spreading of active biomass and EPS is implemented through a cellular automata scheme. Once the total density in a compartment exceeds a threshold value, the ``mother'' cell divides into two ``daughter'' cells, where one of them moves to a neighboring compartment. Were all neighboring compartments occupied, the daughter cell would push their contents to an unoccupied compartment along the shortest path identified by the algorithm. The threshold density for each compartment grows with time. This fact models the consolidation process due to physical forces acting on the biofilm, for example, hydrodynamic forces tend to induce the solid components to pack more densely.

An alternative class of models considers the individual microbial cell, or small groups thereof, as the elementary computational unit~\cite{kreftMB1998,kreftMB2001,kreftWatSciTech2001,picioreanuApplEnvMB2004,xavierEnvMB2005,picioreanuWatSciTech2007}. They differ from cellular automata models as there is no underlying lattice partitioning space. They are called individual-based models, particle-based models, or more generally agent-based models.
The same basic rules as the cellular automata models for metabolism, substrate uptake, cell multiplication and death are used but no global rules on the growth are imposed. The advantages of agent based simulations are multiple. First, the absence of a lattice grid avoids artificial results such as fixed growth directions. Second, no global laws (for the growth, e.g.) are prescribed, but must emerge from the microscopic dynamics. Third, cells are modeled as hard spheres, and their overlap produces repulsive forces that induces naturally the spreading of the biofilm. Furthermore, since individual agents are resolved, it is also natural to introduce genetic variability in the cell's parameters, cell motility~\cite{picioreanuWatSciTech2007}, coupling with hydrodynamic flow~\cite{vonderSchulenburgAICHEJ2009},  inclusion of metabolic reactions~\cite{lardonEnvMB2011}, or a  viscous fluid model for the EPS~\cite{alpkvistBiotechBioeng2006}.  However, there are also drawbacks~\cite{laspidouDesal2010}: for example, the computational cost of this class of models is typically larger than cellular automata's, and the large-scale structure of biofilm, such as porosity, are less faithfully represented.

\section{Conclusions}
\label{sec:concl}

Biofilms are among the most ancient and most common forms of multicellular organization. They exhibit astounding complexity both at microscopic and mesoscopic length scales. Their importance has been recognized only in recent decades and research in this field is growing tremendously, fueled by the possibility of  treating infectious diseases or of benefiting from their appetite for organic materials in wastewater applications. 

We have described the main physical processes involved in the  inception, growth and dispersal of biofilms. A hydrodynamic description is necessary for the initial transition of microbial cells from the planktonic to the sessile state. Details of the physico-chemical nature of the cellular membrane and its appendages are key for the process of adhesion. Biofilms are macroscopically viscoelastic materials with a very complex, history dependent behavior. Progress in the rheological and mechanical characterization will come from studies in conditions mimicking the natural environment. 

We have also reviewed modern experimental techniques of investigation such as microfluidics and lab-on-a-chip technologies that use an integrated approach to biofilms with the aim at capturing their heterogeneous and complex nature. These methods can experimentally ``simulate'' the natural environment of a biofilm and test different hypotheses. Realistic environments are however essential to understand and control the dispersal of biofilms, for example, which has tremendous consequences for ecology and in the context of infectious diseases. Studies of biofilm dispersal, however, are still largely conducted \emph{in vitro}~\cite{kaplanJDentRes2010}.

Finally, we have reviewed the theoretical and computational modeling of biofilms. Computational approaches at simulating a biofilm are now finally coming of age~\cite{kreftPNAS2013} and some predictions quantitatively match the experimental data, but much remains to be done. Ahead lies a particularly daunting challenge when considering all the chemical, biological and physical processes involved in a biofilm, which evolve and interact from the nanometer to the millimeter length scale, and from the nanosecond time scale to hours or days. Future lines of development should include realistic models for the EPS, either microscopically with a polymer network theory, or mesoscopically with a viscoelastic model. Some attempts are already available~\cite{alpkvistWatSciTech2007} where the biofilm is modeled with breakable springs in an immersed boundary method, or through a nonlinear, continuum description of worm-like chains~\cite{ehretJRoySocInterf2013}. Once integrated theoretical and computational models will be developed, they will allow to test and manipulate biofilms under multiple physical, chemical and biological cues, both intrinsic and extrinsic.


 We now want to enter a more speculative realm to discuss some mechanisms at play in the biofilm that could be recognized in the future. The important fact that biofilms represent the most complex effort of simple microorganisms at multicellular organization points to the existence of `communications' channels among the individual cells. We know from the biology of multicellular eukaryotes how a complex, well-regulated and specialized structure may exist. Biofilms are more dynamical and history-dependent in their structure and properties. However, interactions that lead to specialization must be there also in biofilms. Obvious candidates
are quorum sensing and cell signaling. Although the role of quorum sensing has been linked multiple times with the growth, structure and function of biofilms, controversial results abound (see Ref.~\cite{parsekTrMB2005} for a critical review of quorum sensing in relation to biofilms).  The reason for the controversy is probably the inadequacy of our experimental tools of investigation to the task: more subtle tools for \emph{in vivo} and \emph{in situ} analysis are required; another factor that complicates the picture is that the study of mutants elicits the risk of pleiotropy. We speculate that quorum sensing and cell signaling provide the communication channels required to generate complex organization. 
A second physical aspect that is underrepresented in the study of biofilms is the role of motility. We have reviewed the available evidence on the function of swimming, swarming and twitching within a biofilm. We speculate, however, that these motility modes should have a more important role than currently known. While under some physical forces, such as hydrodynamic shear, a biofilm might respond by forming more compact structures with a sessile population, consortia of different species coexisting, possibly in  syntrophic interaction, may utilize motility to assist the spatial segregation often associated with phenotypic differentiation. Biofilms represent  still largely unexplored microcosmoses, that will certainly provide numerous surprises in the future.

In this Review, we have omitted the discussion of some crucial aspects in the life-cycle of a biofilm: its ecology and genetic regulation. There is certainly ample space for physical theories of these different levels of interaction.

We gratefully acknowledge Gal Schkolnik for helpful conversations and for critically reading the manuscript. We thank Paul Stoodley for kindly providing the images in Fig.~\ref{fig:biofilm-examples}.


\begin{thebibliography}{100}
\expandafter\ifx\csname url\endcsname\relax
  \def\url#1{{\tt #1}}\fi
\expandafter\ifx\csname urlprefix\endcsname\relax\def\urlprefix{URL }\fi
\providecommand{\eprint}[2][]{\url{#2}}

\bibitem{noffkeAstrobiology2013}
Noffke N, Christian D, Wacey D and Hazen R~M 2013 {\em Astrobiology\/} {\bf 13}
  1103--1124

\bibitem{VanKranendonkPreRea2008}
Van~Kranendonk M~J, Philippot P, Lepot K, Bodorkos S and Pirajno F 2008 {\em
  Precambrian Res.\/} {\bf 167} 93--124

\bibitem{flemmingNatRM2010}
Flemming H~C and Wingender J 2010 {\em Nat. Rev. Microbiol.\/} {\bf 8} 623--633

\bibitem{hallNatRM2004}
Hall-Stoodley L, Costerton J~W and Stoodley P 2004 {\em Nat. Rev. Microbiol.\/}
  {\bf 2} 95--108

\bibitem{orellARMB2013}
Orell A, Fr{\"o}ls S and Albers S~V 2013 {\em Annu. Rev. Microbiol.\/} {\bf 67}
  337--354

\bibitem{Leadbeater1992}
Leadbeater B~S~C and Callow M~E 1992 Formation, composition and physiology of
  algal biofilms {\em Biofilms -- Science and Technology\/} ({\em NATO ASI
  Series\/} vol 223) ed Melo L~F, Bott T~R, Fletcher M and Capdeville B
  (Springer Netherlands) pp 149--162

\bibitem{fanningPLOSPATHOG2012}
Fanning S and Mitchell A~P 2012 {\em PLoS Pathog.\/} {\bf 8} e1002585

\bibitem{ramageCRMB2009}
Ramage G, Mowat E, Jones B, Williams C and Lopez-Ribot J 2009 {\em Crit. Rev.
  Microbiol.\/} {\bf 35} 340--355

\bibitem{porterBacterRev1976}
Porter J~R 1976 {\em Bacteriol. Rev.\/} {\bf 40} 260

\bibitem{henriciJBacter1933}
Henrici A~T 1933 {\em J. Bacteriol.\/} {\bf 25} 277

\bibitem{zobellJBacter1943}
Zobell C~E 1943 {\em J. Bacteriol.\/} {\bf 46} 39

\bibitem{mccoyCanJMB1981}
McCoy W~F, Bryers J~D, Robbins J and Costerton J~W 1981 {\em Can. J.
  Microbiol.\/} {\bf 27} 910--917

\bibitem{otooleARMB2000}
O'Toole G, Kaplan H~B and Kolter R 2000 {\em Annu. Rev. Microbiol.\/} {\bf 54}
  49--79

\bibitem{vertPAChem2012}
Vert M, Hellwich K~H, Hess M, Hodge P, Kubisa P, Rinaudo M and Schu{\'e} F 2012
  {\em Pure Appl. Chem.\/} {\bf 84} 377--410

\bibitem{watnickJBacter2000}
Watnick P and Kolter R 2000 {\em J. Bacteriol.\/} {\bf 182} 2675--2679

\bibitem{berkScience2012}
Berk V, Fong J~C~N, Dempsey G~T, Develioglu O~N, Zhuang X, Liphardt J, Yildiz
  F~H and Chu S 2012 {\em Science\/} {\bf 337} 236--239

\bibitem{reysenbachTrendsMB2001}
Reysenbach A~L and Cady S~L 2001 {\em Trends Microbiol.\/} {\bf 9} 79--86

\bibitem{taylorAEMB1999}
Taylor C~D, Wirsen C~O and Gaill F 1999 {\em Appl. Environ. Microbiol.\/} {\bf
  65} 2253--2255

\bibitem{wotton2005surface}
Wotton R~S and Preston T~M 2005 {\em BioScience\/} {\bf 55} 137--145

\bibitem{atekwanaRGeophysics2009}
Atekwana E~A and Slater L~D 2009 {\em Rev. Geophys.\/} {\bf 47}

\bibitem{SihorkarPharmaReas2001}
Sihorkar V and Vyas S~P 2001 {\em Pharm. Res.\/} {\bf 18} 1247--1254

\bibitem{pavithraBiomedMaterials2008}
Pavithra D and Doble M 2008 {\em Biomed. Mater.\/} {\bf 3} 034003

\bibitem{haoCREST1996}
Hao O~J, Chen J~M, Huang L and Buglass R~L 1996 {\em Crit. Rev. Env. Sci.
  Tec.\/} {\bf 26} 155--187

\bibitem{hamiltonAnnRevMB1985}
Hamilton W~A 1985 {\em Annu. Rev. Microbiol.\/} {\bf 39} 195--217

\bibitem{muyzerNatRevMB2008}
Muyzer G and Stams A~J~M 2008 {\em Nat. Rev. Microbiol.\/} {\bf 6} 441--454

\bibitem{nicolellaJBiotech2000}
Nicolella C, Van~Loosdrecht M~C~M and Heijnen J~J 2000 {\em J. Biotechnol.\/}
  {\bf 80} 1--33

\bibitem{duBiotechAdv2007}
Du Z, Li H and Gu T 2007 {\em Biotechnol. Adv.\/} {\bf 25} 464--482

\bibitem{DeBeer2013}
De~Beer D and Stoodley P 2013 Microbial biofilms {\em The Prokaryotes\/} ed
  Rosenberg E, DeLong E, Lory S, Stackebrandt E and Thompson F (Springer Berlin
  Heidelberg) pp 343--372

\bibitem{dielsRevEnvSciTech2002}
Diels L, van~der Lelie N and Bastiaens L 2002 {\em Rev. Env. Sci.
  Biotechnol.\/} {\bf 1} 75--82

\bibitem{gaddGeoder2004}
Gadd G~M 2004 {\em Geoderma\/} {\bf 122} 109--119

\bibitem{sandApplMBBiotech1995}
Sand W, Gerke T, Hallmann R and Schippers A 1995 {\em Appl. Microbiol. Biot.\/}
  {\bf 43} 961--966

\bibitem{suzukiBiotechAdv2001}
Suzuki I 2001 {\em Biotechnol. Adv.\/} {\bf 19} 119--132

\bibitem{boseckerFEMSMBRev1997}
Bosecker K 1997 {\em FEMS Microbiol. Rev.\/} {\bf 20} 591--604

\bibitem{lawrenceJBact1991}
Lawrence J~R, Korber D~R, Hoyle B~D, Costerton J~W and Caldwell D~E 1991 {\em
  J. Bacteriol.\/} {\bf 173} 6558--6567

\bibitem{caldwellJApplBact1993}
Caldwell D~E, Korber D~R and Lawrence J~R 1993 {\em J. Appl. Bacteriol.\/} {\bf
  74} 52S--66S

\bibitem{costertonJBact1994}
Costerton J~W, Lewandowski Z, DeBeer D, Caldwell D, Korber D and James G 1994
  {\em J. Bacteriol.\/} {\bf 176} 2137

\bibitem{gjaltemaBiotechBioeng1994}
Gjaltema A, Arts P~A~M, Van~Loosdrecht M~C~M, Kuenen J~G and Heijnen J~J 1994
  {\em Biotechnol. Bioeng.\/} {\bf 44} 194--204

\bibitem{stoodleyBiotechBioeng1999}
Stoodley P, Lewandowski Z, Boyle J~D and Lappin-Scott H~M 1999 {\em Biotechnol.
  Bioeng.\/} {\bf 65} 83--92

\bibitem{stoodleyAnnRevMB2002}
Stoodley P, Sauer K, Davies D~G and Costerton J~W 2002 {\em Annu. Rev.
  Microbiol.\/} {\bf 56} 187--209

\bibitem{blauertBiotechBioeng2015}
Blauert F, Horn H and Wagner M 2015 {\em Biotechnol. Bioeng.\/}

\bibitem{madsenFEMSImmun2012}
Madsen J~S, Burm{\o}lle M, Hansen L~H and S{\o}rensen S~J 2012 {\em FEMS
  Immunol. Med. Microbiol.\/} {\bf 65} 183--195

\bibitem{wingenderMEthEnzym2001}
Wingender J, Strathmann M, Rode A, Leis A and Flemming H~C 2001 {\em Methods
  Enzymol.\/}  302--14

\bibitem{flemmingJBact2007}
Flemming H~C, Neu T~R and Wozniak D~J 2007 {\em J. Bacteriol.\/} {\bf 189}
  7945--7947

\bibitem{christensenJBiotech1989}
Christensen B~E 1989 {\em J. Biotechnol.\/} {\bf 10} 181--202

\bibitem{sutherland1977}
Sutherland I~W and Suthe I 1977 {\em Surface Carbohydrates of the Prokaryotic
  Cell\/} (Academic Press London)

\bibitem{sutherlandMB2001}
Sutherland I~W 2001 {\em Microbiology\/} {\bf 147} 3--9

\bibitem{wingender2012}
Wingender J, Neu T~R and Flemming H~C 2012 {\em Microbial Extracellular
  Polymeric Substances: Characterization, Structure and Function\/} (Springer
  Science \& Business Media)

\bibitem{lawrenceAEMB2003}
Lawrence J~R, Swerhone G~D~W, Leppard G~G, Araki T, Zhang X, West M~M and
  Hitchcock A~P 2003 {\em Appl. Environ. Microb.\/} {\bf 69} 5543--5554

\bibitem{ryderCurrOpMB2007}
Ryder C, Byrd M and Wozniak D~J 2007 {\em Curr. Opin. Microbiol.\/} {\bf 10}
  644--648

\bibitem{colvinEnvirMB2012}
Colvin K~M, Irie Y, Tart C~S, Urbano R, Whitney J~C, Ryder C, Howell P~L,
  Wozniak D~J and Parsek M~R 2012 {\em Environ. Microbiol.\/} {\bf 14}
  1913--1928

\bibitem{maJBacter2006}
Ma L, Jackson K~D, Landry R~M, Parsek M~R and Wozniak D~J 2006 {\em J.
  Bacteriol.\/} {\bf 188} 8213--8221

\bibitem{cooleySoftMatt2013}
Cooley B~J, Thatcher T~W, Hashmi S~M, L'Her G, Le H~H, Hurwitz D~A, Provenzano
  D, Touhami A and Gordon V~D 2013 {\em Soft Matter\/} {\bf 9} 3871--3876

\bibitem{donlanEID2002}
Donlan R~M 2002 {\em Emerg. Infect. Dis.\/} {\bf 8} 881--890

\bibitem{spiersMolMB2003}
Spiers A~J, Bohannon J, Gehrig S~M and Rainey P~B 2003 {\em Mol. Microbiol.\/}
  {\bf 50} 15--27

\bibitem{gotzMolMB2002}
G{\"o}tz F 2002 {\em Mol. Microbiol.\/} {\bf 43} 1367--1378

\bibitem{mackJBacter1996}
Mack D, Fischer W, Krokotsch A, Leopold K, Hartmann R, Egge H and Laufs R 1996
  {\em J. Bacteriol.\/} {\bf 178} 175--183

\bibitem{seminaraPNAS2012}
Seminara A, Angelini T~E, Wilking J~N, Vlamakis H, Ebrahim S, Kolter R, Weitz
  D~A and Brenner M~P 2012 {\em Proc. Natl. Acad. Sci. USA\/} {\bf 109}
  1116--1121

\bibitem{wilkingMRS2011}
Wilking J~N, Angelini T~E, Seminara A, Brenner M~P and Weitz D~A 2011 {\em MRS
  Bull.\/} {\bf 36} 385--391

\bibitem{rubinstein2003}
Rubinstein M and Colby R~H 2003 {\em Polymer Physics\/} (Oxford University
  Press, Oxford)

\bibitem{orVadose2007}
Or D, Phutane S and Dechesne A 2007 {\em Vadose Zone J.\/} {\bf 6} 298--305

\bibitem{freemanLimnolOcean1995}
Freeman C and Lock M~A 1995 {\em Limnol. Oceanogr.\/} {\bf 40} 273--278

\bibitem{norbergBiotechBioeng1984}
Norberg A~B and Persson H 1984 {\em Biotechnol. Bioeng.\/} {\bf 26} 239--246

\bibitem{teitzelApplEnvMB2003}
Teitzel G~M and Parsek M~R 2003 {\em Appl. Environ. Microbiol.\/} {\bf 69}
  2313--2320

\bibitem{hentzer2001alginate}
Hentzer M, Teitzel G~M, Balzer G~J, Heydorn A, Molin S, Givskov M and Parsek
  M~R 2001 {\em J Bacteriol.\/} {\bf 183} 5395--5401

\bibitem{colvinPLOS2011}
Colvin K~M, Gordon V~D, Murakami K, Borlee B~R, Wozniak D~J, Wong G~C and
  Parsek M~R 2011 {\em PLoS Pathog.\/} {\bf 7} e1001264--e1001264

\bibitem{billingsPLOS2013}
Billings N, Ramirez~Millan M, Caldara M, Rusconi R, Tarasova Y, Stocker R and
  Ribbeck K 2013 {\em PLoS Pathog.\/} {\bf 9} e1003526

\bibitem{wingender1999}
Wingender J, Jaeger K~E and Flemming H~C 1999 Interaction between extracellular
  polysaccharides and enzymes {\em Microbial Extracellular Polymeric
  Substances\/} (Springer) pp 231--251

\bibitem{lockOikos1984}
Lock M~A, Wallace R~R, Costerton J~W, Ventullo R~M and Charlton S~E 1984 {\em
  Oikos\/} {\bf 42} 10--22

\bibitem{oteroJPhyco2004}
Otero A and Vincenzini M 2004 {\em J. Phycol.\/} {\bf 40} 74--81

\bibitem{steinbergerApplEnvMB2005}
Steinberger R and Holden P 2005 {\em Appl. Environ. Microbiol.\/} {\bf 71}
  5404--5410

\bibitem{yangMB2007}
Yang L, Barken K~B, Skindersoe M~E, Christensen A~B, Givskov M and
  Tolker-Nielsen T 2007 {\em Microbiology\/} {\bf 153} 1318--1328

\bibitem{whitchurchScience2002}
Whitchurch C~B, Tolker-Nielsen T, Ragas P~C and Mattick J~S 2002 {\em
  Science\/} {\bf 295} 1487--1487

\bibitem{vilainApplEnvMB2009}
Vilain S, Pretorius J~M, Theron J and Br{\"o}zel V~S 2009 {\em Appl. Environ.
  Microbiol.\/} {\bf 75} 2861--2868

\bibitem{mulcahyPLOSPath2008}
Mulcahy H, Charron-Mazenod L and Lewenza S 2008 {\em PLoS Pathog.\/} {\bf 4}
  e1000213

\bibitem{ahimouEnzMBTech2000}
Ahimou F, Jacques P and Deleu M 2000 {\em Enzyme Microb. Tech.\/} {\bf 27}
  749--754

\bibitem{irie2005pseudomonas}
Irie Y, O'toole G~A and Yuk M~H 2005 {\em FEMS Microbiol. Lett.\/} {\bf 250}
  237--243

\bibitem{rodriguesJAMB2006}
Rodrigues L~R, Banat I~M, Mei H~C, Teixeira J~A and Oliveira R 2006 {\em J.
  Appl. Microbiol.\/} {\bf 100} 470--480

\bibitem{RivardoApplMBBiotech2009}
Rivardo F, Turner R~J, Allegrone G, Ceri H and Martinotti M~G 2009 {\em Appl.
  Microbiol. Biot.\/} {\bf 83} 541--553

\bibitem{romaniMBEcol2008}
Roman{\'\i} A~M, Fund K, Artigas J, Schwartz T, Sabater S and Obst U 2008 {\em
  Microb. Ecol.\/} {\bf 56} 427--436

\bibitem{zhangChemosph2003}
Zhang X and Bishop P~L 2003 {\em Chemosphere\/} {\bf 50} 63--69

\bibitem{elkinsApplEnvironMB1999}
Elkins J~G, Hassett D~J, Stewart P~S, Schweizer H~P and McDermott T~R 1999 {\em
  Appl. Environ. Microb.\/} {\bf 65} 4594--4600

\bibitem{barondesScience1984}
Barondes S~H 1984 {\em Science\/} {\bf 223} 1259--1264

\bibitem{diggleEnvMB2006}
Diggle S~P, Stacey R~E, Dodd C, C{\'a}mara M, Williams P and Winzer K 2006 {\em
  Environ. Microb.\/} {\bf 8} 1095--1104

\bibitem{perezIntJMB2009}
P{\'e}rez-Gim{\'e}nez J, Mongiardini E~J, Althabegoiti M~J, Covelli J, Quelas
  J~I, L{\'o}pez-Garc{\'\i}a S~L and Lodeiro A~R 2009 {\em Int. J.
  Microbiol.\/} {\bf 2009} 719367

\bibitem{karatanMBMolBiolRev2009}
Karatan E and Watnick P 2009 {\em Microbiol. Mol. Biol. R.\/} {\bf 73} 310--347

\bibitem{sauerJBact2004}
Sauer K, Cullen M~C, Rickard A~H, Zeef L~A~H, Davies D~G and Gilbert P 2004
  {\em J Bacteriol.\/} {\bf 186} 7312--7326

\bibitem{bolesMolMB2005}
Boles B~R, Thoendel M and Singh P~K 2005 {\em Mol. Microbiol.\/} {\bf 57}
  1210--1223

\bibitem{jacksonJBact2002}
Jackson D~W, Suzuki K, Oakford L, Simecka J~W, Hart M~E and Romeo T 2002 {\em
  Journal of bacteriology\/} {\bf 184} 290--301

\bibitem{mcdougaldNatRevMB2012}
McDougald D, Rice S~A, Barraud N, Steinberg P~D and Kjelleberg S 2012 {\em Nat.
  Rev. Microbiol.\/} {\bf 10} 39--50

\bibitem{crespiTrendsEcolEvol2001}
Crespi B~J 2001 {\em Trends Ecol. Evol.\/} {\bf 16} 178--183

\bibitem{shapiroAnnRevMB1998}
Shapiro J~A 1998 {\em Annu. Rev. Microbiol.\/} {\bf 52} 81--104

\bibitem{benAdvPhys2000}
Ben-Jacob E, Cohen I and Levine H 2000 {\em Adv. Phys.\/} {\bf 49} 395--554

\bibitem{kaplanJDentRes2010}
Kaplan J~B 2010 {\em J. Dent. Res.\/} {\bf 89} 205--218

\bibitem{teschlerNRMB2015}
Teschler J~K, Zamorano-S{\'a}nchez D, Utada A~S, Warner C~J~A, Wong G~C~L,
  Linington R~G and Yildiz F~H 2015 {\em Nat. Rev. Microbiol.\/} {\bf 13}
  255--268

\bibitem{sauerJBact2002}
Sauer K, Camper A~K, Ehrlich G~D, Costerton J~W and Davies D~G 2002 {\em J.
  Bacteriol.\/} {\bf 184} 1140--1154

\bibitem{tolkerJBact2000}
Tolker-Nielsen T, Brinch U~C, Ragas P~C, Andersen J~B, Jacobsen C~S and Molin S
  2000 {\em J. Bacteriol.\/} {\bf 182} 6482--6489

\bibitem{barkenEnvMB2008}
Barken K~B, Pamp S~J, Yang L, Gjermansen M, Bertrand J~J, Klausen M, Givskov M,
  Whitchurch C~B, Engel J~N and Tolker-Nielsen T 2008 {\em Environ.
  Microbiol.\/} {\bf 10} 2331--2343

\bibitem{vlamakisGeneDevel2008}
Vlamakis H, Aguilar C, Losick R and Kolter R 2008 {\em Genes Dev.\/} {\bf 22}
  945--953

\bibitem{henrichsenBactRev1972}
Henrichsen J 1972 {\em Bacteriol Rev.\/} {\bf 36} 478

\bibitem{turnerJBact2000}
Turner L, Ryu W~S and Berg H~C 2000 {\em J. Bacteriol.\/} {\bf 182} 2793--2801

\bibitem{bergCurrBiol2008}
Berg H~C 2008 {\em Curr. Biol.\/} {\bf 18} R689--R691

\bibitem{parkinsonCell1993}
Parkinson J~S 1993 {\em Cell\/} {\bf 73} 857--871

\bibitem{blockNature1989}
Block S~M, Blair D~F and Berg H~C 1989 {\em Nature\/} {\bf 338} 514--518

\bibitem{bergAnnRevBiochem2003}
Berg H~C 2003 {\em Ann. Rev. Biochem.\/} {\bf 72} 19

\bibitem{laugaRepProgPhys2009}
Lauga E and Powers T~R 2009 {\em Rep. Prog. Phys.\/} {\bf 72} 096601

\bibitem{kochARFM2011}
Koch D~L and Subramanian G 2011 {\em Annu. Rev. Fluid Mech.\/} {\bf 43}
  637--659

\bibitem{happel2012}
Happel J and Brenner H 2012 {\em Low Reynolds Number Hydrodynamics: with
  special applications to particulate media\/} (Springer Science \& Business
  Media)

\bibitem{copelandSoftMatt2009}
Copeland M~F and Weibel D~B 2009 {\em Soft Matter\/} {\bf 5} 1174--1187

\bibitem{guyonBookhydrodyn}
Guyon E, Hulin J~P, Petit L and Mitescu C~D 2001 {\em Physical Hydrodynamics\/}
  (Springer, Berlin Heidelberg, New York)

\bibitem{purcellAJP1977}
Purcell E~M 1977 {\em Am. J. Phys\/} {\bf 45} 3--11

\bibitem{baskaranPNAS2009}
Baskaran A and Marchetti M~C 2009 {\em Proc. Natl. Acad. Sci. USA\/} {\bf 106}
  15567--15572

\bibitem{SimhaPRL2002}
Simha R~A and Ramaswamy S 2002 {\em Phys. Rev. Lett.\/} {\bf 89}(5) 058101

\bibitem{SaintillanPRL2007}
Saintillan D and Shelley M~J 2007 {\em Phys. Rev. Lett.\/} {\bf 99}(5) 058102

\bibitem{kesslerBookChap1997}
Kessler J~O and Wojciechowski M~F 1997 Collective behavior and dynamics of
  swimming bacteria {\em Bacteria as multicellular organisms\/} ed Shapiro J~A
  and Dworkin M (Oxford University Press) pp 417--450

\bibitem{DombrowskiPRL2004}
Dombrowski C, Cisneros L, Chatkaew S, Goldstein R~E and Kessler J~O 2004 {\em
  Phys. Rev. Lett.\/} {\bf 93}(9) 098103

\bibitem{sokolovPRL2012}
Sokolov A and Aranson I~S 2012 {\em Phys. Rev. Lett.\/} {\bf 109} 248109

\bibitem{dunkelPRL2013}
Dunkel J, Heidenreich S, Drescher K, Wensink H~H, B{\"a}r M and Goldstein R~E
  2013 {\em Phys. Rev. Lett.\/} {\bf 110} 228102

\bibitem{zottlPRL2014}
Z{\"o}ttl A and Stark H 2014 {\em Phys. Rev. Lett.\/} {\bf 112} 118101

\bibitem{wensinkPNAS2012}
Wensink H~H, Dunkel J, Heidenreich S, Drescher K, Goldstein R~E, L{\"o}wen H
  and Yeomans J~M 2012 {\em Proc. Natl. Acad. Sci. USA\/} {\bf 109}
  14308--14313

\bibitem{drescherPNAS2011}
Drescher K, Dunkel J, Cisneros L~H, Ganguly S and Goldstein R~E 2011 {\em Proc.
  Natl. Acad. Sci. USA\/} {\bf 108} 10940--10945

\bibitem{lushiPNAS2014}
Lushi E, Wioland H and Goldstein R~E 2014 {\em Proc. Natl. Acad. Sci. USA\/}
  201405698

\bibitem{katzJFM1974}
Katz D~F 1974 {\em J. Fluid Mech.\/} {\bf 64} 33--49

\bibitem{brennenARFM1977}
Brennen C and Winet H 1977 {\em Annu. Rev. Fluid Mech.\/} {\bf 9} 339--398

\bibitem{diluzioNature2005}
DiLuzio W~R, Turner L, Mayer M, Garstecki P, Weibel D~B, Berg H~C and
  Whitesides G~M 2005 {\em Nature\/} {\bf 435} 1271--1274

\bibitem{kudoFEMSMicrobioLett2005}
Kudo S, Imai N, Nishitoba M, Sugiyama S and Magariyama Y 2005 {\em FEMS
  Microbiol. Lett.\/} {\bf 242} 221--225

\bibitem{magariyamaBiophysJ2005}
Magariyama Y, Ichiba M, Nakata K, Baba K, Ohtani T, Kudo S and Goto T 2005 {\em
  Biophys. J.\/} {\bf 88} 3648--3658

\bibitem{liPNAS2008}
Li G, Tam L~K and Tang J~X 2008 {\em Proc. Natl. Acad. Sci. USA\/} {\bf 105}
  18355--18359

\bibitem{rothschildNature1963}
Rothschild L 1963 {\em Nature\/} {\bf 198} 1221

\bibitem{cossonCellMotCytosk2003}
Cosson J, Huitorel P and Gagnon C 2003 {\em Cell Motil. Cytoskel.\/} {\bf 54}
  56--63

\bibitem{berkePRL2008}
Berke A~P, Turner L, Berg H~C and Lauga E 2008 {\em Phys. Rev. Lett.\/} {\bf
  101} 038102

\bibitem{blakeJEngMath1974}
Blake J~R and Chwang A~T 1974 {\em J. Eng. Math.\/} {\bf 8} 23--29

\bibitem{kearnsNatRevMicrobiol2010}
Kearns D~B 2010 {\em Nature Rev. Microbiol.\/} {\bf 8} 634--644

\bibitem{darntonBiophysJ2010}
Darnton N~C, Turner L, Rojevsky S and Berg H~C 2010 {\em Biophys. J.\/} {\bf
  98} 2082--2090

\bibitem{shinodaJBacter1977}
Shinoda S and Okamoto K 1977 {\em Journal of bacteriology\/} {\bf 129}
  1266--1271

\bibitem{mccarterJMMB1999}
McCarter L 1999 {\em J Mol. Microbiol. Biotechnol.\/} {\bf 1} 51--57

\bibitem{merinoFEMSMicroLett2006}
Merino S, Shaw J~G and Tom{\'a}s J~M 2006 {\em FEMS Microbiol. Lett.\/} {\bf
  263} 127--135

\bibitem{gavinMolecMB2002}
Gavin R, Rabaan A~A, Merino S, Tom{\'a}s J~M, Gryllos I and Shaw J~G 2002 {\em
  Mol Microbiol.\/} {\bf 43} 383--397

\bibitem{kirovJBact2002}
Kirov S~M, Tassell B~C, Semmler A~B~T, O'Donovan L~A, Rabaan A~A and Shaw J~G
  2002 {\em J. Bacteriol.\/} {\bf 184} 547--555

\bibitem{jonesJGMB1967}
Jones H~E and Park R~W~A 1967 {\em J. Gen. Microbiol.\/} {\bf 47} 369--378

\bibitem{harsheyAnnRevMB2003}
Harshey R~M 2003 {\em Ann. Rev. Microbiol.\/} {\bf 57} 249--273

\bibitem{kearnsMolMB2003}
Kearns D~B and Losick R 2003 {\em Mol. Microbiol.\/} {\bf 49} 581--590

\bibitem{julkowskaJBacter2005}
Julkowska D, Obuchowski M, Holland I~B and S{\'e}ror S~J 2005 {\em Journal of
  bacteriology\/} {\bf 187} 65--76

\bibitem{lindumJBacter1998}
Lindum P~W, Anthoni U, Christophersen C, Eberl L, Molin S and Givskov M 1998
  {\em J. Bacteriol.\/} {\bf 180} 6384--6388

\bibitem{eberlJBacter1999}
Eberl L, Molin S and Givskov M 1999 {\em J. Bacteriol.\/} {\bf 181} 1703--1712

\bibitem{copelandAEnvMB2010}
Copeland M~F, Flickinger S~T, Tuson H~H and Weibel D~B 2010 {\em Appl. Environ.
  Microbiol.\/} {\bf 76} 1241--1250

\bibitem{morrisonNat1966}
Morrison R~B and Scott A 1966 {\em Nature\/} {\bf 211} 255--257

\bibitem{jonesInfImm2004}
Jones B~V, Young R, Mahenthiralingam E and Stickler D~J 2004 {\em Infect.
  Immun.\/} {\bf 72} 3941--3950

\bibitem{zhangPNAS2010}
Zhang H~P, Be'er A, Florin E~L and Swinney H~L 2010 {\em Proc. Natl. Acad. Sci.
  USA\/} {\bf 107} 13626--13630

\bibitem{bonabeauPNAS1999}
Bonabeau E, Dagorn L and Fr\'eon P 1999 {\em Proc. Natl. Acad. Sci. USA\/} {\bf
  96} 4472--4477

\bibitem{narayanScience2007}
Narayan V, Ramaswamy S and Menon N 2007 {\em Science\/} {\bf 317} 105--108

\bibitem{dasJTheoBiol2012}
Das D, Das D and Prasad A 2012 {\em J. Theor. Biol.\/} {\bf 308} 96--104

\bibitem{ShimadaJPSJ2004}
Shimada H, Ikeda T, ichi Wakita J, Itoh H, Kurosu S, Hiramatsu F, Nakatsuchi M,
  Yamazaki Y, Matsuyama T and Matsushita M 2004 {\em J. Phys. Soc. Japan\/}
  {\bf 73} 1082--1089

\bibitem{hiramatsuMBEnv2005}
Hiramatsu F, Wakita J~i, Kobayashi N, Yamazaki Y, Matsushita M and Matsuyama T
  2005 {\em Microbes Environ.\/} {\bf 20} 120--125

\bibitem{arouhPRE2001}
Arouh S 2001 {\em Phys. Rev. E\/} {\bf 63} 031908

\bibitem{czirokPRE2001}
Czir{\'o}k A, Matsushita M and Vicsek T 2001 {\em Phys. Rev. E\/} {\bf 63}
  031915

\bibitem{goldingPhysica1998}
Golding I, Kozlovsky Y, Cohen I and Ben-Jacob E 1998 {\em Physica A\/} {\bf
  260} 510--554

\bibitem{wakanoPRL2003}
Wakano J~Y, Maenosono S, Komoto A, Eiha N and Yamaguchi Y 2003 {\em Phys. Rev.
  Lett.\/} {\bf 90} 258102--258102

\bibitem{benPhysA1997}
Ben-Jacob E, Cohen I, Czir{\'o}k A, Vicsek T and Gutnick D~L 1997 {\em Physica
  A\/} {\bf 238} 181--197

\bibitem{benNature1994}
Ben-Jacob E, Schochet O, Tenenbaum A, Cohen I, Czirok A and Vicsek T 1994 {\em
  Nature\/} {\bf 368} 46--49

\bibitem{benPRL1995}
Ben-Jacob E, Cohen I, Shochet O, Tenenbaum A, Czir{\'o}k A and Vicsek T 1995
  {\em Phys. Rev. Lett.\/} {\bf 75} 2899

\bibitem{wensinkJPhysCM2012}
Wensink H~H and L{\"o}wen H 2012 {\em J. Phys.: Condens. Matter\/} {\bf 24}
  464130

\bibitem{shroutMolMB2006}
Shrout J~D, Chopp D~L, Just C~L, Hentzer M, Givskov M and Parsek M~R 2006 {\em
  Mol. Microbiol.\/} {\bf 62} 1264--1277

\bibitem{verstraetenTrendsMB2008}
Verstraeten N, Braeken K, Debkumari B, Fauvart M, Fransaer J, Vermant J and
  Michiels J 2008 {\em Trends Microbiol.\/} {\bf 16} 496--506

\bibitem{parsekTrMB2005}
Parsek M~R and Greenberg E~P 2005 {\em Trends Microbiol.\/} {\bf 13} 27--33

\bibitem{hahnGene1997}
Hahn H~P 1997 {\em Gene\/} {\bf 192} 99--108

\bibitem{craigNatRevMB2004}
Craig L, Pique M~E and Tainer J~A 2004 {\em Nat. Rev. Microbiol.\/} {\bf 2}
  363--378

\bibitem{skerkerPNAS2001}
Skerker J~M and Berg H~C 2001 {\em Proc. Natl. Acad. Sci. USA\/} {\bf 98}
  6901--6904

\bibitem{craigCurrOpStrBio2008}
Craig L and Li J 2008 {\em Curr. Opin. Struc. Biol.\/} {\bf 18} 267--277

\bibitem{mattickARMB2002}
Mattick J~S 2002 {\em Annu. Rev. Microbiol.\/} {\bf 56} 289--314

\bibitem{merzNature2000}
Merz A~J, So M and Sheetz M~P 2000 {\em Nature\/} {\bf 407} 98--102

\bibitem{sunCurrBio2000}
Sun H, Zusman D~R and Shi W 2000 {\em Curr. Biol.\/} {\bf 10} 1143--1146

\bibitem{maierPNAS2002}
Maier B, Potter L, So M, Seifert H~S and Sheetz M~P 2002 {\em Proc. Natl. Acad.
  Sci. USA\/} {\bf 99} 16012--16017

\bibitem{clausenBiophysJ2009}
Clausen M, Koomey M and Maier B 2009 {\em Biophys. J.\/} {\bf 96} 1169--1177

\bibitem{clausenJBacter2009}
Clausen M, Jakovljevic V, S{\o}gaard-Andersen L and Maier B 2009 {\em J.
  Bacteriol.\/} {\bf 191} 4633--4638

\bibitem{holzPRL2010}
Holz C, Opitz D, Greune L, Kurre R, Koomey M, Schmidt M~A and Maier B 2010 {\em
  Phys. Rev. Lett.\/} {\bf 104} 178104

\bibitem{conradResMB2012}
Conrad J~C 2012 {\em Res. Microbiol.\/} {\bf 163} 619--629

\bibitem{mullerPNAS2008}
M{\"u}ller M~J~I, Klumpp S and Lipowsky R 2008 {\em Proc. Natl. Acad. Sci.
  USA\/} {\bf 105} 4609--4614

\bibitem{maratheNatComm2014}
Marathe R, Meel C, Schmidt N~C, Dewenter L, Kurre R, Greune L, Schmidt M~A,
  M{\"u}ller M~J~I, Lipowsky R, Maier B and Klumpp S 2014 {\em Nat. Comm.\/}
  {\bf 5}

\bibitem{jinPNAS2011}
Jin F, Conrad J~C, Gibiansky M~L and Wong G~C 2011 {\em Proc. Natl. Acad. Sci.
  USA\/} {\bf 108} 12617--12622

\bibitem{gibianskyScience2010}
Gibiansky M~L, Conrad J~C, Jin F, Gordon V~D, Motto D~A, Mathewson M~A, Stopka
  W~G, Zelasko D~C, Shrout J~D and Wong G~C~L 2010 {\em Science\/} {\bf 330}
  197--197

\bibitem{conradBiophysJ2011}
Conrad J~C, Gibiansky M~L, Jin F, Gordon V~D, Motto D~A, Mathewson M~A, Stopka
  W~G, Zelasko D~C, Shrout J~D and Wong G~C~L 2011 {\em Biophys. J.\/} {\bf
  100} 1608--1616

\bibitem{zaburdaevRMP2015}
Zaburdaev V, Denisov S and Klafter J 2015 {\em Rev. Mod. Phys.\/} {\bf 87} 483

\bibitem{montrollJMathPhys1965}
Montroll E~W and Weiss G~H 1965 {\em J. Math. Phys.\/} {\bf 6} 167--181

\bibitem{bouchaudPhysRep1990}
Bouchaud J~P and Georges A 1990 {\em Phys. Rep.\/} {\bf 195} 127--293

\bibitem{viswanathan2011}
Viswanathan G~M, Da~Luz M~G~E, Raposo E~P and Stanley H~E 2011 {\em The physics
  of foraging: an introduction to random searches and biological encounters\/}
  (Cambridge University Press)

\bibitem{mendez2013}
M{\'e}ndez V, Campos D and Bartumeus F 2013 {\em Stochastic foundations in
  movement ecology: anomalous diffusion, front propagation and random
  searches\/} (Springer Science \& Business Media)

\bibitem{taktikosPLOS2013}
Taktikos J, Stark H and Zaburdaev V 2013 {\em PLoS ONE\/} {\bf 8} e81936

\bibitem{korobkovaNat2004}
Korobkova E, Emonet T, Vilar J~M~G, Shimizu T~S and Cluzel P 2004 {\em
  Nature\/} {\bf 428} 574--578

\bibitem{burrowsARMB2012}
Burrows L~L 2012 {\em Annu. Rev. Microbiol.\/} {\bf 66} 493--520

\bibitem{klausenMolMB2003}
Klausen M, Aaes-J{\o}rgensen A, Molin S and Tolker-Nielsen T 2003 {\em Mol.
  Microbiol.\/} {\bf 50} 61--68

\bibitem{otooleMolecMB1998}
O'Toole G~A and Kolter R 1998 {\em Mol. Microbiol.\/} {\bf 30} 295--304

\bibitem{mengJBacter2005}
Meng Y, Li Y, Galvani C~D, Hao G, Turner J~N, Burr T~J and Hoch H~C 2005 {\em
  J. Bacteriol.\/} {\bf 187} 5560--5567

\bibitem{liMB2007}
Li Y, Hao G, Galvani C~D, Meng Y, De~La~Fuente L, Hoch H and Burr T~J 2007 {\em
  Microbiology\/} {\bf 153} 719--726

\bibitem{watnickMolMB1999}
Watnick P~I and Kolter R 1999 {\em Mol. Microbiol.\/} {\bf 34} 586--595

\bibitem{geigerNatRevMolCellBio2001}
Geiger B, Bershadsky A, Pankov R and Yamada K~M 2001 {\em Nat. Rev. Mol. Cell
  Bio.\/} {\bf 2} 793--805

\bibitem{nanPNAS2011}
Nan B, Chen J, Neu J~C, Berry R~M, Oster G and Zusman D~R 2011 {\em Proc. Natl.
  Acad. Sci. USA\/} {\bf 108} 2498--2503

\bibitem{zamirJCellSci2001}
Zamir E and Geiger B 2001 {\em J. Cell Sci.\/} {\bf 114} 3583--3590

\bibitem{mcbrideARMB2001}
McBride M~J 2001 {\em Annu. Rev. Microbiol.\/} {\bf 55} 49--75

\bibitem{mignotScience2007}
Mignot T, Shaevitz J~W, Hartzell P~L and Zusman D~R 2007 {\em Science\/} {\bf
  315} 853--856

\bibitem{peruaniPRL2012}
Peruani F, Starru{\ss} J, Jakovljevic V, S{\o}gaard-Andersen L, Deutsch A and
  B{\"a}r M 2012 {\em Phys. Rev. Lett.\/} {\bf 108} 098102

\bibitem{GinelliPRL2010}
Ginelli F, Peruani F, B\"ar M and Chat\'e H 2010 {\em Phys. Rev. Lett.\/} {\bf
  104}(18) 184502

\bibitem{peruaniPRE2006}
Peruani F, Deutsch A and B\"ar M 2006 {\em Phys. Rev. E\/} {\bf 74}(3) 030904

\bibitem{yangPRE2010}
Yang Y, Marceau V and Gompper G 2010 {\em Phys. Rev. E\/} {\bf 82}(3) 031904

\bibitem{martinezJBacter1999}
Mart{\'\i}nez A, Torello S and Kolter R 1999 {\em J. Bacteriol.\/} {\bf 181}
  7331--7338

\bibitem{rechtJBacter2001}
Recht J and Kolter R 2001 {\em J. Bacteriol.\/} {\bf 183} 5718--5724

\bibitem{chenJBacter2006}
Chen J~M, German G~J, Alexander D~C, Ren H, Tan T and Liu J 2006 {\em J.
  Bacteriol.\/} {\bf 188} 633--641

\bibitem{wuMBPath2009}
Wu C~w, Schmoller S~K, Bannantine J~P, Eckstein T~M, Inamine J~M, Livesey M,
  Albrecht R and Talaat A~M 2009 {\em Microb. Pathogenesis\/} {\bf 46} 222--230

\bibitem{mayaBioMedResInt2015}
Maya-Hoyos M, Leguizam{\'o}n J, Mari{\~n}o-Ram{\'\i}rez L and Soto C~Y 2015
  {\em BioMed Research International\/} {\bf 2015} 419549

\bibitem{daviesAppleEnvMB1993}
Davies D~G, Chakrabarty A~M and Geesey G~G 1993 {\em Appl. Environ.
  Microbiol.\/} {\bf 59} 1181--1186

\bibitem{christensen1985}
Christensen G~D, Simpson W~A and Beachey E~H 1985 Adhesion of bacteria to
  animal tissues {\em Bacterial adhesion: mechanisms and physiological
  significance\/} ed Savage D and Fletcher M (Springer US) pp 279--305

\bibitem{gristinaScience1987}
Gristina A~G 1987 {\em Science\/} {\bf 237} 1588--1595

\bibitem{marshallJGenMB1971}
Marshall K~C, Stout R and Mitchell R 1971 {\em J. Gen. Microbiol.\/} {\bf 68}
  337--348

\bibitem{dunneClinMBRev2002}
Dunne W~M 2002 {\em Clin. Microbiol. Rev.\/} {\bf 15} 155--166

\bibitem{carpentierJApplMB1993}
Carpentier B and Cerf O 1993 {\em J. Appl. Bacteriol.\/} {\bf 75} 499--511

\bibitem{juckerJBacter1996}
Jucker B~A, Harms H and Zehnder A~J 1996 {\em J. Bacteriol.\/} {\bf 178}
  5472--5479

\bibitem{derjaguin1941}
Derjaguin B and Landau L 1941 {\em Acta Phys-Chim USSR\/} {\bf 14} 331--354

\bibitem{derjaguinProgSurSci1993}
Derjaguin B and Landau L 1993 {\em Prog Surf. Sci.\/} {\bf 43} 30--59

\bibitem{verwey1948}
Verwey E~J~W and Overbeek J~T~G 1948 {\em Theory of the stability of lyophobic
  colloids\/} (Elsevier)

\bibitem{israelachviliAccChemRes1987}
Israelachvili J 1987 {\em Accounts Chem. Res.\/} {\bf 20} 415--421

\bibitem{chandlerNature2005}
Chandler D 2005 {\em Nature\/} {\bf 437} 640--647

\bibitem{israelachvili2011}
Israelachvili J~N 2011 {\em Intermolecular and surface forces\/} (Academic
  Press)

\bibitem{loskillLangmuir2012}
Loskill P, H\"ahl H, Thewes N, Kreis C~T, Bischoff M, Herrmann M and Jacobs K
  2012 {\em Langmuir\/} {\bf 28} 7242--7248

\bibitem{horiBiochemEngJ2010}
Hori K and Matsumoto S 2010 {\em Biochem. Eng. J.\/} {\bf 48} 424--434

\bibitem{dufreneNatRevMB2008}
Dufr{\^e}ne Y~F 2008 {\em Nat. Rev. Microbiol.\/} {\bf 6} 674--680

\bibitem{kotraJACS1999}
Kotra L~P, Golemi D, Amro N~A, Liu G~Y and Mobashery S 1999 {\em J. Am. Chem.
  Soc.\/} {\bf 121} 8707--8711

\bibitem{walkerLangmuir2004}
Walker S~L, Redman J~A and Elimelech M 2004 {\em Langmuir\/} {\bf 20}
  7736--7746

\bibitem{atabekJBacter2007}
Atabek A and Camesano T~A 2007 {\em J. Bacteriol.\/} {\bf 189} 8503--8509

\bibitem{juckerCollSurfB1997}
Jucker B~A, Harms H, Hug S~J and Zehnder A~J~B 1997 {\em Colloids Surf. B\/}
  {\bf 9} 331--343

\bibitem{makinMB1996}
Makin S~A and Beveridge T~J 1996 {\em Microbiology\/} {\bf 142} 299--307

\bibitem{genevauxArchMB1999}
Genevaux P, Bauda P, DuBow M~S and Oudega B 1999 {\em Arch. Microbiol.\/} {\bf
  172} 1--8

\bibitem{abuEnvSciTech2003}
Abu-Lail N~I and Camesano T~A 2003 {\em Environ. Sci. \& Technol.\/} {\bf 37}
  2173--2183

\bibitem{thewesSoftMatt2015}
Thewes N, Thewes A, Loskill P, Peisker H, Bischoff M, Herrmann M, Santen L and
  Jacobs K 2015 {\em Soft Matter\/} {\bf XXX} XXX

\bibitem{tsunedaFEMSMBLett2003}
Tsuneda S, Aikawa H, Hayashi H, Yuasa A and Hirata A 2003 {\em FEMS Microbiol.
  Lett.\/} {\bf 223} 287--292

\bibitem{vanLoosdrechtAquaSci1990}
van Loosdrecht M~C~M, Norde W, Lyklema J and Zehnder A~J~B 1990 {\em Aquat.
  Sci.\/} {\bf 52} 103--114

\bibitem{azeredoBiofoul2000}
Azeredo J and Oliveira R 2000 {\em Biofouling\/} {\bf 16} 59--67

\bibitem{vanLoosdrechtExperientia1990}
van Loosdrecht M~C~M and Zehnder A~J~B 1990 {\em Experientia\/} {\bf 46}
  817--822

\bibitem{azeredoCollSurfB1999}
Azeredo J, Visser J and Oliveira R 1999 {\em Colloids Surf. B\/} {\bf 14}
  141--148

\bibitem{rosenberg1986}
Rosenberg M and Kjelleberg S 1986 Hydrophobic interactions: role in bacterial
  adhesion {\em Advances in microbial ecology\/} (Springer US) pp 353--393

\bibitem{vanLoosdrechtApplEnvMB1987}
van Loosdrecht M~C~M, Lyklema J, Norde W, Schraa G and Zehnder A~J~B 1987 {\em
  Appl. Environ. Microbiol.\/} {\bf 53} 1893--1897

\bibitem{thewesBeilstein2014}
Thewes N, Loskill P, Jung P, Peisker H, Bischoff M, Herrmann M and Jacobs K
  2014 {\em Beilstein J. Nanotechnol.\/} {\bf 5} 1501--1512

\bibitem{stillingerJSolChem1973}
Stillinger F~H 1973 {\em J. Solution Chem.\/} {\bf 2} 141--158

\bibitem{GiovambattistaPRE2006}
Giovambattista N, Rossky P~J and Debenedetti P~G 2006 {\em Phys. Rev. E\/} {\bf
  73}(4) 041604

\bibitem{giovambattistaPNAS2008}
Giovambattista N, Lopez C~F, Rossky P~J and Debenedetti P~G 2008 {\em Proc.
  Natl. Acad. Sci. USA\/} {\bf 105} 2274--2279

\bibitem{tsangPNAS2006}
Tsang P~H, Li G, Brun Y~V, Freund L~B and Tang J~X 2006 {\em Proc. Natl. Acad.
  Sci. USA\/} {\bf 103} 5764--5768

\bibitem{ofekNature1977}
Ofek I, Mirelman D and Sharon N 1977 {\em Nature\/} {\bf 265} 623--625

\bibitem{hansonNature1988}
Hanson M~S and Brinton~Jr C~C 1988 {\em Nature\/} {\bf 332} 265--268

\bibitem{kuehnNature1992}
Kuehn M~J, Heuser J, Normark S and Hultgren S~J 1992 {\em Nature\/} {\bf 356}
  252--255

\bibitem{kangScience2007}
Kang H~J, Coulibaly F, Clow F, Proft T and Baker E~N 2007 {\em Science\/} {\bf
  318} 1625--1628

\bibitem{nobbsMBMolBiolRev2009}
Nobbs A~H, Lamont R~J and Jenkinson H~F 2009 {\em Microbiol. Mol. Biol. Rev.\/}
  {\bf 73} 407--450

\bibitem{veenstraJBActer1996}
Veenstra G~J~C, Cremers F~F~M, van Dijk H and Fleer A 1996 {\em J.
  Bacteriol.\/} {\bf 178} 537--541

\bibitem{heilmannMolMB1997}
Heilmann C, Hussain M, Peters G and Gotz F 1997 {\em Mol. Microbiol.\/} {\bf
  24} 1013--1024

\bibitem{heilmannMB2003}
Heilmann C, Thumm G, Chhatwal G~S, Hartleib J, Uek{\"o}tter A and Peters G 2003
  {\em Microbiology\/} {\bf 149} 2769--2778

\bibitem{loebAdvChem1975}
Loeb G~I and Neihof R~A 1975 {\em Adv. Chem.\/} {\bf 145} 319--335

\bibitem{bosFEMSMBRev1999}
Bos R, Van~der Mei H~C and Busscher H~J 1999 {\em FEMS Microbiol. Rev.\/} {\bf
  23} 179--230

\bibitem{schneiderCollSurfB1994}
Schneider R~P and Marshall K~C 1994 {\em Colloids Surf. B\/} {\bf 2} 387--396

\bibitem{schneiderJCollInterf1996}
Schneider R~P 1996 {\em J. Colloid Interface Sci.\/} {\bf 182} 204--213

\bibitem{deKerchove2007ApplEnvMB}
de~Kerchove A~J and Elimelech M 2007 {\em Appl. Environ. Microbiol.\/} {\bf 73}
  5227--5234

\bibitem{loriteJCollInterfSci2011}
Lorite G~S, Rodrigues C~M, De~Souza A~A, Kranz C, Mizaikoff B and Cotta M~A
  2011 {\em J. Colloid Interface Sci.\/} {\bf 359} 289--295

\bibitem{larson1999}
Larson R~G 1999 {\em The structure and rheology of complex fluids\/} (Oxford
  University Press New York)

\bibitem{klapperBiotechBioeng2002}
Klapper I, Rupp C, Cargo R, Purvedorj B and Stoodley P 2002 {\em Biotechnol.
  Bioeng.\/} {\bf 80} 289--296

\bibitem{towlerBiofoul2003}
Towler B~W, Rupp C~J, Cunningham A~L~B and Stoodley P 2003 {\em Biofouling\/}
  {\bf 19} 279--285

\bibitem{vinogradovBiofilms2004}
Vinogradov A~M, Winston M, Rupp C~J and Stoodley P 2004 {\em Biofilms\/} {\bf
  1} 49--56

\bibitem{shawPRL2004}
Shaw T, Winston M, Rupp C~J, Klapper I and Stoodley P 2004 {\em Phys. Rev.
  Lett.\/} {\bf 93}(9) 098102

\bibitem{ruppApplEnvMB2005}
Rupp C~J, Fux C~A and Stoodley P 2005 {\em Appl. Environ. Microbiol.\/} {\bf
  71} 2175--2178

\bibitem{lauBiophysJ2009}
Lau P~C~Y, Dutcher J~R, Beveridge T~J and Lam J~S 2009 {\em Biophys. J.\/} {\bf
  96} 2935--2948

\bibitem{hohneLangmuir2009}
Hohne D~N, Younger J~G and Solomon M~J 2009 {\em Langmuir\/} {\bf 25}
  7743--7751

\bibitem{lielegSoftMatt2011}
Lieleg O, Caldara M, Baumg{\"a}rtel R and Ribbeck K 2011 {\em Soft Matter\/}
  {\bf 7} 3307--3314

\bibitem{billingsRepPrgPhys2015}
Billings N, Birjiniuk A, Samad T~S, Doyle P~S and Ribbeck K 2015 {\em Rep.
  Prog. Phys.\/} {\bf 78} 036601

\bibitem{korstgensJMBMeth2001}
K{\"o}rstgens V, Flemming H~C, Wingender J and Borchard W 2001 {\em J.
  Microbiol. Meth.\/} {\bf 46} 9--17

\bibitem{wlokaCollPolySci2004}
Wloka M, Rehage H, Flemming H~C and Wingender J 2004 {\em Colloid Polym.
  Sci.\/} {\bf 282} 1067--1076

\bibitem{joseph2013}
Joseph D~D 2013 {\em Fluid dynamics of viscoelastic liquids\/} (Springer
  Science \& Business Media)

\bibitem{morozov2015}
Morozov A and Spagnolie S~E 2015 Introduction to complex fluids {\em Complex
  Fluids in Biological Systems\/} (Springer) pp 3--52

\bibitem{stoodleyJIndMBBiotech2002}
Stoodley P, Cargo R, Rupp C~J, Wilson S and Klapper I 2002 {\em J. Ind.
  Microbiol. Biot.\/} {\bf 29} 361--367

\bibitem{cheongRheolActa2009}
Cheong F~C, Duarte S, Lee S~H and Grier D~G 2009 {\em Rheol. Acta\/} {\bf 48}
  109--115

\bibitem{wilemanBioresTech2012}
Wileman A, Ozkan A and Berberoglu H 2012 {\em Bioresource Technol.\/} {\bf 104}
  432--439

\bibitem{vandykApplEnvMB1994}
Van~Dyk T~K, Majarian W~R, Konstantinov K~B, Young R~M, Dhurjati P~S and
  La{R}ossa R~A 1994 {\em Appl. Environ. Microbiol.\/} {\bf 60} 1414--1420

\bibitem{ptitsynApplEnvMB1997}
Ptitsyn L~R, Horneck G, Komova O, Kozubek S, Krasavin E~A, Bonev M and Rettberg
  P 1997 {\em Appl Environ. Microbiol.\/} {\bf 63} 4377--4384

\bibitem{rusconiInterface2010}
Rusconi R, Lecuyer S, Guglielmini L and Stone H~A 2010 {\em J. R. Soc.
  Interface\/} {\bf 7} 1293--1299

\bibitem{rusconiBiophysJ2011}
Rusconi R, Lecuyer S, Autrusson N, Guglielmini L and Stone H~A 2011 {\em
  Biophys. J.\/} {\bf 100} 1392--1399

\bibitem{drescherPNAS2013}
Drescher K, Shen Y, Bassler B~L and Stone H~A 2013 {\em Proc. Natl. Acad. Sci.
  USA\/} {\bf 110} 4345--4350

\bibitem{stoodleyEnvMB1999}
Stoodley P, Lewandowski Z, Boyle J~D and Lappin-Scott H~M 1999 {\em Environ
  Microbiol.\/} {\bf 1} 447--455

\bibitem{thomasPhilTransRoySocLond2013}
Thomas K, Herminghaus S, Porada H and Goehring L 2013 {\em Phil. Trans. R. Soc.
  A\/} {\bf 371} 20120362

\bibitem{walkerInfImm2005}
Walker T~S, Tomlin K~L, Worthen G~S, Poch K~R, Lieber J~G, Saavedra M~T,
  Fessler M~B, Malcolm K~C, Vasil M~L and Nick J~A 2005 {\em Infect. Immun.\/}
  {\bf 73} 3693--3701

\bibitem{andesInfImm2004}
Andes D, Nett J, Oschel P, Albrecht R, Marchillo K and Pitula A 2004 {\em
  Infect. Immun.\/} {\bf 72} 6023--6031

\bibitem{phillipsAICHEJ1989}
Phillips R~J, Deen W~M and Brady J~F 1989 {\em AIChE J.\/} {\bf 35} 1761--1769

\bibitem{masaroProgPolySci1999}
Masaro L and Zhu X~X 1999 {\em Prog. Polym. Sci.\/} {\bf 24} 731--775

\bibitem{amsdenMacromol1998}
Amsden B 1998 {\em Macromolecules\/} {\bf 31} 8382--8395

\bibitem{woolSoftMatt2008}
Wool R~P 2008 {\em Soft Matter\/} {\bf 4} 400--418

\bibitem{wuProgPolymSci2008}
Wu D~Y, Meure S and Solomon D 2008 {\em Prog. Polym. Sci.\/} {\bf 33} 479--522

\bibitem{stewartSciRep2015}
Stewart E~J, Ganesan M, Younger J~G and Solomon M~J 2015 {\em Sci. Rep.\/} {\bf
  5}

\bibitem{simon1991}
Simon H~A 1991 The architecture of complexity {\em Facets of Systems Science\/}
  ({\em International Federation for Systems Research International Series on
  Systems Science and Engineering\/} vol~7) (Springer US) pp 457--476

\bibitem{morgenrothRevEnvSciBiotech2009}
Morgenroth E and Milferstedt K 2009 {\em Rev. Environ. Sci. Biotechnol.\/} {\bf
  8} 203--208

\bibitem{vertesAnalChem2012}
Vertes A, Hitchins V and Phillips K~S 2012 {\em Anal Chem.\/} {\bf 84}
  3858--3866

\bibitem{karimiLOC2015}
Karimi A, Karig D, Kumar A and Ardekani A~M 2015 {\em Lab Chip\/} {\bf 15}
  23--42

\bibitem{delafuente2007}
De~La~Fuente L, Montanes E, Meng Y, Li Y, Burr T~J, Hoch H~C and Wu M 2007 {\em
  Appl. Environ. Microbiol.\/} {\bf 73} 2690--2696

\bibitem{baharFEMSMBLett2010}
Bahar O, De~La~Fuente L and Burdman S 2010 {\em FEMS Microbiol. Lett.\/} {\bf
  312} 33--39

\bibitem{wrightLabChip2014}
Wright E, Neethirajan S, Warriner K, Retterer S and Srijanto B 2014 {\em Lab
  Chip\/} {\bf 14} 938--946

\bibitem{leeBiomedDev2008}
Lee J~H, Kaplan J~B and Lee W~Y 2008 {\em Biomed. Microdevices\/} {\bf 10}
  489--498

\bibitem{kimLabChip2010}
Kim K~P, Kim Y~G, Choi C~H, Kim H~E, Lee S~H, Chang W~S and Lee C~S 2010 {\em
  Lab Chip\/} {\bf 10} 3296--3299

\bibitem{richterLabChip2007}
Richter L, Stepper C, Mak A, Reinthaler A, Heer R, Kast M, Br{\"u}ckl H and
  Ertl P 2007 {\em Lab Chip\/} {\bf 7} 1723--1731

\bibitem{hongNatCom2012}
Hong S~H, Hegde M, Kim J, Wang X, Jayaraman A and Wood T~K 2012 {\em Nature
  Comm.\/} {\bf 3} 613

\bibitem{groismanNatMeth2005}
Groisman A, Lobo C, Cho H, Campbell J~K, Dufour Y~S, Stevens A~M and Levchenko
  A 2005 {\em Nat. Methods\/} {\bf 2} 685--689

\bibitem{valiei2LabChip012}
Valiei A, Kumar A, Mukherjee P~P, Liu Y and Thundat T 2012 {\em Lab Chip\/}
  {\bf 12} 5133--5137

\bibitem{liBiotechBioeng2011}
Li Z, Zhang Y, LeDuc P~R and Gregory K~B 2011 {\em Biotechnol. Bioeng.\/} {\bf
  108} 2061--2069

\bibitem{nealsonAnnRevMB1994}
Nealson K~H and Saffarini D 1994 {\em Annu. Rev. Microbiol.\/} {\bf 48}
  311--343

\bibitem{gorbyPNAS2006}
Gorby Y~A, Yanina S, McLean J~S, Rosso K~M, Moyles D, Dohnalkova A, Beveridge
  T~J, Chang I~S, Kim B~H, Kim K~S, Culley D~E, Reed S~B, Romine M~F, Saffarini
  D~A, Hill E~A, Shi L, Elias D~A, Kennedy D~W, Pinchuk G, Watanabe K, Ishii S,
  Logan B, Nealson K~H and Fredrickson J~K 2006 {\em Proc. Natl. Acad. Sci.
  USA\/} {\bf 103} 11358--11363

\bibitem{schkolnikPLOS2015}
Schkolnik G, Schmidt M, Mazza M~G, Harnisch F and Musat N 2015 {\em PLoS One\/}
  {\bf 10} e0145871

\bibitem{weibelNAtRevMB2007}
Weibel D~B, DiLuzio W~R and Whitesides G~M 2007 {\em Nat. Rev. Microbiol.\/}
  {\bf 5} 209--218

\bibitem{holScience2014}
Hol F~J~H and Dekker C 2014 {\em Science\/} {\bf 346} 1251821

\bibitem{rusconiAnnRevBiophys2014}
Rusconi R, Garren M and Stocker R 2014 {\em Annu. Rev. Biophys.\/} {\bf 43} 65

\bibitem{friedlanderPNAS2013}
Friedlander R~S, Vlamakis H, Kim P, Khan M, Kolter R and Aizenberg J 2013 {\em
  Proc. Natl. Acad. Sci. USA\/} {\bf 110} 5624--5629

\bibitem{nadellFEMSMBRev2009}
Nadell C~D, Xavier J~B and Foster K~R 2009 {\em FEMS Microbiol. Rev.\/} {\bf
  33} 206--224

\bibitem{keymerPNAS2006}
Keymer J~E, Galajda P, Muldoon C, Park S and Austin R~H 2006 {\em Proc. Natl.
  Acad. Sci. USA\/} {\bf 103} 17290--17295

\bibitem{kimPNAS2008}
Kim H~J, Boedicker J~Q, Choi J~W and Ismagilov R~F 2008 {\em Proc. Natl. Acad.
  Sci. USA\/} {\bf 105} 18188--18193

\bibitem{connellPNAS2013}
Connell J~L, Ritschdorff E~T, Whiteley M and Shear J~B 2013 {\em Proc. Natl.
  Acad. Sci. USA\/} {\bf 110} 18380--18385

\bibitem{holBMCResNot2015}
Hol F~J, Galajda P, Woolthuis R~G, Dekker C and Keymer J~E 2015 {\em BMC Res.
  Notes\/} {\bf 8} 245

\bibitem{herbertJGenMB1956}
Herbert D, Elsworth R and Telling R~C 1956 {\em J. Gen. Microbiol.\/} {\bf 14}
  601--622

\bibitem{vanUden1967}
Van~Uden N 1967 {\em Arch. Mikrobiol.\/} {\bf 58} 155--168

\bibitem{monod1942}
Monod J 1942 {\em Recherches sur la croissance des cultures bacteriennes\/}
  (Paris, Hermann et Cie)

\bibitem{rittmannBiotechBioeng1980}
Rittmann B~E and McCarty P~L 1980 {\em Biotechnol. Bioeng.\/} {\bf 22}
  2343--2357

\bibitem{kisselJEnvEng1984}
Kissel J~C, McCarty P~L and Street R~L 1984 {\em J. Environ. Eng.\/} {\bf 110}
  393--411

\bibitem{wannerBiotechBioeng1986}
Wanner O and Gujer W 1986 {\em Biotechnol. Bioeng.\/} {\bf 28} 314--328

\bibitem{chaudhryChemEng1998}
Chaudhry M~A~S and Beg S~A 1998 {\em Chem. Eng. Technol.\/} {\bf 21} 701--710

\bibitem{costertonARMB1995}
Costerton J~W, Lewandowski Z, Caldwell D~E, Korber D~R and Lappin-Scott H~M
  1995 {\em Annu. Rev. Microbiol.\/} {\bf 49} 711--745

\bibitem{eberlJTheoMed2001}
Eberl H~J, Parker D~F and Van~Loosdrecht M~C~M 2001 {\em J. Theor. Med.\/} {\bf
  3} 161--175

\bibitem{picioreanuBiotechBioeng1998}
Picioreanu C, Van~Loosdrecht M~C~M and Heijnen J~J 1998 {\em Biotechnol.
  Bioeng.\/} {\bf 58} 101--116

\bibitem{eberlchemEngSc2000}
Eberl H~J, Picioreanu C, Heijnen J~J and Van~Loosdrecht M~C~M 2000 {\em Chem.
  Eng. Sci.\/} {\bf 55} 6209--6222

\bibitem{nogueraWatSciTech1999}
Noguera D~R, Pizarfo G, Stahl D~A and Rittmann B~E 1999 {\em Water Sci.
  Technol.\/} {\bf 39} 123--130

\bibitem{fuxJBacter2004}
Fux C~A, Wilson S and Stoodley P 2004 {\em J. Bacteriol.\/} {\bf 186}
  4486--4491

\bibitem{klapperPRE2006}
Klapper I and Dockery J 2006 {\em Phys. Rev. E\/} {\bf 74} 031902

\bibitem{klapperSIAMRev2010}
Klapper I and Dockery J 2010 {\em SIAM Rev.\/} {\bf 52} 221--265

\bibitem{wangSolStComm2010}
Wang Q and Zhang T 2010 {\em Solid State Commun.\/} {\bf 150} 1009--1022

\bibitem{tierraJRoySocInterf2015}
Tierra G, Pavissich J~P, Nerenberg R, Xu Z and Alber M~S 2015 {\em J. R. Soc.
  Interface\/} {\bf 12} 20150045

\bibitem{lindleyPRE2012}
Lindley B, Wang Q and Zhang T 2012 {\em Phys. Rev. E\/} {\bf 85}(3) 031908

\bibitem{fujikawaJPhysSocJap1989}
Fujikawa H and Matsushita M 1989 {\em J. Phys. Soc. Japan\/} {\bf 58}
  3875--3878

\bibitem{matsushitaPhysA1990}
Matsushita M and Fujikawa H 1990 {\em Physica A\/} {\bf 168} 498--506

\bibitem{fujikawaFEMSMBEco1994}
Fujikawa H 1994 {\em FEMS Microbiol. Ecol.\/} {\bf 13} 159--168

\bibitem{wittenPRL1981}
Witten T~A and Sander L~M 1981 {\em Phys. Rev. Lett.\/} {\bf 47} 1400--1403

\bibitem{eden1961}
Eden M 1961 A two-dimensional growth process {\em Proceedings of the Fourth
  Berkeley Symposium on Mathematical Statistics and Probability, Volume 4:
  Contributions to Biology and Problems of Medicine\/} (Berkeley, Calif.:
  University of California Press) pp 223--239

\bibitem{benPRL1986}
Ben-Jacob E, Deutscher G, Garik P, Goldenfeld N~D and Lareah Y 1986 {\em Phys.
  Rev. Lett.\/} {\bf 57} 1903--1906

\bibitem{meakin1998fractals}
Meakin P 1998 {\em Fractals, scaling and growth far from equilibrium\/} vol~5
  (Cambridge University Press)

\bibitem{stanley2012}
Stanley H~E and Ostrowsky N 2012 {\em Random fluctuations and pattern growth:
  experiments and models\/} vol 157 (Springer Science \& Business Media)

\bibitem{hayakawaPRL1997}
Hayakawa Y and Sato S 1997 {\em Phys. Rev. Lett.\/} {\bf 79} 95--98

\bibitem{halleyOikos1994}
Halley J~M, Comins H~N, Lawton J~H and Hassell M~P 1994 {\em Oikos\/} {\bf 70}
  435--442

\bibitem{wimpennyFEMSMBEco1997}
Wimpenny J~W~T and Colasanti R 1997 {\em FEMS Microbiol. Ecol.\/} {\bf 22}
  1--16

\bibitem{hermanowiczWatSciTech1999}
Hermanowicz S~W 1999 {\em Water Sci. Technol.\/} {\bf 39} 107--114

\bibitem{picioreanuBiotechBioeng2000}
Picioreanu C, Van~Loosdrecht M~C~M and Heijnen J~J 2000 {\em Biotechnol.
  Bioeng.\/} {\bf 69} 504--515

\bibitem{pizarroJEnvEng2001}
Pizarro G, Griffeath D and Noguera D~R 2001 {\em J. Environ. Eng.\/} {\bf 127}
  782--789

\bibitem{tangWatRes2013}
Tang Y and Valocchi A~J 2013 {\em Water Research\/} {\bf 47} 5729--5742

\bibitem{wolframNature1984}
Wolfram S 1984 {\em Nature\/} {\bf 311} 419--424

\bibitem{laspidouWatRes2004a}
Laspidou C~S and Rittmann B~E 2004 {\em Water Res.\/} {\bf 38} 3349--3361

\bibitem{laspidouWatRes2004b}
Laspidou C~S and Rittmann B~E 2004 {\em Water Res.\/} {\bf 38} 3362--3372

\bibitem{laspidouWatRes2002b}
Laspidou C~S and Rittmann B~E 2002 {\em Water Res.\/} {\bf 36} 2711--2720

\bibitem{laspidouWatRes2002a}
Laspidou C~S and Rittmann B~E 2002 {\em Water Res.\/} {\bf 36} 1983--1992

\bibitem{kreftMB1998}
Kreft J~U, Booth G and Wimpenny J~W~T 1998 {\em Microbiology\/} {\bf 144}
  3275--3287

\bibitem{kreftMB2001}
Kreft J~U, Picioreanu C, Wimpenny J~W~T and van Loosdrecht M~C~M 2001 {\em
  Microbiology\/} {\bf 147} 2897--2912

\bibitem{kreftWatSciTech2001}
Kreft J~U and Wimpenny J~W~T 2001 {\em Water Sci. Technol.\/} {\bf 43} 135--142

\bibitem{picioreanuApplEnvMB2004}
Picioreanu C, Kreft J~U and van Loosdrecht M~C~M 2004 {\em Appl. Environ.
  Microbiol.\/} {\bf 70} 3024--3040

\bibitem{xavierEnvMB2005}
Xavier J~B, Picioreanu C and Van~Loosdrecht M 2005 {\em Environ. Microbiol.\/}
  {\bf 7} 1085--1103

\bibitem{picioreanuWatSciTech2007}
Picioreanu C, Kreft J~U, Klausen M, Haagensen J~A~J, Tolker-Nielsen T and Molin
  S 2007 {\em Water Sci. Technol.\/} {\bf 55} 337--343

\bibitem{vonderSchulenburgAICHEJ2009}
Graf von~der Schulenburg D~A, Pintelon T~R~R, Picioreanu C, Van~Loosdrecht
  M~C~M and Johns M~L 2009 {\em AIChE J.\/} {\bf 55} 494--504 ISSN 1547-5905

\bibitem{lardonEnvMB2011}
Lardon L~A, Merkey B~V, Martins S, D{\"o}tsch A, Picioreanu C, Kreft J~U and
  Smets B~F 2011 {\em Environ. Microbiol.\/} {\bf 13} 2416--2434

\bibitem{alpkvistBiotechBioeng2006}
Alpkvist E, Picioreanu C, van Loosdrecht M and Heyden A 2006 {\em Biotechnol.
  Bioeng.\/} {\bf 94} 961--979

\bibitem{laspidouDesal2010}
Laspidou C~S, Kungolos A and Samaras P 2010 {\em Desalination\/} {\bf 250}
  390--394

\bibitem{kreftPNAS2013}
Kreft J~U, Plugge C~M, Grimm V, Prats C, Leveau J~H~J, Banitz T, Baines S,
  Clark J, Ros A, Klapper I, Topping C~J, Field A~J, Schuler A, Litchman E and
  Hellweger F~L 2013 {\em Proc. Natl. Acad. Sci. USA\/} {\bf 110} 18027--18028

\bibitem{alpkvistWatSciTech2007}
Alpkvist E and Klapper I 2007 {\em Water Sci. Technol.\/} {\bf 55} 265--273

\bibitem{ehretJRoySocInterf2013}
Ehret A~E and B{\"o}l M 2013 {\em J. R. Soc. Interface\/} {\bf 10} 20120676

\end{thebibliography}
\providecommand{\newblock}{}

\end{document}